\documentclass[11pt,a4paper]{article}

\usepackage{slashed}
\usepackage{hyperref}
\usepackage{color}
\usepackage{amsthm}
\usepackage{mathtools}
\usepackage[normalem]{ulem} 
\usepackage{graphicx}
\usepackage{amssymb,latexsym,cite}
\usepackage{amsmath}
\usepackage{amsfonts}
\usepackage{mathrsfs}
\usepackage{bbm}
\usepackage{bm}
\usepackage[T1]{fontenc}
\usepackage{multicol}
\usepackage{multirow}

\usepackage{tikz-cd}

\usepackage{simplewick}

\usepackage[matrix,arrow,color]{xy}

\usepackage{enumitem}

\def\slasha#1{\setbox0=\hbox{$#1$}#1\hskip-\wd0\hbox to\wd0{\hss\sl/\/\hss}}

\def\periodb#1{\setbox0=\hbox{$#1$}#1\hskip-\wd0\hbox to\wd0{-}}

\newcommand{\unit}{\mathbbm{1}}   			
\newcommand{\e}{{\mathrm{e}}}   			
\newcommand{\ii}{{\mathrm{i}}}   			
\newcommand{\id}{\mathrm{id}}   			

\newcommand{\CA}{\mathcal{A}}    			

\newcommand{\CCL}{\mathscr{L}}

\newcommand{\CCD}{\mathscr{D}}

\newcommand{\CF}{\mathcal{F}}

\newcommand{\CG}{\mathcal{G}}

\newcommand{\CR}{\mathcal{R}}
\newcommand{\CS}{\mathcal{S}}
\newcommand{\CCS}{\mathscr{S}}

\newcommand{\CZ}{\mathcal{Z}}

\newcommand{\frv}{\mathfrak{v}}

\newcommand{\mink}{{\mathbbm{R}^{1,3}}}

\newcommand{\mbf}[1]{{\boldsymbol {#1} }}

\newcommand{\Sym}{{\mathrm{Sym}}}

\newcommand{\BV}{{\textrm{\tiny BV}}}
\newcommand{\BVL}{{\mathsf\Delta}_{\textrm{\tiny BV}}}

\newcommand{\FR}{\mathbbm{R}}     			
\newcommand{\FC}{\mathbbm{C}}     			
\newcommand{\RZ}{\mathbbm{Z}}     			

\newcommand{\dd}{\mathrm{d}}     			

\newcommand{\eand}{{\qquad\mbox{and}\qquad}}     		

\newcommand{\iso}{\mathfrak{iso}}

\newcommand{\sU}{\mathsf{U}}     			

\newcommand{\sH}{\mathsf{H}}

\newcommand{\sP}{\mathsf{P}}

\newcommand{\sSpin}{\mathsf{Spin}}

\newcommand{\sEnd}{\mathrm{End}}

\newcommand{\comment}[1]{}     				
\def\tyng(#1){\hbox{\tiny$\yng(#1)$}}			
\def\tyoung(#1){\hbox{\tiny$\young(#1)$}}			

\newcommand{\beq}{\begin{eqnarray}}
\newcommand{\eeq}{\end{eqnarray}}

\newcommand{\sgreen}{\mathsf{G}}
\newcommand{\mgreen}{\mathsf{D}}
\newcommand{\fgreen}{\mathsf{S}}
\newcommand{\sPi}{\mathsf{\Pi}}

\newcommand{\sfp}{{\sf p}}

\newcommand{\sfh}{{\sf h}}

\newcommand{\sff}{{\sf f}}

\newcommand{\tte}{\mathtt{e}}
\newcommand{\ttv}{\mathtt{v}}
\newcommand{\ttu}{\mathtt{u}}

\newcommand{\sfR}{\mathsf{R}}

\definecolor{outrageousorange}{rgb}{1.0, 0.43, 0.29}

\newenvironment{myitemize}{\begin{itemize}[itemsep=-0.05cm, leftmargin=*, topsep=0.1cm]}{\end{itemize}}

\newcommand{\Tr}{\mathrm{Tr}}

\theoremstyle{plain}

\theoremstyle{definition}

\theoremstyle{remark}

\newcommand{\nn}{\nonumber}

\newcommand{\dsf}{{\mathsf{d}}}

\def\RR{{\mathcal R}}

\def\d{{\rm d}}
\def\swone{{\textrm{\tiny $(1)$}}}
\def\swtwo{{\textrm{\tiny $(2)$}}}
\def\swthree{{\textrm{\tiny $(3)$}}}
\def\swzero{{\textrm{\tiny $(0)$}}}

\def\swk{{\textrm{\tiny $(k)$}}}

\def\hodge{{\textrm{\tiny H}}}

\numberwithin{equation}{section}
\catcode`@=12

\begin{document}


\renewcommand{\thefootnote}{\fnsymbol{footnote}}

\begin{flushright}
		\small
		{\sf EMPG--23--02}
	\end{flushright}
	
\vspace{1cm}

\begin{center}

{\LARGE{\bf Braided Quantum Electrodynamics}}

\vspace*{1.5cm}

\baselineskip=14pt
		
		{\large\bf Marija Dimitrijevi\'c \'Ciri\'c}${}^{\,(a)\,,\,}$\footnote{Email: \ {\tt
				dmarija@ipb.ac.rs}} \ \ \ \ \ {\large\bf Nikola Konjik}${}^{\,(a)\,,\,}$\footnote{Email: \ {\tt
				konjik@ipb.ac.rs}} \\[2mm] {\large\bf Voja Radovanovi\'c}${}^{\,(a)\,,\,}$\footnote{Email: \ {\tt
				rvoja@ipb.ac.rs}} \ \ \ \ \ {\large\bf Richard
			J. Szabo}${}^{\,(b)\,,\,}$\footnote{Email: \ {\tt R.J.Szabo@hw.ac.uk}} 
		\\[6mm]
		
		\noindent  ${}^{(a)}$ {\it Faculty of Physics, University of
			Belgrade}\\ {\it Studentski trg 12, 11000 Beograd, Serbia}
		\\[3mm]
		
		\noindent  ${}^{(b)}$ {\it Department of Mathematics, Heriot-Watt University\\ Colin Maclaurin Building,
			Riccarton, Edinburgh EH14 4AS, U.K.}\\ and {\it Maxwell Institute for
			Mathematical Sciences, Edinburgh, U.K.} \\ and {\it Higgs Centre
			for Theoretical Physics, Edinburgh, U.K.}
		\\[30mm]

\end{center}

\begin{abstract}
\noindent
The homotopy algebraic formalism of braided noncommutative field theory is used to define the explicit example of braided electrodynamics, that is, $\sU(1)$ gauge theory minimally coupled to a Dirac fermion. We construct the braided $L_\infty$-algebra of this field theory and obtain the braided equations of motion, action functional and conserved matter current. The modifications of the electric charge conservation law due to the braided noncommutative deformation are described. We develop a braided generalization of Wick's theorem, and use it to compute correlation functions of the braided quantum field theory using homological perturbation theory. Our putative calculations indicate that the braided theory does not contain the non-planar Feynman diagrams of conventional noncommutative quantum field theory, and that correlators do not exhibit UV/IR mixing.
\end{abstract}

\vspace*{2cm}



{\baselineskip=10pt
\tableofcontents
}

\setcounter{footnote}{0}
\renewcommand{\thefootnote}{\arabic{footnote}}

\bigskip

\setcounter{page}{1}
\newcommand{\Section}[1]{\setcounter{equation}{0}\section{#1}}
\renewcommand{\theequation}{\arabic{section}.\arabic{equation}}

\section{Introduction}

Homotopical methods based on $L_\infty$-algebras and $A_\infty$-algebras have been playing an increasingly significant role in our understanding of the algebraic and kinematic structures inherent in scattering amplitudes and correlation functions of quantum field theory; for an incomplete sample of recent works see e.g.~\cite{HohmZwiebach,BVChristian,Doubek:2017naz,Arvanitakis:2019ald,Macrelli:2019afx,Jurco:2019yfd,Arvanitakis:2020rrk,Saemann:2020oyz,LInfMatter,Borsten:2021hua,SzaboAlex,Chiaffrino:2021pob,Gaunt:2022elo,Okawa:2022sjf} and references therein. At a given order of perturbation theory, these can be calculated in a purely algebraic fashion without resorting to canonical quantization or path integral techniques: the quantum Batalin--Vilkovisky (BV) formalism gives an explicit homological construction of correlators which algebraically generates Feynman diagrams. The diagrammatics of homological perturbation theory is discussed in e.g.~\cite{Saemann:2020oyz,SzaboAlex,Gaunt:2022elo}.

In this paper we are interested in what these techniques can teach us about noncommutative quantum field theory. Noncommutative field theories arise in many scenarios as effective theories; see~\cite{Douglas:2001ba,U1Reviewa} for early reviews of the subject as well as its relevance in open string theory with $B$-fields. Their $L_\infty$-algebra formulation at the classical level was discussed in~\cite{Munich18}; see~\cite{Giotopoulos:2021ieg} for a recent exposition. There are two main problems that this paper aims to address. 

The standard noncommutative quantum field theories are famously plagued by the notorious problem of `UV/IR mixing', which is related to the appearance of non-planar loop diagrams in perturbation theory~\cite{Minwalla:1999px}. The naive ultraviolet regulator $\theta$ provided by the noncommutative deformation becomes effective at energies $E$ with $E\,\sqrt\theta\ll1$. UV/IR mixing occurs in a loop correlator when the regularisation entangles ultraviolet and infrared regimes: an ultraviolet cutoff $\Lambda$ induces an effective infrared cutoff $\Lambda_0=1/\theta\,\Lambda$. While non-planar graphs are generically well-defined, they can lead to uncontrollable divergences when inserted as subgraphs into higher order graphs. These divergences increase with the order of perturbation theory, and all correlation functions are affected and diverge. As a consequence, the quantum field theory cannot be renormalized.

For  noncommutative $\phi^4$-theory with the Moyal--Weyl star-product, the UV/IR mixing problem can be cured by adding a background harmonic oscillator potential to the free part of the classical action functional. The quantum field theory is then covariant under Fourier transformation of the fields~\cite{Langmann:2002cc}, which renders the interchange of ultraviolet and infrared regimes a symmetry. This is the celebrated Grosse--Wulkenhaar model~\cite{Grosse:2004yu}, which is renormalizable to all orders in perturbation theory. But it is not understood how to achieve this in gauge theories.

In this paper we explore a new approach to renormalizable noncommutative quantum field theory by instead modifying the path integral directly, rather than the classical action functional. Our approach is firmly rooted in the homological techniques developed by~\cite{Ciric:2020eab,BraidedLinf,SzaboAlex,Giotopoulos:2021ieg}: By deforming the $L_\infty$-structure of a field theory to a \emph{braided $L_\infty$-algebra}, one constructs field theories which are covariant under the action of a triangular Hopf algebra of symmetries, with braided noncommutative fields. Quantum correlation functions are then computed via a braided deformation of the BV formalism and homological perturbation theory: This is called \emph{braided quantum field theory}. 
The renormalization properties of braided quantum field theory turn out to be very different, and UV/IR mixing seems to be less severe and maybe even absent. This is due in part to the absence of non-planar diagrams, which we demonstrate explicitly in the example of noncommutative $\phi^4$-theory.

Our formulation of braided quantum field theory realises a special case of Oeckl's  approach~\cite{Oeckl:1999zu,Oeckl} (see also~\cite{Sasai:2007me}), which formulates the path integral  of ordinary quantum field theory in a purely algebraic language  and then generalizes it to `braided spaces' of fields which are objects in the braided monoidal representation category of a quasi-triangular Hopf algebra. This algebraic approach is based on normalised Gaussian integration over braided spaces, which leads to a braided generalization of Wick’s theorem. In this way braided quantum field theory follows the traditional path integral approach, going from Gaussian path integrals via perturbation theory to Feynman diagrams. The braided generalization of the quantum BV formalism~\cite{SzaboAlex} does exactly this, while also having the advantage of going beyond the classes of theories immediately covered by Oeckl's approach: our formalism also naturally treats theories with gauge symmetries.

For Moyal--Weyl twists, our systematic treatment of noncommutative $\phi^4$-theory makes precise some older claims from the literature, many of which were reached through somewhat ad hoc lines of reasoning and constructions. Based on Oeckl's computation of  deformed Green's functions for scalar fields in terms of undeformed ones~\cite{Oeckl}, braided deformations of free quantum field commutation and anti-commutation relations, i.e. of the oscillator algebras of creation and annihilation operators, were suggested in several works, see e.g.~\cite{Balachandran:2005pn,Bu:2006ha,Fiore:2007vg,Balachandran:2008gr,Lukierski:2011mj}. In particular, from this it was argued in~\cite{Balachandran:2005pn} that noncommutative quantum field theory with braided symmetry dispels with UV/IR mixing in S-matrix elements of scalar field theories. These treatments employ braided tensor products, which also twist the propagators of the field theory, in contrast to our approach. In fact, our more systematic calculations generally reach somewhat different conclusions: braided quantum field theory is \emph{not} the same as its undeformed counterpart~\cite{SzaboAlex}.

The second main goal of this paper is to understand the homotopy algebraic approach to quantization of braided field theories with gauge symmetries. As a first step towards understanding the more elaborate non-abelian gauge theories, here we undertake a detailed study of the simplest example of a $\sU(1)$ gauge theory coupled to a Dirac fermion. We call this theory braided quantum electrodynamics~(QED); a preliminary investigation of this model was announced in~\cite{CiricDimitrijevic:2022eei}. This theory is markedly different from that of the standard noncommutative QED, whose photon field is self-interacting, contrary to the photon of braided~QED. We develop the classical braided field theory in detail, and in particular demonstrate how the homotopy Noether identity associated to the $\sU(1)$ gauge symmetry naturally implies the electric charge conservation law.
In the quantized theory we demonstrate the absence of non-planar diagrams as well as UV/IR mixing in one-loop two-point correlators. This is also in agreement with earlier calculations~\cite{Balachandran:2008gr} which found that no UV/IR mixing occurs in S-matrix elements of  $\sU(1)$ gauge theory coupled to matter. The implications of this for the renormalizability of braided QED is left for future work. 

\paragraph{Outline.}
A central purpose of this paper is to demonstrate how to compute correlation functions for perturbative braided quantum field theory using braided quantum $L_\infty$-algebras (equivalently the braided BV formalism), and in particular to present explicit expressions for correlation functions obtained through these algebraic techniques. The structure of this paper is as follows. 

In Section~\ref{sec:LinftybraidedQFT} we briefly review braided $L_\infty$-algebras and their application in developing a novel particular class of examples of noncommutative field theories, called braided field theory. We further describe their perturbative quantization by developing a braided version of Wick's theorem and applying the BV formalism. We treat the noncommutative $\phi^4$-theory in some detail, demonstrating the absence of UV/IR mixing in the one-loop self-energy.

In Section~\ref{sec:braidedED} we present the explicit example of braided electrodynamics. We first construct the corresponding braided $L_\infty$-algebra, and then use it to formulate the action functional and the equations of motion. Using the braided Noether identity we find the associated conserved matter current and discuss the modifications to the electric charge conservation law. 

In Section~\ref{sec:braidedQED} we explore the perturbative expansion of quantized braided electrodynamics defined by the braided Wick's theorem and homological perturbation theory. In particular, we find that there are no non-planar Feynman diagrams as well as no UV/IR mixing through explicit computations of the vacuum polarization and the fermion self-energy at one-loop.

Two appendices at the end of the paper contain some technical details which are used in the main text: Appendix~\ref{app:Drinfeld} briefly reviews the basics of twist deformations that are used in our constructions of braided field theories, while Appendix~\ref{app:Dirac} summarises our conventions for Dirac spinors.

\paragraph{Acknowledgements.}
We are grateful to Paolo Aschieri, Martin Cederwall, Branislav Jur\v{c}o, Denjoe O'Connor, Biljana Nikoli\'c, Christian S\"amann, Mi\v sa Toman, Francesco Toppan and Guillaume Trojani for helpful discussions and correspondence. M.D.C. and R.J.S. thank the Mainz Institute for Theoretical Physics (MITP) of the Cluster of Excellence PRISMA${}^+$ (Project ID 39083149) for hospitality and support during part of this work. The work of
M.D.C., N.K. and V.R. is supported by Project 451-03-47/2023-01/200162 of the Serbian Ministry
of Education, Science and Technological Development. The work of M.D.C and R.J.S. was partially supported by the Croatian Science Foundation Project IP-2019-04-4168. The work of R.J.S. was supported by the
Consolidated Grant ST/P000363/1 from the UK Science and Technology Facilities Council.

\section{$L_{\infty}$-algebras and braided quantum field theory}
\label{sec:LinftybraidedQFT}

\subsection{$L_{\infty}$-algebras and classical field theory}
\label{sec:Linftyclassical}

Let us begin by briefly recalling the definition of a classical $L_{\infty}$-algebra.
An $L_\infty$-algebra $\CCL$ is a $\RZ $-graded real vector space $L=\bigoplus_{k\in \RZ }\, L^{k}$ equipped with graded antisymmetric multilinear maps
\begin{align*}
\ell_n: L^{\otimes n} \longrightarrow L \ , \quad  a_1\otimes \dots\otimes a_n \longmapsto \ell_n (a_1,\dots,a_n)
\end{align*}
for each $n\geq1$, which have degree $|\ell_n|=2-n$. The graded antisymmetry translates to 
\begin{equation}\label{eq:gradedantisym}
\ell_n (\dots, a,a',\dots) = -(-1)^{|a|\,|a'|}\, \ell_n (\dots, a',a,\dots) \ ,
\end{equation}
where $|a|$ denotes the degree of a homogeneous element $a\in L$. We write $\ell:=\{\ell_n\}_{n\geq1}$ for the collection of all multilinear maps, which are also called multibrackets.

The $n$-brackets $\ell_n$ are required to fulfill infinitely many
homotopy relations, for each $n\geq1$. The first $L_\infty$-relation
\begin{align*}
\ell_1\big(\ell_1(a)\big) = 0 
\end{align*}
says that underlying any $L_\infty$-algebra $\CCL$ is a cochain complex $(L,\ell_1)$:
\begin{align*}
\cdots \xrightarrow{ \ \ \ell_1 \ \ } L^k \xrightarrow{ \ \ \ell_1 \ \ }
  L^{k+1} \xrightarrow{ \ \ \ell_1 \ \ } \cdots \ .
\end{align*}
The second relation
\begin{align*}
\ell_1\big(\ell_2(a_1,a_2)\big) = \ell_2\big(\ell_1(a_1),a_2\big) + (-1)^{|a_1|}\, \ell_2\big(a_1, \ell_1(a_2)\big)
\end{align*}
says that the differential $\ell_1$ is a graded derivation with respect to the $2$-bracket $\ell_2$, or in other words that $\ell_2$ is a cochain map. The third homotopy relation
\begin{align*}
\begin{split}
&\ell_2\big(\ell_2(a_1,a_2),a_3\big) - (-1)^{|a_2|\,|a_3|}\,
  \ell_2\big(\ell_2(a_1,a_3),a_2\big) +  (-1)^{(|a_2|+|a_3|)\,|a_1|}\,
  \ell_2\big(\ell_2(a_2,a_3),a_1\big) \nn\\[4pt]
\!&= -\ell_3\big(\ell_1(a_1),a_2,a_3\big) - (-1)^{|a_1|}
  \ell_3\big(a_1, \ell_1(a_2), a_3\big) - (-1)^{|a_1|+|a_2|}
  \ell_3\big(a_1,a_2, \ell_1(a_3)\big) -\ell_1\big(\ell_3(a_1,a_2,a_3)\big)
\end{split}
\end{align*}
says that the graded Jacobi identity for $\ell_2$ is violated by a cochain homotopy determined by $\ell_3$, and so on for $n>3$.

In applications to Lagrangian field theory, one additionally asks that an $L_\infty$-algebra $\CCL$ is endowed with a graded symmetric non-degenerate bilinear pairing $\langle-,-\rangle:L\otimes L\to\FR $ which is cyclic in the sense that
\begin{align*}
\langle a_0,\ell_n(a_1,a_2,\dots,a_n)\rangle = \pm \, \langle a_n,\ell_n(a_0,a_1,\dots,a_{n-1})\rangle
\end{align*}
for all $n\geq1$. Here and in the following we write $\pm$ for the sign factors determined by the grading of the elements $a_i$ involved through the Koszul sign rule. 

It was shown in \cite{HohmZwiebach,Munich18,BVChristian, LinfUs} (see~\cite{Giotopoulos:2021ieg} for a review) that any classical field theory with irreducible gauge symmetries (that is, with independent gauge transformations) can be completely
encoded in $4$-term $L_{\infty}$-algebras $\CCL$ with underlying graded vector space
$$
L=L^0\oplus L^1\oplus L^2\oplus L^3 \ .
$$
Given a gauge parameter $ c \in L^{0}$ and a dynamical field $A\in L^{1}$, the gauge
variations are given by
\begin{align} \label{gaugetransfA}
\delta_{ c}A=\ell_1( c) + \ell_2( c,A) - \tfrac12\,\ell_3( c,A,A) + \cdots \ ,
\end{align}
where the ellipses designate higher brackets involving tensor powers $A^{\otimes n}$ for $n\geq3$, which are not needed in the applications considered in this paper.
The equations of motion $F_A=0$ in $L^2$ are encoded through homotopy Maurer--Cartan equations
\begin{align}\label{EOM} 
F_A=\ell_1(A) - \tfrac12\,\ell_2(A,A) - \tfrac1{6}\,\ell_3(A,A,A) + \cdots \ .
\end{align}
Under gauge transformations (\ref{gaugetransfA}) the homotopy Maurer--Cartan equations transform covariantly
\begin{align} \label{gaugetransfF}
\delta_{ c} F_A=\ell_2( c,F_A) + \ell_3( c,F_A,A) + \cdots \ .
\end{align}

For the purposes of this paper, we may assume for simplicity that the algebra of gauge variations closes off-shell (that is, when $F_A\neq0$). We further assume that the brackets of $\CCL$ satisfy
\begin{align}\label{eq:vanishingells}
\ell_{n+2}(c_1,c_2,A_1,\dots,A_n)=0 \qquad \mbox{and} \qquad \ell_{n+2}(c_1,c_2,A^+,A_1,\dots,A_{n-1})=0
\end{align}
for all $n\geq1$, $c_1,c_2\in L^0$, $A_1,\dots,A_n\in L^1$ and $A^+\in L^2$. 
Then the homotopy relations imply that the closure relation for the gauge algebra has the form
\begin{align}\label{eq:closure}
[\delta_{ c_1},\delta_{ c_2}]_\circ 
= \delta_{-\ell_2( c_1, c_2)} \ ,
\end{align}
where $[\delta_{ c_1},\delta_{ c_2}]_\circ :=\delta_{ c_1}\circ\delta_{ c_2} - \delta_{ c_2}\circ\delta_{ c_1}$ is the commutator of gauge variations.
The Noether identities in $L^3$ corresponding to the gauge symmetry are encoded by
\begin{align}\label{eq:Noether}
\dsf_AF_A := \ell_1(F_A) + \ell_2(F_A,A) - \tfrac12\,\ell_3(F_A,A,A)+\cdots = 0 \ ,
\end{align}
which vanishes identically as a consequence of the homotopy relations on $A^{\otimes n}$ for all $n\geq1$. 

The action functional of a Lagrangian field theory can be written
via a symmetric non-degenerate bilinear pairing $\langle -,-\rangle :
L \otimes L\to\FR $ of degree~$-3$ which makes $\CCL$ into a cyclic
$L_\infty$-algebra. Then the equations of motion
$F_A = 0$ follow from varying the homotopy Maurer--Cartan action functional  
\begin{align} \label{action}
S(A) := \tfrac12\,\langle A,\ell_1(A)\rangle - \tfrac16\,\langle A,\ell_2(A,A)\rangle - \tfrac1{24}\,\langle A,\ell_3(A,A,A)\rangle +\cdots \ ,
\end{align}
since cyclicity implies $\delta S(A)=\langle F_A,\delta A\rangle$.
Cyclicity also implies
\begin{align}\label{eq:gtNoether}
\delta_ c S(A)=\langle F_A,\delta_ c A\rangle=-\langle\dsf_AF_A,  c\rangle \ ,
\end{align}
so that gauge invariance of the action functional $\delta_ c
S(A)=0$ is then equivalent to the Noether identities
$\dsf_AF_A=0$. 

Note that $\ell_1(A)$ is associated with the free equations of motion, whereas $\ell_n(A^{\otimes n})$ for $n\geq2$ correspond to interaction vertices in the Lagrangian. In this formalism, free fields (that is, solutions to the linearised equations of motion) are in the cohomology $H^\bullet(L)$ of the underlying cochain complex~$(L,\ell_1)$.

\subsection{Braided $L_\infty$-algebras and braided field theory}
\label{sec:braidedLinfty}

Starting from a suitable classical $L_\infty$-algebra $\CCL$, using Drinfel'd twist
deformation techniques one can construct a braided $L_\infty$-algebra $\CCL^\star$ in the sense of~\cite{Ciric:2020eab,BraidedLinf} (see~\cite{Giotopoulos:2021ieg} for a review). A brief review of the Drinfel'd twist deformation formalism is presented in  Appendix~\ref{app:Drinfeld}, while more details can be found in \cite{MajidBook,SLN}. Let $\frv:=\Gamma(TM)$ be the Lie algebra of vector fields on a manifold $M$, and let  $U\frv$ be its enveloping algebra. For a twist $\CF\in U\frv[[\nu]]\otimes U\frv[[\nu]]$, we write ${\cal F} = \sff^\alpha \otimes \sff_\alpha$ and $\CF^{-1} = \bar\sff^\alpha \otimes \bar\sff_\alpha$ for its inverse. The corresponding triangular $\RR$-matrix is $\RR = \CF_{21}\, \CF^{-1}=:\sfR^\alpha\otimes\sfR_\alpha$, with inverse $\RR^{-1} = \RR_{21} = \sfR_\alpha\otimes \sfR^\alpha$.\footnote{Throughout this paper, repeated upper and lower indices are always implicitly summed over.}

To apply the twist deformation formalism, we start from a classical $L_\infty$-algebra $\CCL$ whose underlying graded vector space 
$L=\bigoplus_{k\in\RZ }\,L^k$ is a $\RZ $-graded (left)
$U\frv$-module and the $n$-brackets $\ell_n:L^{\otimes n}\to L$ are
equivariant maps, that is, they all commute with the action of
$\frv=\Gamma(TM)$ on $L$ via the trivial coproduct $\Delta$. Given any 
Drinfel'd twist $\CF\in U\frv[[\nu]]\otimes U\frv[[\nu]]$, we
deform the brackets $\ell_n$ to twisted brackets $\ell_n^\star$ which
commute with the action of $\Gamma(TM)$ on $L[[\nu]]$ via the
twisted coproduct $\Delta_\CF$. Following the
standard prescription (\ref{eq:mustar}), we set
$\ell_1^\star:=\ell_1$ and
\begin{align}\label{eq:ellnstardef}
	\ell_n^\star(a_1,\dots,a_n) :=
	\ell_n(a_1\otimes_\star\cdots\otimes_\star a_n)
\end{align}
for $n\geq2$, where $a\otimes_\star a':=\CF^{-1}(a\otimes
a')=\bar\sff^\alpha(a)\otimes\bar\sff_\alpha(a')$ for $a,a'\in L[[\nu]]$. These define multilinear maps $\ell_n^\star:L[[\nu]]^{\otimes n}\to L[[\nu]]$ which are braided graded antisymmetric:
\begin{align*}
\ell_n^\star (\dots, a,a',\dots) = -(-1)^{|a|\,|a'|}\, \ell_n^\star \big(\dots,
  \sfR_\alpha(a'),\sfR^\alpha(a),\dots\big) \ .
\end{align*}

The first and second homotopy relations are unchanged with respect to the corresponding classical homotopy relations; that is, the braided $L_\infty$-algebra $\CCL^\star$ still has underlying cochain complex $(L[[\nu]],\ell_1)$ and $\ell_2^\star$ is again a cochain map. The third homotopy relation is given by
\begin{align}
\begin{split}
& \ell^\star_2\big(\ell^\star_2(a_1,a_2),a_3\big) - (-1)^{|a_2|\,|a_3|}\,
  \ell^\star_2\big(\ell^\star_2(a_1,\sfR_\alpha(a_3)),\sfR^\alpha(a_2)\big) \\ 
&\hspace{5.5cm} +  (-1)^{(|a_2|+|a_3|)\,|a_1|}\,
  \ell^\star_2\big(\ell^\star_2(\sfR_\alpha(a_2),\sfR_\beta(a_3)),\sfR^\beta\,\sfR^\alpha(a_1)\big) \\[4pt]
& \hspace{1.5cm} = -\ell^\star_3\big(\ell_1(a_1),a_2,a_3\big) - (-1)^{|a_1|}\,
  \ell^\star_3\big(a_1, \ell_1(a_2), a_3\big) - (-1)^{|a_1|+|a_2|}\,
  \ell^\star_3\big(a_1,a_2, \ell_1(a_3)\big) \\
& \hspace{5.5cm} -\ell_1\big(\ell^\star_3(a_1,a_2,a_3)\big) \ .
\end{split}
\label{I3braided} 
\end{align}
We observe that the non-trivial braiding now appears in this relation, which says that the braided graded Jacobi identity for $\ell_2^\star$ is violated by the cochain homotopy $\ell_3^\star$.

If $\CCL$ is a cyclic $L_\infty$-algebra with a $U\frv$-invariant inner product, then its cyclic structure $\langle-,-\rangle:L\otimes L\to\FR$ is twist
deformed to a new inner product
$\langle-,-\rangle_\star: L[[\nu]]\otimes L[[\nu]]\to\FR[[\nu]]$ defined by
\begin{align}\label{eq:twistpairing}
\langle a_1,a_2\rangle_\star := \langle
  \bar\sff^\alpha(a_1),\bar\sff_\alpha(a_2) \rangle \ .
\end{align}
In general, graded symmetry of the cyclic pairing $\langle-,-\rangle$
implies that the twisted pairing $\langle-,-\rangle_\star$ is naturally
braided graded symmetric 
\begin{equation}
\langle
  a_2,a_1\rangle_\star  = (-1)^{|a_1|\,|a_2|} \, \langle \sfR_\alpha(a_1),\sfR^\alpha(a_2)\rangle_\star \  ,  \nn 
\end{equation}
and also braided cyclic
\begin{align*}
\langle a_0\,,\,\ell_n^\star(a_1,a_2,\dots,a_n)\rangle_\star
= \pm\,\langle\sfR_{\alpha_0}\,\sfR_{\alpha_1}\cdots\sfR_{\alpha_{n-1}}(a_n)\,,\, \ell_n^\star(\sfR^{\alpha_0}(a_0),\sfR^{\alpha_1}(a_1),\dots,\sfR^{\alpha_{n-1}}(a_{n-1}))\rangle_\star \ .
\end{align*}
However, for applications to field theory, we have to restrict to \emph{compatible} Drinfel'd twists~\cite{BraidedLinf} that result in a strictly graded symmetric pairing
\begin{align*}
\langle
  a_2,a_1\rangle_\star  = (-1)^{|a_1|\,|a_2|} \, \langle
  a_1,a_2\rangle_\star 
\end{align*}
for all homogeneous $a_1,a_2\in L[[\nu]]$. In this case, $\CCL^\star$ becomes a \emph{strictly} cyclic braided $L_\infty$-algebra.

Following the classical case, a braided field theory is built as a noncommutative deformation of a classical field theory which is completely defined in terms of its braided $L_\infty$-algebra. Let $\CCL^\star = (L[[\nu]],\ell^\star)$ be a $4$-term braided $L_\infty$-algebra, obtained by twist deformation of an $L_\infty$-algebra $\CCL=(L,\ell)$ which organises the symmetries and dynamics of a classical field theory. For a gauge parameter $ c\in L^0[[\nu]]$, we define the braided gauge variation\footnote{In general, one can define both left and right braided gauge transformations. In this paper we focus only on left braided gauge transformations for simplicity. More details can be found in~\cite{BraidedLinf}.} of a dynamical field $A\in L^1[[\nu]]$ by
\begin{align}\label{eq:LgtL} 
\delta_ c^\star A = \ell_1( c) + \ell_2^\star( c,A) - \tfrac12\,\ell_3^\star( c,A,A) + \cdots
 \ .
\end{align}
Braided covariant dynamics is described by the equations of motion $F_A^\star=0$, where the braided homotopy Maurer--Cartan equations
\begin{align}\label{eq:braidedeom}
F_A^\star =  \ell_1(A) - \tfrac12\,\ell_2^\star(A,A) - \tfrac1{6}\,\ell_3^\star(A,A,A) + \cdots 
\end{align}
transform covariantly as
\begin{align}\label{eq:lefteomcov}
\delta_ c^\star F_A^\star = \ell_2^\star( c,F_A^\star) + \tfrac12\,\big(\ell_3^\star( c,F_A^\star,A) - \ell_3^\star( c,A,F_A^\star)\big) + \cdots   \ ,
\end{align}
for all gauge parameters $ c \in L^0[[\nu]]$. 

The braided gauge transformations obey the off-shell closure relation in terms of the braided commutator:
\begin{align}\label{eq:braidedclosure}
\big[\delta_{ c_1}^\star,\delta_{ c_2}^\star\big]_\circ^\star :=  \delta_{ c_1}^\star\circ\delta_{ c_2}^\star - \delta_{\sfR_\alpha( c_2)}^\star\circ\delta_{\sfR^\alpha( c_1)}^\star
= \delta_{-\ell^\star_2( c_1, c_2)}^\star \ .
\end{align}
Corresponding to the braided gauge symmetry, a suitable combination of the braided homotopy relations leads to an identity
\begin{align}\label{eq:braidedNoether}
\begin{split}
\dsf^\star_A F_A^\star :\!& = \ell_1(F^\star_A) + \tfrac12\,\big(\ell^\star_2(F^\star_A,A) - \ell^\star_2(A,F^\star_A)\big) +  \tfrac1{6} \, \ell_1\big(\ell^\star_{3}(A,A,A)\big)+ \cdots  \\
& \hspace{1cm} \quad \, + \tfrac18\,\big(\ell_2^\star(\ell_2^\star(A,A),A) - \ell_2^\star(A,\ell_2^\star(A,A))\big) \\
& \hspace{3cm} \quad \, + \tfrac1{12}\,\big( \ell_2^\star(\ell_3^\star(A,A,A),A) - \ell_2^\star(A,\ell_3^\star(A,A,A))\big) + \cdots \ = \ 0 \ . 
\end{split}
\end{align}
Unlike the classical Noether identity \eqref{eq:Noether}, the braided Noether identity (\ref{eq:braidedNoether}) is no longer linear in the equations of motion $F^\star_A$ and contains inhomogeneous terms involving brackets of the fields $A$ themselves. This is related to the violations of the Bianchi identities in braided gauge theories~\cite{BraidedLinf}. In the classical limit $\nu=0$, where $\CR=1\otimes1$, the braided homotopy formulas \eqref{eq:LgtL}--\eqref{eq:braidedNoether} all reduce to the classical formulas in \eqref{gaugetransfA}--\eqref{eq:Noether}.

For a Lagrangian field theory, using the (strictly) cyclic inner product one can define an analogue of the homotopy Maurer--Cartan action functional for the braided field theory as
\begin{align}\label{eq:braidedaction}
S_\star(A) := \tfrac12\,\langle A,\ell_1(A)\rangle_\star - \tfrac16\,\langle A,\ell^\star_2(A,A)\rangle_\star - \tfrac1{24}\,\langle A,\ell^\star_3(A,A,A)\rangle_\star +\cdots \ ,
\end{align}
whose variational principle yields the braided equations of motion $F_A^\star=0$.
This action functional is invariant under braided gauge transformations:
\begin{align}\label{eq:braidedgaugeinv}
\delta_ c^\star S_\star(A)=0 \ ,
\end{align}
for all $ c\in L^0[[\nu]]$ and $A\in L^1[[\nu]]$. However, unlike the classical case, the braided Noether identity for the braided gauge symmetry cannot be derived from the variational principle, because braided gauge variations and Euler--Lagrange variations behave very differently, see~\cite{BraidedLinf}. 

Note that the free fields of braided field theory are unchanged from the classical field theory: they are still the degree~$1$ elements of the cohomology $H^\bullet(L[[\nu]])$ of the underlying cochain complex $(L[[\nu]],\ell_1)$. Only the interaction vertices, corresponding to the higher brackets $\ell_n^\star$ for $n\geq2$, are modified by the braided noncommutative deformation.

\subsection{Braided Wick's theorem}
\label{sec:braidedWick}

$L_\infty$-algebras are the natural algebraic structure underlying the Batalin--Vilkovisky (BV) formalism~\cite{BVChristian}, which may be used for the quantization of classical field theories. Similarly, we expect that braided $L_\infty$-algebras should be related to a braided generalization of BV quantization. This was described explicitly in~\cite{SzaboAlex} for field theories with finitely many degrees of freedom, and applied to an example of braided fuzzy scalar field theory; the braided BV formalism is also discussed in~\cite{Giotopoulos:2021ieg}. One of the purposes of the present paper is to formulate braided quantum field theory and study its features in a simple example of a continuum field theory with gauge symmetry. 

In order to set up and illustrate the general framework for this, we consider here the simple example of scalar field theory on $d$-dimensional Minkowski spacetime $\FR^{1,d-1}$. We show how to recover, using heuristic field theory arguments, Oeckl's approach to (symmetric) braided quantum field theory which relies upon a braided generalization of Wick's theorem based on purely algebraic arguments~\cite{Oeckl:1999zu,Oeckl}. This result will then be substantiated in Section~\ref{sec:braidedBV}, where we develop the braided BV quantization of scalar field theory.

The action functional for a free real scalar field $\phi$ on $\FR^{1,d-1}$ with mass $m$ is given by
\begin{align}\label{eq:S0scalar}
S_0(\phi) = \int_{\FR^{1,d-1}} \, \dd^dx \ \frac12\, \phi\,\big(-\square - m^2\big)\,\phi
\end{align}
where
\begin{align*}
\dd^dx = \dd x^0\wedge\dd x^1\wedge\cdots\wedge\dd x^{d-1}
\end{align*}
is the standard volume form on Minkowski spacetime and $\square$ is the d'Alembertian operator. The differential of the underlying abelian $L_\infty$-algebra $\CCL_0$ is given by the Klein--Gordon operator as
$\ell_1 = -(\square + m^2)$. 
Since there are no gauge symmetries, only $L^1=L^2=\Omega^0(\FR^{1,d-1})$ are non-trivial and given by two copies of the space of functions (regarded as $0$-forms) on $\FR^{1,d-1}$. 

The $L_\infty$-algebra $\CCL_0$ is completely described by the 2-term cochain complex
\begin{align}\label{eq:scalarcomplex}
\Omega^0(\FR^{1,d-1})[-1]  \xrightarrow{ \ -(\square+m^2) \ }\Omega^0(\FR^{1,d-1})[-2]
\end{align}
concentrated in degrees~$1$ and~$2$, where the square brackets indicate shifts of the
cohomological degree.\footnote{For any vector space $W$ and integer $p\in\RZ$, elements of $W[p]$ are of degree~$-p$.}
Elements of the degree~$1$ cohomology of this complex $H^1(L)=\ker(\square+m^2)$ are the states $\phi^{\textrm{\tiny(0)}}\in\Omega^0(\FR^{1,d-1})$ that solve the Klein--Gordon equation. Elements $\phi^+\in L^2$ correspond to the BV antifields of the physical fields $\phi\in L^1$.

The cyclic structure of degree $-3$ is given by the non-zero inner product
\begin{align}\label{eq:scalarpairing}
\langle\phi,\phi^+\rangle = \int_{\FR^{1,d-1}} \, \dd^dx \ \phi \ \phi^+ \ ,
\end{align}
for $\phi\in L^1$ and $\phi^+\in L^2$, which is cyclic because the Klein--Gordon operator is formally self-adjoint with respect to this inner product.
Interaction vertices are incorporated by including the non-zero higher brackets~\cite{BVChristian,Giotopoulos:2021ieg}
\begin{align}\label{eq:intbrackets}
\ell_n(\phi_1,\dots,\phi_n) = \lambda_n \, \phi_1\cdots\phi_n \ ,
\end{align}
for $n\geq2$, $\lambda_n\in\FR$ and $\phi_1,\dots,\phi_n\in L^1$, and writing the homotopy Maurer--Cartan action functional \eqref{action}. The homotopy relations follow trivially for degree reasons, while cyclicity of \eqref{eq:scalarpairing} with respect to $\ell_n$ is a trivial consequence of commutativity of pointwise multiplication of functions.

The perturbative $n$-point functions of the free quantum field theory are defined by the formal normalized functional integral
\begin{align}\label{eq:npointfn}
\begin{split}
G_n(x_1,\dots,x_n)^{(0)} :=& \ \langle 0| {\rm T}[\phi(x_1)\cdots \phi(x_n)]|0\rangle^{(0)}\\[4pt] =& \ \frac1{\CZ} \, \int_{\Omega^0(\FR^{1,d-1})} \, \CCD\phi \ \phi(x_1) \cdots \phi(x_n) \ {\rm e}^{\,\frac{\rm i}\hbar \, S_0(\phi)} \ ,
\end{split}
\end{align}
where $\rm T$ implements the time-ordered product of fields, and $\CZ$ is a normalization factor such that $\langle0|0\rangle^{(0)}=1$. This is non-zero only when $n=2k$ is even, and Wick's theorem expresses it as the Hafnian of the two-point correlation matrix:
\begin{align*}
G_{2k}(x_1,\dots,x_{2k})^{(0)} = \frac1{k! \, 2^k} \, \sum_{\sigma\in S_{2k}} \ \prod_{a=1}^{k} \, \langle 0|{\rm T}[\phi(x_{\sigma(2a-1)})\,\phi(x_{\sigma(2a)})]|0\rangle^{(0)} \ ,
\end{align*}
where $S_n$ denotes the symmetric group of all permutations of degree $n$.  The free two-point functions are related to the scalar Feynman propagator by
\begin{align}\label{eq:Feynmanprop}
\begin{split}
\frac{\ii}{\hbar} \, \langle 0|{\rm T}[\phi(x)\,\phi(y)]|0\rangle^{(0)} = -\frac1{\square+m^2 - \ii\,\epsilon}(x,y) = \int_{(\FR^{1,d-1})^*} \, \frac{\d ^d k}{(2\pi)^d} \ \frac{{\rm e}^{-{\rm i}\,k\cdot(x-y)}}{k^2-m^2+\ii\,\epsilon} \ ,
\end{split}
\end{align}
for $\epsilon\in\FR_{>0}$. In the following we will drop the $\mathrm{i}\,\epsilon$-prescription in the Feynman propagators, and abbreviate momentum space integrals as 
\begin{align*}
\int_{k_1,\dots,k_v} := \int_{(\FR^{1,d-1})^*} \, \frac{\d ^d k_1}{(2\pi)^d} \ \cdots \ \int_{(\FR^{1,d-1})^*} \, \frac{\d ^d k_v}{(2\pi)^d} \ , 
\end{align*}
in order to simplify the presentation.

The braided noncommutative deformation follows the twist formalism discussed in Section~\ref{sec:braidedLinfty}. The classical scalar field theory is Poincar\'e invariant (but not diffeomorphism invariant), so its $L_\infty$-algebra $\CCL_0$ consists of modules and equivariant brackets for the universal  enveloping algebra $U\mathfrak{iso}(1,d-1)\subset U\frv$ of the Poincar\'e algebra $\mathfrak{iso}(1,d-1)$. Hence we have to restrict to twists $\CF\in U\iso(1,d-1)[[\nu]]\otimes U\iso(1,d-1)[[\nu]]$. For simplicity, we will work with abelian twists, for which $\RR=\CF_{21}\,\CF^{-1}=\CF^{-2}$; the standard Moyal--Weyl twist and also the angular twist of~\cite{DimitrijevicCiric:2018blz} are examples of such twists. We will further restrict to abelian twists $\CF\in U\iso(d-1)[[\nu]]\otimes U\iso(d-1)[[\nu]]$ constructed from the spatial isometries of $\FR^{d-1}\subset\FR^{1,d-1}$, as this simplifies some of the analysis in the quantum field theory, such as the treatment of time-ordering, as well as avoiding potential issues with unitarity. For definiteness, and for the sake of illustration, let us choose here the Moyal--Weyl twist
\begin{align}\label{eq:MWtwist}
\CF =
  \exp\big(-\tfrac{\mathrm{i}\,\nu}2\,\theta^{ij}\,\partial_i\otimes\partial_j\big) 
  \ ,
\end{align}
where $(\theta^{ij})$ is a $(d-1){\times}(d-1)$ antisymmetric real-valued
matrix, and $\partial_i=\frac\partial{\partial x^i}\in\Gamma(T\FR^{d-1})$ for $i=1,\dots,d-1$ are vector fields generating spatial translations in \smash{$\FR^{1,d-1}$}.

Since $\CCL_0^\star=\CCL_0$, and since the twist $\CF$ is compatible with the cyclic inner product \eqref{eq:scalarpairing}, the free braided scalar field theory is unchanged from its commutative version. In particular, because $\langle-,-\rangle_\star=\langle-,-\rangle$, the action functional \eqref{eq:S0scalar} is unchanged: 
\begin{align*}
S_{0\star}(\phi)=\tfrac12\,\langle\phi,\ell_1(\phi)\rangle_\star=S_0(\phi) \ .
\end{align*} 
Interaction vertices are included by twisting the brackets \eqref{eq:intbrackets} to 
\begin{align*}
\ell_n^\star(\phi_1,\dots,\phi_n) = \lambda_n \, \phi_1\star\cdots\star\phi_n \ , 
\end{align*}
for $n\geq2$, $\lambda_n\in\FR$ and $\phi_1,\dots,\phi_n\in L^1[[\nu]]$, and writing the braided version of the homotopy Maurer--Cartan action functional \eqref{eq:braidedaction}. 

It follows that, at the classical level, even the standard noncommutative scalar field theory is organised by a braided $L_\infty$-algebra~\cite{Giotopoulos:2021ieg}. However, the braided symmetry makes a crucial difference in the quantum field theory, particularly in the application of a braided Wick expansion, rather than the standard one, and in the different interaction vertices arising from the braided symmetry. Correlation functions in the interacting quantum field theory will be discussed in Section~\ref{sec:braidedBV} below.

Given this braided $L_\infty$-algebra structure, we would now like to define the free braided $n$-point functions. While the traditional path integral and canonical quantization methods are not readily available for braided quantum field theory, we can quantize the theory in a purely algebraic fashion using modern techniques from homological algebra, as we explain in Section~\ref{sec:braidedBV}. Here we shall define them operationally by a heuristic noncommutative deformation of the Feynman representation of \eqref{eq:npointfn} in the following way, which also leads to a purely algebraic prescription, while at the same time elucidating the physical meaning of the braiding.\footnote{Since the free braided field theory is the same as its commutative counterpart, this can also be formulated via the operator formalism in the Dyson representation of \eqref{eq:npointfn}.}

Firstly, whereas in conventional noncommutative field theory the functional integral would still be taken over the commutative space of fields $\Omega^0(\FR^{1,d-1})$, in braided quantum field theory the domain of integration is a `braided space' of fields $\Omega^0_\star(\FR^{1,d-1})$~\cite{Oeckl}, that is, the algebra of fields with the star-product $\star$ which can be thought of as endowing them with `braided statistics'. The braided $n$-point function is thus denoted as
\begin{align*}
G_n^\star(x_1,\dots,x_n)^{(0)} = \langle0| {\rm T}[\phi(x_1)\star\cdots\star\phi(x_n)]|0\rangle_\star^{(0)} \ ,
\end{align*}
with the star-product canonically extended to the tensor product $\phi^{\otimes n}\in\Omega^0\big((\FR^{1,d-1})^{\times n}\big)[[\nu]]=\Omega^0(\FR^{1,d-1})[[\nu]]^{\otimes n}$ as~\cite{Oeckl,U1Reviewa}
\begin{align*}
\phi(x_1)\star\cdots\star\phi(x_n) =  \exp\Big(\frac{\mathrm{i}\,\nu}2 \, \sum_{a<b} \, \theta^{ij}\,\frac\partial{\partial x_a^i}\,\frac\partial{\partial x_b^j}\Big)\,\phi(x_1)\cdots\phi(x_n) \ .
\end{align*}
More generally, the twist \eqref{eq:MWtwist} can be lifted to the space of functionals of the fields as \cite{Aschieri:2007sq} 
\begin{align}\nn
\begin{split}
\CF = \exp\bigg(-\frac{\ii\,\nu}2 \, \theta^{ij} \, & \int_{\FR^{d-1}} \, \d^{d-1} x \ \Big(\partial_i\phi \, \frac\delta{\delta\phi(x)} + \partial_i\varPi \, \frac\delta{\delta\varPi(x)} \Big) \\ & \hspace{2cm} \otimes \ \int_{\FR^{d-1}} \, \d^{d-1} y \ \Big(\partial_j\phi \, \frac\delta{\delta\phi(y)} + \partial_j\varPi \, \frac\delta{\delta\varPi(y)} \Big)\bigg) \ ,
\end{split}
\end{align}
where $\varPi$ is the conjugate momentum to the field $\phi$.

Secondly, the integration measure $\CCD_\star\phi$ is taken to be $U_\CF\iso(d-1)$-invariant, so that the operation \smash{$\langle0|{\rm T}[-]|0\rangle^{(0)}_\star$} defines a $U_\CF\iso(d-1)$-equivariant map. This implies that the twist can be factored out of the functional integration and taken to act on the $n$-point function of the commutative scalar field theory, which may then be expanded using the usual Wick theorem. 

Altogether, for the non-vanishing correlation functions we prescribe the simple expression
\begin{align}\label{eq:G2kstar}
\begin{split}
& G_{2k}^\star(x_1,\dots,x_{2k})^{(0)} \\[4pt]
&\hspace{2cm} = \frac1{k!\,2^k} \, \sum_{\sigma\in S_{2k}} \, \exp\Big(\frac{\mathrm{i}\,\nu}2 \, \sum_{a<b} \, \theta^{ij}\,\frac\partial{\partial x_a^i}\,\frac\partial{\partial x_b^j}\Big)\, \prod_{c=1}^k \, \langle0|{\rm T}[\phi(x_{\sigma(2c-1)})\,\phi(x_{\sigma(2c)})]|0\rangle^{(0)} \ .
\end{split}
\end{align}
This formula will be derived in a more precise way in Section~\ref{sec:braidedBV}.

Unravelling the combinatorics of the formula \eqref{eq:G2kstar} reveals the following general statement of the braided Wick theorem:
\begin{myitemize}
\item The braided $n$-point function is defined by replacing the pointwise products of fields in the commutative $n$-point function with star-products.
\item Only nearest neighbouring fields can contract. To contract we therefore have to first permute fields, which introduces corresponding $\RR$-matrices.
\item After the contractions we use simplifications to remove some $\RR$-matrices, such as the identities from Appendix~\ref{app:Drinfeld} together with relativistic invariance of the commutative two-point functions.
\item The final result is the braided Wick theorem, which agrees with the results from \cite{Oeckl}.
\end{myitemize}
Note that no star-products will appear between contractions, due to $\iso(d-1)$-invariance of the commutative two-point functions: noncommutativity enters only through the permutations of fields, which produce $\RR$-matrices. Let us give a few explicit examples to illustrate how the theorem works.

\paragraph{Two-point function.}
The braided two-point function is defined as
\begin{equation}
G_2^\star (x_1,x_2)^{(0)} = \langle0| {\rm T}[\phi(x_1)\star \phi(x_2) ]|0\rangle_\star^{(0)} \ .
\label{2Point}
\end{equation}
Using \eqref{eq:G2kstar} along with \eqref{eq:Feynmanprop} we find
\begin{align}\label{2Pointcalc}
\begin{split}
G_2^\star(x_1,x_2)^{(0)} &=  \exp\Big(\frac{\ii\,\nu}{2}\,\theta^{ij}\, \frac\partial{\partial x_1^i}\,\frac\partial{\partial x^j_2}\Big) \langle 0| {\rm T}[\phi(x_1)\,\phi(x_2)] |0\rangle^{(0)} \\[4pt]
&= -\ii \, \hbar \, \int_k \ \e^{\,\frac{\ii\,\nu}2\,\theta^{ij}\,k_i\,k_j} \ \frac{\e^{-\ii\,k\cdot(x_1-x_2)}}{k^2-m^2}\\[4pt]
&= -\ii \, \hbar \, \int_k \ \frac{\e^{-\ii\,k\cdot(x_1-x_2)}}{k^2-m^2} = \langle 0| {\rm T}[\phi(x_1) \phi(x_2)] |0\rangle^{(0)} \ ,
\end{split}
\end{align}
where the noncommutative phase factor vanishes due to the antisymmetry of $\theta^{ij}$. Thus the two-point function remains unchanged, as expected. This same result is explained in~\cite{Oeckl}. In the following we will abbreviate Wick contractions as
\begin{align*} 
 \bcontraction{}{\phi_a}{}{\phi_b} \phi_a \, \phi_b
:=\langle0| {\rm T}[\phi(x_a) \, \phi(x_b) ]|0\rangle^{(0)} \ ,
\end{align*}
and denote \smash{$\partial_i^a:=\frac\partial{\partial x_a^i}$} for brevity.

\paragraph{Four-point function.}
The braided four-point function is defined as
\begin{align}
G_4^\star (x_1,x_2, x_3,x_4)^{(0)} = \langle 0| {\rm T} [\phi(x_1)\star \phi(x_2) \star \phi(x_3) \star \phi(x_4)]|0\rangle_\star^{(0)} \ .\label{4Point}
\end{align}
Applying the first biderivative operation from \eqref{eq:G2kstar} results in
\begin{align*}
\begin{split}
& \e^{\,\frac{\ii\,\nu}{2}\,\theta^{ij}\, \partial_i^1\, \partial_j^2} \big(
\bcontraction{}{\phi_1}{}{\phi_2} \phi_1\, \phi_2 \, \bcontraction{}{\phi_3}{}{\phi_4} \phi_3\, \phi_4
+ \bcontraction{}{\phi_1}{}{\phi_4} \phi_1\, \phi_3 \, \bcontraction{}{\phi_2}{}{\phi_4} \phi_2 \, \phi_4 + \bcontraction{}{\phi_1}{}{\phi_4} \phi_1 \, \phi_4 \, \bcontraction{}{\phi2}{}{\phi_3} \phi_2 \, \phi_3
\big) \\[4pt] & \hspace{5cm} = \bcontraction{}{\phi_1}{}{\phi_2} \phi_1\, \phi_2 \, \bcontraction{}{\phi_3}{}{\phi_4} \phi_3\, \phi_4 + \e^{\,\frac{\ii\,\nu}{2}\,\theta^{ij}\, \partial_i^1\, \partial_j^2} \big(\bcontraction{}{\phi_1}{}{\phi_3} \phi_1\, \phi_3 \, \bcontraction{}{\phi_2}{}{\phi_4} \phi_2 \, \phi_4 + \bcontraction{}{\phi_1}{}{\phi_4} \phi_1 \, \phi_4 \, \bcontraction{}{\phi_2}{}{\phi_3} \phi_2 \, \phi_3
\big) \ ,
\end{split}
\end{align*}
where we used
\begin{align*}
\partial_i^1\big(\bcontraction{}{\phi_1}{}{\phi_2} \phi_1\,\phi_2\big) = -\partial_i^2\big(\bcontraction{}{\phi_1}{}{\phi_2} \phi_1\,\phi_2\big) \ , 
\end{align*}
which follows from \eqref{eq:Feynmanprop}. 
The remaining biderivative operations from (\ref{eq:G2kstar}) follow in a similar way and the end result is
\begin{align}
G_4^\star (x_1,x_2, x_3,x_4)^{(0)} = \bcontraction{}{\phi_1}{}{\phi_2} \phi_1\, \phi_2 \, \bcontraction{}{\phi_3}{}{\phi_4} \phi_3\, \phi_4
+ \e^{-\ii\,\nu\,\theta^{ij}\, \partial_i^3\, \partial_j^2}\big(\bcontraction{}{\phi_1}{}{\phi_3} \phi_1\, \phi_3 \, \bcontraction{}{\phi_2}{}{\phi_4} \phi_2 \, \phi_4 + \bcontraction{}{\phi_1}{}{\phi_4} \phi_1 \, \phi_4 \, \bcontraction{}{\phi_2}{}{\phi_3} \phi_2 \, \phi_3\big) \ . \label{4PointIOrder}
\end{align}

From \eqref{4PointIOrder} we recognise the appearance of the inverse $\RR$-matrix and we can finally write
\begin{align}
G_4^\star (x_1,x_2, x_3,x_4)^{(0)} = \bcontraction{}{\phi_1}{}{\phi_2} \phi_1\, \phi_2 \, \bcontraction{}{\phi_3}{}{\phi_4} \phi_3\,\phi_4 
+ \bcontraction{}{\phi_1}{}{\sfR_\alpha(\phi_3)} \phi_1\, \sfR_\alpha(\phi_3)\, \bcontraction{}{\sfR^\alpha(\phi_2)}{}{\phi_4} \sfR^\alpha (\phi_2)\, \phi_4 + \bcontraction{}{\phi_1}{}{\phi_4} \phi_1 \, \phi_4\,\bcontraction{}{\phi_2}{}{\phi_3} \phi_2\, \phi_3 \ .\label{4PointFull}
\end{align}
In arriving at \eqref{4PointFull} we used 
\begin{align}\label{eq:4ptid}
\begin{split}
\bcontraction{}{\phi_1}{}{\sfR_\alpha \, \sfR_\beta(\phi_4)} \phi_1\, \sfR_\alpha \, \sfR_\beta(\phi_4)\, \bcontraction{}{\sfR^\alpha(\phi_2)}{}{\sfR^\beta(\phi_3)} \sfR^\alpha(\phi_2)\, \sfR^\beta(\phi_3) &= \bcontraction{}{\phi_1}{}{ \sfR_\alpha(\phi_4)} \phi_1\, \sfR_\alpha(\phi_4)\,\bcontraction{}{\sfR^\alpha_{\bar{\textrm{\tiny(1)}}}(\phi_2)}{}{\sfR^\alpha_{\bar{\textrm{\tiny(2)}}}(\phi_3)} \sfR^\alpha_{\bar{\textrm{\tiny(1)}}}(\phi_2) \, \sfR^\alpha_{\bar{\textrm{\tiny(2)}}}(\phi_3) \\[4pt]
&= \bcontraction{}{\phi_1}{}{\sfR_\alpha(\phi_4)} \phi_1\, \sfR_\alpha(\phi_4)\,\varepsilon(\sfR^\alpha)\,\bcontraction{}{\phi_2}{}{\phi_3} \phi_2\, \phi_3
= \bcontraction{}{\phi_1}{}{\phi_4} \phi_1\, \phi_4\,\bcontraction{}{\phi_2}{}{\phi_3} \phi_2\, \phi_3 \ ,
\end{split}
\end{align}
where in the first equality we used the $\RR$-matrix identities \eqref{eq:Rmatrixidsw}, in the second equality we used $U\iso(d-1)$-invariance of the two-point function, and in the last equality we used the normalization \eqref{eq:twistnorm} of the twist; this can also be checked by explicitly computing the left-hand side of \eqref{eq:4ptid} using~\eqref{eq:Feynmanprop}.

\paragraph{Six-point function.}
As a final illustration of our statement of the braided Wick theorem, we look at the six-point function defined as
\begin{equation}
G_6^\star (x_1,x_2, x_3,x_4, x_5,x_6)^{(0)} = \langle 0| {\rm T}[ \phi(x_1)\star \phi(x_2) \star \phi(x_3) \star \phi(x_4)\star \phi(x_5)\star \phi(x_6)]|0\rangle^{(0)}_\star \ .\label{6Point}
\end{equation}
Applying \eqref{eq:G2kstar}, a calculation similar to that for the four-point function gives the slightly cumbersome result
\begin{align}\label{6PointFull}
\begin{split}
G_6^\star (x_1,\dots,x_6)^{(0)} & = \bcontraction{}{\phi_1}{}{\phi_2} \phi_1\, \phi_2\, \bcontraction{}{\phi_3}{}{\phi_4} \phi_3\, \phi_4 \, \bcontraction{}{\phi_5}{}{\phi_6} \phi_5\, \phi_6
+ \bcontraction{}{\phi_1}{}{\phi_2} \phi_1\, \phi_2\, \bcontraction{}{\phi_3}{}{sfR_\alpha (\phi_5)}  \phi_3\,\sfR_\alpha (\phi_5) \, \bcontraction{}{\sfR^\alpha(\phi_4)}{}{\phi_6} \sfR^\alpha(\phi_4)\, \phi_6 \\
& \quad \, 
+ \bcontraction{}{\phi_1}{}{\phi_2} \phi_1\, \phi_2 \, \bcontraction{}{\phi_3}{}{\phi_6} \phi_3 \, \phi_6\, \bcontraction{}{\phi_4}{}{\phi_5} \phi_4\, \phi_5 + \bcontraction{}{\phi_1}{}{\sfR_\alpha(\phi_3)} \phi_1\, \sfR_\alpha(\phi_3)\, \bcontraction{}{\sfR^\alpha(\phi_2)}{}{\phi_4} \sfR^\alpha(\phi_2)\, \phi_4 \,\bcontraction{}{\phi_5}{}{\phi_6} \phi_5\, \phi_6 \\
& \quad \, 
+ \bcontraction{}{\phi_1}{}{\sfR_\alpha(\phi_3)} \phi_1\, \sfR_\alpha(\phi_3)\, \bcontraction{}{\sfR^\alpha(\phi_2)}{}{\sfR_\beta(\phi_5)} \sfR^\alpha(\phi_2)\, \sfR_\beta(\phi_5)\, \bcontraction{}{\sfR^\beta(\phi_4)}{}{\phi_6} \sfR^\beta(\phi_4)\, \phi_6 + \bcontraction{}{\phi_1}{}{\phi_6} \phi_1\, \phi_6 \, \bcontraction{}{\phi_2}{}{\phi_5} \phi_2\, \phi_5\, \bcontraction{}{\phi_3}{}{\phi_4} \phi_3\, \phi_4
\\
& \quad \, + \bcontraction{}{\phi_1}{}{\sfR_\alpha(\phi_3)} \phi_1\, \sfR_\alpha(\phi_3)\, \bcontraction{}{\sfR^\alpha(\phi_2)}{}{\phi_6} \sfR^\alpha(\phi_2)\, \phi_6\, \bcontraction{}{\phi_4}{}{\phi_5} \phi_4\, \phi_5 + \bcontraction{}{\phi_1}{}{\phi_4} \phi_1 \,\phi_4\, \bcontraction{}{\phi_2}{}{\phi_3} \phi_2\, \phi_3\, \bcontraction{}{\phi_5}{}{\phi_6} \phi_5\, \phi_6 \\
& \quad \, 
+ \bcontraction{}{\phi_1}{}{\sfR_\alpha\,\sfR_\beta(\phi_4)} \phi_1\, \sfR_\alpha\,\sfR_\beta(\phi_4)\, \bcontraction{}{\sfR^\alpha(\phi_2)}{}{\sfR_\sigma(\phi_5)} \sfR^\alpha(\phi_2)\, \sfR_\sigma(\phi_5)\, \bcontraction{}{\sfR^\sigma\,\sfR^\beta(\phi_3)}{}{\phi_6}  \sfR^\sigma\,\sfR^\beta(\phi_3)\, \phi_6 \\
& \quad \, 
+ \bcontraction{}{\phi_1}{}{\sfR_\alpha(\phi_5)} \phi_1\, \sfR_\alpha(\phi_5)\, \bcontraction{}{\phi_2}{}{\phi_3}  \phi_2 \, \phi_3\, \bcontraction{}{\sfR^\alpha(\phi_4)}{}{\phi_6} \sfR^\alpha(\phi_4)\, \phi_6 
+ \bcontraction{}{\phi_1}{}{\sfR_\alpha\,\sfR_\beta (\phi_4)} \phi_1\, \sfR_\alpha\,\sfR_\beta (\phi_4)\, \bcontraction{}{\sfR^\alpha(\phi_2)}{}{\phi_6}  \sfR^\alpha(\phi_2)\, \phi_6\, \bcontraction{}{\sfR^\beta(\phi_3)}{}{\phi_5} \sfR^\beta(\phi_3)\, \phi_5 \\
& \quad \, + \bcontraction{}{\phi_1}{}{\sfR_\alpha(\phi_5)} \phi_1\, \sfR_\alpha(\phi_5)\, \bcontraction{}{\phi_2}{}{\sfR_\beta(\phi_4)}  \phi_2\, \sfR_\beta(\phi_4)\, \bcontraction{}{\sfR^\beta\,\sfR^\alpha(\phi_3)}{}{\phi_6} \sfR^\beta\,\sfR^\alpha(\phi_3)\, \phi_6 
+ \bcontraction{}{\phi_1}{}{\sfR_\alpha(\phi_5)} \phi_1\, \sfR_\alpha(\phi_5)\, \bcontraction{}{\sfR^\alpha(\phi_2)}{}{\phi_6} \sfR^\alpha(\phi_2)\, \phi_6\, \bcontraction{}{\phi_3}{}{\phi_4} \phi_3\, \phi_4 \\
& \quad \,  + \bcontraction{}{\phi_1}{}{\phi_6} \phi_1\, \phi_6 \, \bcontraction{}{\phi_2}{}{\phi_3} \phi_2\, \phi_3\, \bcontraction{}{\phi_4}{}{\phi_5} \phi_4 \, \phi_5 
+ \bcontraction{}{\phi_1}{}{\phi_6} \phi_1\, \phi_6\, \bcontraction{}{\phi_2}{}{\sfR_\alpha(\phi_4)} \phi_2\, \sfR_\alpha(\phi_4)\, \bcontraction{}{\sfR^\alpha(\phi_3)}{}{\phi_5} \sfR^\alpha(\phi_3)\, \phi_5 
 \ .
\end{split}
\end{align}
In the commutative limit, when the $\RR$-matrix reduces to the identity operator, our results (\ref{4PointFull}) and (\ref{6PointFull}) reduce to the well-known expressions for the four-point and the six-point functions in standard free scalar field theory. The noncommutative deformation enters through the permutations of fields before contracting them.

\subsection{Braided homological perturbation theory}
\label{sec:braidedBV}

Following~\cite{Doubek:2017naz,Macrelli:2019afx,SzaboAlex,Okawa:2022sjf} we now explain how to compute correlation functions of the interacting braided scalar field theory via the technique of `homotopy transfer'. 
We start from the cohomology $H^\bullet(\CCL_0^\star)$ of the abelian $L_\infty$-algebra $\CCL^\star_0$, which describes the classical vacua of the free scalar field theory on $\FR^{1,d-1}$. This is also an abelian $L_\infty$-algebra, and from \eqref{eq:scalarcomplex} it follows that it is also concentrated in degrees~$1$ and~$2$, given by the solution space $H^1(L)=\ker(\ell_1)$ of the massive Klein--Gordon equation $\square\,\phi+m^2\,\phi=0$ and the space $H^2(L)={\rm coker}(\ell_1)$ of on-shell Maurer--Cartan expansions. The underlying $2$-term cochain complex of $H^\bullet(\CCL_0^\star)$ is
\begin{align*}
\ker\big(\square+m^2\big)[-1]  \xrightarrow{ \ 0 \ }  {\rm coker}\big(\square+m^2\big)[-2] \ .
\end{align*}

To describe correlation functions in the path integral framework, we need to define a $U\iso(d-1)$-equivariant projection $\sfp:L\to H^\bullet(L)$ of degree~$0$ and a $U\iso(d-1)$-invariant contracting homotopy $\sfh:L\to L$ of degree~$-1$. For this, we denote the scalar Feynman propagator $\sgreen:\Omega^0(\FR^{1,d-1})\to\Omega^0(\FR^{1,d-1})$ by
\begin{align*}
\sgreen = -\frac1{\square+m^2} \qquad \mbox{with} \quad \tilde \sgreen(k) = \frac1{k^2-m^2} \ ,
\end{align*}
where $\tilde \sgreen(k)$ are the eigenvalues of the Green operator $\sgreen$ when acting on plane wave eigenfunctions of the form $\e^{\,\ii\, k\cdot x}$. It satisfies
\begin{align*}
\ell_1\circ \sgreen = -\big(\square+m^2\big)\circ \sgreen = \id_{\Omega^0(\FR^{1,3})} \ .
\end{align*}

If we were to compute scattering amplitudes, then the first component of the projection $\sfp^\swone:L^1\to H^1(L)$ should be taken to be the projection $\id_{\Omega^0(\FR^{1,3})} - \sgreen\circ\ell_1$ to on-shell states. However, for the purposes of computing correlation functions, we should project to the trivial vacuum $\phi=0$. This is a consequence of the fact that correlators can be equivalently computed by Wick rotating to Euclidean signature, where the kernel and cokernel of the kinetic operator $\ell_1$ become trivial, and this was already used in Section~\ref{sec:braidedWick}. Similarly, whereas the second component $\sfp^\swtwo:L^2\to H^2(L)$ could be taken to be the natural projection induced by the quotient map to ${\rm coker}(\ell_1)$, we use the trivial projection to $\phi^+=0$. Hence we define
\begin{align*}
\sfp^\swone = 0 = \sfp^\swtwo \ ,
\end{align*}
or more accurately we restrict the cochain complex of $H^\bullet(\CCL_0^\star)$ to its trivial subspaces.

With these choices, the only non-vanishing component of the contracting homotopy $\sfh^\swtwo:L^2\to L^1$ is given by the propagator $\sfh^\swtwo = \sgreen$. Explicitly
\begin{align}\label{eq:htwo}
\sfh^{\swtwo}(\phi^+)(x) = -\frac1{\square+m^2}\, \phi^+(x) = \int_{\FR^{1,d-1}} \, \d^d y \ \int_k \, \frac{\e^{-\ii\,k\cdot(x-y)}}{k^2-m^2} \ \phi^+(y) \ ,
\end{align}
for $\phi^+\in L^2$.

We apply the braided homological perturbation theory developed by~\cite{SzaboAlex}. For this, we need to extend the maps $\sfp$ and $\sfh$ to the space of functionals on $L$; in this paper we will only compute correlation functions of polynomial observables, hence we restrict to the braided symmetric algebra $\Sym_\RR L[2]=\bigoplus_{n\geq0} \, \Sym_\RR^n L[2]$ over $\FR[[\nu]]$. The data above induce a trivial projection $\sP:\Sym_\RR L[2]\to \Sym_\RR H^\bullet(L[2])$ by
\begin{align*}
\sP(1)=1 \qquad \mbox{and} \qquad \sP(\varphi_1\odot_\star\cdots\odot_\star\varphi_n) = 0 \ ,
\end{align*}
along with a contracting homotopy $\sH:\Sym_\RR L[2]\to\Sym_\RR L[2]$ through
\begin{align}\label{eq:sfH}
\begin{split}
\sH(1)&=0 \ , \\[4pt]
\sH(\varphi_1\odot_\star\cdots\odot_\star\varphi_n) &= \frac1n \, \sum_{a=1}^n \, \pm \ \varphi_1\odot_\star\cdots\odot_\star\varphi_{a-1}\odot_\star\sfh(\varphi_a) \odot_\star\varphi_{a+1} \odot_\star\cdots\odot_\star \varphi_n \ ,
\end{split}
\end{align}
for all $\varphi_a\in L[2]$, with $a=1,\dots,n$; we used $U\mathfrak{iso}(d-1)$-invariance of $\sfh$ in \eqref{eq:sfH} which trivializes the actions of $\RR$-matrices. Note that on generators the twisted symmetric product $\odot_\star$ is braided graded commutative:
\begin{align*}
\varphi_a\odot_\star\varphi_b = (-1)^{|\varphi_a|\,|\varphi_b|} \ \sfR_\alpha(\varphi_b)\odot_\star\sfR^\alpha(\varphi_a) \ .
\end{align*}

We perturb the free differential $\ell_1$ to the `quantum' differential
\begin{align*}
Q_{\mbf\delta} = \ell_1 + {\mbf\delta}
\end{align*}
on $L[2]$, where the formal $U\iso(d-1)$-invariant perturbation $\mbf\delta$ will be specified below. The braided extension of the homological perturbation lemma~\cite{SzaboAlex} then constructs the perturbed  projection map $\sP+\sP_{\mbf\delta}$, where $\sP_{\mbf\delta}:\Sym_\RR L[2]\to\Sym_\RR H^\bullet(L[2])$ is given by
\begin{align*}
\sP_{\mbf\delta} = \sP\,\big(\id_{\Sym_\RR L[1]} - {\mbf\delta}\,\sH\big)^{-1} \, \mbf\delta \, \sH \ ,
\end{align*}
which in the classical case gives the path integral~\cite{Doubek:2017naz}. 

We thus define the $n$-point correlation functions of the braided quantum field theory by
\begin{align}\label{eq:BQFTnpoint}
\begin{split}
G_n^\star(x_1,\dots,x_n) &= \langle0|{\rm T}[\phi(x_1)\star\cdots\star\phi(x_n)]|0\rangle_\star := \sP_{\mbf\delta}(\delta_{x_1}\odot_\star\cdots\odot_\star\delta_{x_n}) \\[4pt]
&= \sum_{p=1}^\infty \, \sP\big(({\mbf\delta} \,\sH)^p(\delta_{x_1}\odot_\star\cdots\odot_\star\delta_{x_n})\big) \ ,
\end{split}
\end{align}
where $\delta_{x_a}(x):=\delta(x-x_a)$ are Dirac distributions supported at the insertion points $x_a$ of the physical field $\phi\in L^1$. Because only $\sP(1)=1$ is non-zero, this is a function in $\Omega^0\big((\FR^{1,d-1})^{\times n}\big)[[\nu]]$.\footnote{As usual in quantum field theory, the amputated correlation functions are distributions in position space, and so should be properly defined by smearing them with suitable test functions, as done in~\cite{SzaboAlex}. Here we follow the more conventional physics practice of defining `localized' correlation functions.}
We are interested in two perturbations $\mbf\delta$ of $\ell_1$.

\paragraph{Free theory.}
The free braided scalar field theory of Section~\ref{sec:braidedWick} is recovered from the perturbation
\begin{align*}
{\mbf\delta} = \ii\,\hbar\,{\mathsf \Delta}_\BV \ ,
\end{align*}
where $\BVL:\Sym_\RR L[2]\to (\Sym_\RR L[2])[1]$ is the braided BV Laplacian defined by 
\begin{align}\label{eq:BVL}
\begin{split}
\BVL(1)=0 & \qquad , \qquad \BVL(\varphi_1)=0 \qquad , \qquad  \BVL(\varphi_1\odot_\star\varphi_2) = \langle\varphi_1,\varphi_2\rangle_\star \ , \\[4pt]
\BVL(\varphi_1\odot_\star\cdots\odot_\star\varphi_n) & = \sum_{a<b} \, \pm \, \langle \varphi_a,\sfR_{\alpha_{a+1}}\cdots\sfR_{\alpha_{b-1}}(\varphi_b)\rangle_\star \  \varphi_1\odot_\star\cdots\odot_\star\varphi_{a-1}\\ & \hspace{1.2cm} \odot_\star\sfR^{\alpha_{a+1}}(\varphi_{a+1})\odot_\star\cdots\odot_\star\sfR^{\alpha_{b-1}}(\varphi_{b-1})\odot_\star \varphi_{b+1}\odot_\star\cdots\odot_\star\varphi_n \ ,
\end{split}
\end{align}
for all $\varphi_1,\dots,\varphi_n\in L[2]$. The BV Laplacian satisfies the two key properties $(\BVL)^2=0$ and $\BVL\circ\ell_1 = -\ell_1\circ\BVL$ which guarantee that $Q_0=\ell_1+\ii\,\hbar\,\BVL$ is a differential, $(Q_0)^2=0$.

Since the braided BV Laplacian contracts fields pairwise and lowers the symmetric algebra degree from $n$ to $n-2$, it is clear that in this case the correlation functions \eqref{eq:BQFTnpoint} vanish unless $n=2k$ is even, in which case the free braided $2k$-point functions are then defined by
\begin{align}\label{eq:freeBQFT2k}
\begin{split}
G_{2k}^\star(x_1,\dots,x_{2k})^{(0)} = \langle0|{\rm T}[\phi(x_1)\star\cdots\star\phi(x_{2k})]|0\rangle_\star^{(0)} := (\ii\,\hbar\,\BVL\,\sH)^k(\delta_{x_1}\odot_\star\cdots\odot_\star\delta_{x_{2k}}) \ .
\end{split}
\end{align}
It is not difficult to check using \eqref{eq:scalarpairing}, \eqref{eq:htwo}, \eqref{eq:sfH} and \eqref{eq:BVL} that \eqref{eq:freeBQFT2k} reproduces the braided Wick expansion \eqref{eq:G2kstar} by iterating the basic operation
\begin{align*}
\begin{split}
\ii\,\hbar\,\BVL\,\sH(\delta_{x_1}\odot_\star\cdots\odot_\star\delta_{x_{2k}}) &= \frac1{2k} \, \sum_{a<b} \, \bcontraction{}{\phi_a}{}{\sfR_{\alpha_{a+1}}\cdots\sfR_{\alpha_{b-1}}(\phi_b)} \phi_a\,\sfR_{\alpha_{a+1}}\cdots\sfR_{\alpha_{b-1}}(\phi_b) \  \delta_{x_1}\odot_\star\cdots\odot_\star\delta_{x_{a-1}} \\ & \qquad \ \ \odot_\star\sfR^{\alpha_{a+1}}(\delta_{x_{a+1}})\odot_\star\cdots\odot_\star\sfR^{\alpha_{b-1}}(\delta_{x_{b-1}})\odot_\star \delta_{x_{b+1}}\odot_\star\cdots\odot_\star\delta_{x_{2k}} \ ,
\end{split}
\end{align*}
where we again use the notation 
\begin{align*}
\bcontraction{}{\phi_a}{}{\phi_b} \phi_a\,\phi_b:=\langle0| {\rm T}[\phi(x_a) \, \phi(x_b) ]|0\rangle^{(0)}=-\ii\,\hbar\,\sgreen(x_a-x_b)
\end{align*} 
for the free propagator, and the Koszul sign factors are trivial for antifields $\varphi_a=\delta_{x_a}\in L^2[2]$; we used the fact that the only non-zero pairings are $\langle\delta_{x_a},\sgreen(\delta_{x_b})\rangle_\star = \sgreen(x_a-x_b)$.

For the two-point function one finds immediately the free propagator
\begin{align*}
G^\star_2(x_1,x_2)^{(0)} = \ii\,\hbar\,\BVL\,\sH(\delta_{x_1}\odot_\star\delta_{x_2}) = -\ii\,\hbar\, \sgreen(x_1-x_2) = -\ii\,\hbar\, \int_k \, \frac{\e^{-\ii\,k\cdot(x_1-x_2)}}{k^2-m^2} \ ,
\end{align*}
in agreement with \eqref{2Pointcalc}.

For the four-point function we start from
\begin{align*}
\begin{split}
\ii\,\hbar\,\BVL\,\sH(\delta_{x_1}\odot_\star\cdots\odot_\star\delta_{x_{4}}) &= \tfrac12\, \big(\bcontraction{}{\phi_1}{}{\phi_2} \phi_1\,\phi_2\ \ \delta_{x_3}\odot_\star\delta_{x_4} + \bcontraction{}{\phi_1}{}{\sfR_\alpha(\phi_3)} \phi_1\,\sfR_\alpha(\phi_3) \ \sfR^\alpha(\delta_{x_2})\odot_\star\delta_{x_4} \\ & \hspace{1cm} + \bcontraction{}{\phi_1}{}{\sfR_\alpha\,\sfR_\beta(\phi_4)} \phi_1\,\sfR_\alpha\,\sfR_\beta(\phi_4) \ \sfR^\alpha(\delta_{x_2})\odot_\star\sfR^\beta(\delta_{x_3}) + \bcontraction{}{\phi_2}{}{\phi_3} \phi_2\,\phi_3 \ \delta_{x_1}\odot_\star\delta_{x_4} \\
& \hspace{1cm} + \bcontraction{}{\phi_2}{}{\sfR_\alpha(\phi_4)} \phi_2\,\sfR_\alpha(\phi_4) \ \delta_{x_1}\odot_\star\sfR^\alpha(\delta_{x_3}) + \bcontraction{}{\phi_3}{}{\phi_4} \phi_3\,\phi_4 \ \delta_{x_1}\odot_\star\delta_{x_2} \big) \ .
\end{split}
\end{align*}
Using
\begin{align*}
\ii\,\hbar\,\BVL\,\sH(\delta_{x_a}\odot_\star\delta_{x_b}) = \bcontraction{}{\phi_a}{}{\phi_b} \phi_a\,\phi_b
\end{align*}
we then find that the four-point function is given by
\begin{align*}
\begin{split}
G_4^\star(x_1,x_2,x_3,x_4)^{(0)} &= (\ii\,\hbar\,\BVL\,\sH)^2 \, (\delta_{x_1}\odot_\star\delta_{x_2}\odot_\star\delta_{x_3}\odot_\star\delta_{x_4})  \\[4pt]
&= \tfrac12\,\big(2\,\bcontraction{}{\phi_1}{}{\phi_2} \phi_1\,\phi_2\,\bcontraction{}{\phi_3}{}{\phi_4} \phi_3\,\phi_4 + \bcontraction{}{\phi_1}{}{\sfR_\alpha(\phi_3)} \phi_1\,\sfR_\alpha(\phi_3) \, \bcontraction{}{\sfR^\alpha(\phi_2)}{}{\phi_4} \sfR^\alpha(\phi_2)\,\phi_4 + \bcontraction{}{\phi_1}{}{\sfR^\alpha(\phi_3)} \phi_1\,\sfR^\alpha(\phi_3)\,\bcontraction{}{\phi_2}{}{\sfR_\alpha(\phi_4)} \phi_2\,\sfR_\alpha(\phi_4) \\
& \hspace{1cm} + \bcontraction{}{\phi_1}{}{\sfR_\alpha\,\sfR_\beta(\phi_4)} \phi_1\,\sfR_\alpha\,\sfR_\beta(\phi_4) \, \bcontraction{}{\sfR^\alpha(\phi_2)}{}{\sfR^\beta(\phi_3)} \sfR^\alpha(\phi_2)\,\sfR^\beta(\phi_3) + \bcontraction{}{\phi_1}{}{\phi_4} \phi_1\,\phi_4 \, \bcontraction{}{\phi_2}{}{\phi_3} \phi_2\,\phi_3 \big) \ .
\end{split}
\end{align*}
We now employ the same $\RR$-matrix manipulations as in Section~\ref{sec:braidedWick}. Using
\begin{align*}
\partial_i^2\big(\bcontraction{}{\phi_2}{}{\phi_4} \phi_2\,\phi_4\big) = -\partial_i^4\big(\bcontraction{}{\phi_2}{}{\phi_4} \phi_2\,\phi_4\big)
\end{align*}
we conclude that the second and third terms are equal; this also follows abstractly from the triangular Hopf algebra identity $(S_\CF\otimes\id_{U\iso(d-1)})\,\RR=\RR^{-1}=\RR_{21}$ and $U\iso(d-1)$-invariance of the two-point functions. From the identity \eqref{eq:4ptid} we see that the last two terms are also equal. 

Altogether we arrive at
\begin{align*}
G_4^\star(x_1,x_2,x_3,x_4)^{(0)} = \bcontraction{}{\phi_1}{}{\phi_2} \phi_1\,\phi_2\,\bcontraction{}{\phi_3}{}{\phi_4} \phi_3\,\phi_4 + \bcontraction{}{\phi_1}{}{\sfR_\alpha(\phi_3)} \phi_1\,\sfR_\alpha(\phi_3) \, \bcontraction{}{\sfR^\alpha(\phi_2)}{}{\phi_4} \sfR^\alpha(\phi_2)\,\phi_4 + \bcontraction{}{\phi_1}{}{\phi_4} \phi_1\,\phi_4 \, \bcontraction{}{\phi_2}{}{\phi_3} \phi_2\,\phi_3 \ ,
\end{align*}
which agrees with \eqref{4PointFull}, and also with~\cite[Equation~(5.40)]{SzaboAlex}.
It is straightforward, if lengthy, to extend this calculation to compute the six-point function \eqref{6PointFull}, and also to higher order correlation functions. In this way we recover the braided Wick theorem of Section~\ref{sec:braidedWick}.

\paragraph{Interacting theory.} 
An interacting scalar field theory on $\FR^{1,d-1}$ is captured by the perturbation
\begin{align*}
\mbf\delta = \ii\,\hbar\,\BVL + \{\CS _{\rm int},-\}_\star \ ,
\end{align*}
where the operator $\{\CS _{\rm int},-\}_\star$ is constructed in the following way. 

For definiteness, we consider braided $\lambda\,\phi^4$-theory in four dimensions, but the methods easily extend to scalar field theories with arbitrary polynomial interactions in any dimension. This amounts to extending $\CCL_0^\star$ to a non-abelian braided $L_\infty$-algebra $\CCL^\star$ with the single non-vanishing higher bracket $\ell_3^\star$ defined by
\begin{align*}
\ell_3^\star(\phi_1,\phi_2,\phi_3) = -\lambda \, \phi_1\star\phi_2\star\phi_3 
\end{align*}
for all $\phi_1,\phi_2,\phi_3\in L^1$. The braided version of the homotopy Maurer--Cartan action functional \eqref{eq:braidedaction} is then
\begin{align}\label{eq:phi4action}
\begin{split}
S_\star(\phi) &= \tfrac12\,\langle\phi,\ell_1(\phi)\rangle_\star - \tfrac1{24}\,\langle\phi,\ell_3^\star(\phi,\phi,\phi)\rangle_\star =: S_0(\phi) + S_{\rm int}(\phi) \\[4pt]
&= \int_{\FR^{1,3}} \, \dd^4x \ \frac12\, \phi\,\big(-\square - m^2\big)\,\phi + \frac\lambda{4!} \, \phi\star\phi\star\phi\star\phi \ ,
\end{split}
\end{align}
which is just the standard noncommutative scalar $\lambda\,\phi^4$-theory~\cite{U1Reviewa}. We want to ``transfer'' the bracket $\ell_3^\star$ to give a new braided $L_\infty$-algebra on $H^\bullet(L)$, called the \emph{minimal model} of the braided $L_\infty$-algebra $\CCL^\star$, and at the same time ``quantize'' it.

For the computation of interacting correlation functions, we extend the braided $L_\infty$-algebra structure on $L[[\nu]]$ to $(\Sym_\RR L[2])\otimes L[[\nu]]$ via the non-zero brackets~\cite{Giotopoulos:2021ieg}
\begin{align*}
\begin{split}
{\mbf\ell}_1(a_1\otimes\phi_1) &= \pm\,a_1\otimes\ell_1(\phi_1) \ , \\[4pt]
{\mbf\ell}_3^\star(a_1\otimes\phi_1,a_2\otimes\phi_2,a_3\otimes\phi_3) &= \pm\,\big(a_1\odot_\star\sfR_\alpha(a_2)\odot_\star\sfR_\beta\,\sfR_\sigma(a_3)\big)\otimes\ell_3^\star\big(\sfR^\beta\,\sfR^\alpha(\phi_1),\sfR^\sigma(\phi_2),\phi_3\big) \ ,
\end{split}
\end{align*}
for $a_1,a_2,a_3\in\Sym_\RR L[2]$ and $\phi_1,\phi_2,\phi_3\in L^1[[\nu]]$; again we write $\pm$ for the Koszul sign factors determined by the gradings of the elements involved in all operations. Similarly, the cyclic structure is extended via the non-zero $\Sym_\RR L[2]$-valued pairing
\begin{align}
\langle\!\!\langle a_1\otimes\phi,a_2\otimes\phi^+\rangle\!\!\rangle_\star = \pm\, \big(a_1\odot_\star\sfR_\alpha(a_2)\big) \, \langle\sfR^\alpha(\phi),\phi^+\rangle_\star \ ,
\end{align}
for $a_1,a_2\in\Sym_\RR L[2]$, $\phi\in L^1[[\nu]]$ and $\phi^+\in L^2[[\nu]]$.

The antibracket is the braided graded Poisson bracket $\{-,-\}_\star:\Sym_\RR L[2]\otimes \Sym_\RR L[2]\to (\Sym_\RR L[2])[1]$ defined by setting 
\begin{align*}
\{\varphi_a,\varphi_b\}_\star = \langle\varphi_a , \varphi_b\rangle_\star =\pm\,\{\sfR_\alpha(\varphi_b),\sfR^\alpha(\varphi_a)\}_\star
\end{align*}
for $\varphi_a\in L[2]$, and extending this to all of $\Sym_\RR L[2]$ as a braided graded Lie bracket which is a braided graded derivation on $\Sym_\RR L[2]$ in each of its slots. For example
\begin{align}\label{eq:braidedder}
\{\varphi_1,\varphi_2\odot_\star\varphi_3\}_\star = \langle\varphi_1,\varphi_2\rangle_\star\odot_\star\varphi_3 \pm \sfR_\alpha(\varphi_2)\odot_\star\langle\sfR^\alpha(\varphi_1),\varphi_3\rangle_\star \ .
\end{align}
The antibracket is compatible with the differential~$\ell_1$, extended as a graded derivation to all of $\Sym_\RR L[2]$, as a consequence of cyclicity of the inner product $\langle-,-\rangle_\star$. It is also related to the braided BV Laplacian through
\begin{align}\label{eq:BVLantibracket}
\BVL(a_1\odot_\star a_2) = \BVL(a_1)\odot_\star a_2 + (-1)^{\vert a_1\vert}\, 
a_1\odot_\star\BVL(a_2) +  \{a_1,a_2\}_\star \ ,
\end{align}
for all $a_1,a_2\in \Sym_\RR L[2]$.

Via Fourier transformation, we introduce the basis of plane waves $\tte_k(x)=\e^{-\ii\, k\cdot x}$ for $L^1$ and the basis 
\begin{align*}
\tte^k(x)=\tte_k^*(x)=\tte_{-k}(x)=\e^{\,\ii\,k\cdot x}
\end{align*}
for $L^2$. These bases are dual with respect to the inner product \eqref{eq:scalarpairing}, in the sense that
\begin{align*}
\int_p \ \langle \tte_k,\tte^p\rangle_\star \ \tte_p = \tte_k \qquad \mbox{and} \qquad \int_k \ \tte^k \ \langle \tte_k,\tte^p\rangle_\star = \tte^p \ ,
\end{align*}
where throughout we use
\begin{align*}
\int_{\FR^{1,3}} \, \dd^4x \ \e^{\pm\,\ii\,k\cdot x} = (2\pi)^4 \, \delta(k) \ .
\end{align*}
The star-products among basis fields are
\begin{align}\label{eq:ekstarep}
\tte_k\star \tte_p = \e^{-\frac\ii2 \, k\cdot\theta\, p} \ \tte_{k+p} \ ,
\end{align}
where $k\cdot\theta\,p:=\nu\,k_\mu\,\theta^{\mu\lambda}\,p_\lambda=-p\cdot\theta\, k$, while the action of the inverse $\RR$-matrix on them is given by
\begin{align}\label{eq:Rekotimesep}
\RR^{-1}(\tte_k\otimes \tte_p) = \sfR_\alpha(\tte_k)\otimes\sfR^\alpha(\tte_p) = \e^{\,\ii\,k\cdot\theta\,p} \ \tte_k\otimes \tte_p \ .
\end{align}

Using this basis we now define the contracted coordinate functions $\mbf\xi\in(\Sym_\RR L[2])\otimes L[[\nu]]$ in degree~$1$ by the formal $U\mathfrak{iso}(3)$-invariant expression
\begin{align*}
\mbf\xi = \int_k \, \big(\tte_k\otimes \tte^k + \tte^k\otimes \tte_k\big) \ .
\end{align*}
Using the braided Maurer--Cartan action functional \eqref{eq:phi4action}, we define the the interacting part of the BV action functional $\CS _{\rm int}\in\Sym_\RR L[2]$ in degree~$0$ by the $U\mathfrak{iso}(3)$-invariant element
\begin{align}\label{eq:Sintxi}
\CS _{\rm int} := -\tfrac1{24} \, \langle\!\!\langle \mbf\xi \,,\,\mbf\ell^\star_3(\mbf\xi,\mbf\xi,\mbf\xi)\rangle\!\!\rangle_\star \ .
\end{align}
As discussed in~\cite{SzaboAlex,Giotopoulos:2021ieg}, this satisfies the classical master equation
\begin{align*}
\ell_1(\CS _{\rm int}) + \tfrac12\,\{\CS _{\rm int},\CS _{\rm int}\}_\star=0 \ , 
\end{align*}
and it is annihilated by the braided BV Laplacian, $\BVL(\CS _{\rm int})=0$. As a consequence, the operator 
\begin{align*}
Q_{\rm int}=\ell_1 + \ii\,\hbar\,\BVL + \{\CS _{\rm int},-\}_{\star}
\end{align*}
is a differential, $(Q_{\rm int})^2=0$, which describes the correlation functions in terms of a braided quantum $L_\infty$-algebra.

Calculating \eqref{eq:Sintxi} explicitly using \eqref{eq:scalarpairing}, \eqref{eq:ekstarep} and \eqref{eq:Rekotimesep} we find
\begin{align*}
\begin{split}
\CS _{\rm int}&= -\frac1{4!} \, \int_{k_1,\dots,k_4} \, \langle\!\!\langle \tte^{k_1}\otimes \tte_{k_1}\,,\,\mbf\ell_3^\star(\tte^{k_2}\otimes \tte_{k_2},\tte^{k_3}\otimes \tte_{k_3},\tte^{k_4}\otimes \tte_{k_4})\rangle\!\!\rangle_\star \\[4pt]
&= -\frac1{4!} \, \int_{k_1,\dots,k_4} \, {\langle\!\!\langle} \tte^{k_1}\otimes \tte_{k_1}\,,\, \big(\tte^{k_2}\odot_\star\sfR_\alpha(\tte^{k_3})\odot_\star \sfR_\beta\,\sfR_\sigma(\tte^{k_4})\big)\otimes\ell_3^\star\big(\sfR^\beta\,\sfR^\alpha(\tte_{k_2}),\sfR^\sigma(\tte_{k_3}),\tte_{k_4}\big){\rangle\!\!\rangle}_\star \\[4pt]
&= -\frac1{4!} \, \int_{k_1,\dots,k_4} \, \e^{\,\ii\,k_2\cdot\theta\,k_3+\ii\,k_2\cdot\theta\, k_4 + \ii\,k_3\cdot\theta\,k_4} \, \langle\!\!\langle \tte^{k_1}\otimes \tte_{k_1} \,,\, (\tte^{k_2}\odot_\star \tte^{k_3}\odot_\star \tte^{k_4}) \otimes \ell_3^\star(\tte_{k_2},\tte_{k_3},\tte_{k_4})\rangle\!\!\rangle_\star \\[4pt]
&= -\frac1{4!} \, \int_{k_1,\dots,k_4} \, \e^{\,\ii\,k_2\cdot\theta\,k_3+\ii\,k_2\cdot\theta\, k_4+ \ii\,k_3\cdot\theta\,k_4} \\
& \hspace{5cm} \times \tte^{k_1}\odot_\star\sfR_\alpha(\tte^{k_2}\odot_\star \tte^{k_3}\odot_\star \tte^{k_4}) \, \langle\sfR^\alpha(\tte_{k_1}),\ell_3^\star(\tte_{k_2},\tte_{k_3},\tte_{k_4})\rangle_\star \\[4pt]
&= \frac\lambda{4!} \, \int_{k_1,\dots,k_4} \, \e^{\,\ii\,\sum\limits_{a<b} \, k_a\cdot\theta\, k_b} \ \tte^{k_1}\odot_\star \tte^{k_2}\odot_\star \tte^{k_3}\odot_\star \tte^{k_4} \ \langle \tte_{k_1},\tte_{k_2}\star \tte_{k_3}\star \tte_{k_4}\rangle_\star \\[4pt]
&=: \int_{k_1,\dots,k_4} \, V(k_1,k_2,k_3,k_4) \ \tte^{k_1}\odot_\star \tte^{k_2}\odot_\star \tte^{k_3}\odot_\star \tte^{k_4}  \ .
\end{split}
\end{align*}

The interaction vertex 
\begin{align}\label{eq:Vint}
V(k_1,k_2,k_3,k_4) = \frac\lambda{4!} \ \e^{\,\frac\ii2\,\sum\limits_{a<b} \, k_a\cdot\theta\, k_b} \ (2\pi)^4 \, \delta(k_1+k_2+k_3+k_4)
\end{align}
coincides with the vertex of the standard noncommutative $\lambda\,\phi_4^{\star4}$ theory~\cite{U1Reviewa}. It has the braided symmetry
\begin{align}\label{eq:Vbraidedsym}
V(\ \  k_{a+1},k_a\ \ ) = \e^{-\ii\,k_a\cdot\theta\,k_{a+1}} \ V(k_1,k_2,k_3,k_4)
\end{align}
under interchange of any pair of neighbouring momenta, and also the cyclic symmetry
\begin{align}\label{eq:Vcyclicsym}
V(k_1,k_2,k_3,k_4) = V(k_4,k_1,k_2,k_3)
\end{align}
which follows from momentum conservation.

The interacting correlation functions of the braided quantum field theory are now given by
\begin{align}\label{eq:intcorrelationfn}
\begin{split}
G_n^\star(x_1,\dots,x_n)^{\rm int}&=\langle 0|{\rm T}[\phi(x_1)\star\cdots\star\phi(x_n)]|0\rangle^{\rm int} \\[4pt]
:\!&= \sum_{p=1}^\infty \, \sP\big((\ii\,\hbar\,\BVL\,\sH + \{\CS _{\rm int},-\}_\star\,\sH)^p(\delta_{x_1}\odot_\star\cdots\odot_\star\delta_{x_n})\big) \ .
\end{split}
\end{align}
The interaction terms are computed by using the non-zero pairings 
\begin{align}\label{eq:eGdeltapair}
\langle \tte^{k_a},\sgreen(\delta_{x_b})\rangle_\star = \e^{\,\ii\,k_a\cdot x_b} \,\tilde \sgreen(k_a)  = \langle \sgreen(\tte^{k_a}),\delta_{x_b}\rangle_\star \ , 
\end{align}
together with the braided derivation property \eqref{eq:braidedder} of the  antibracket and the  symmetry properties \eqref{eq:Vbraidedsym}--\eqref{eq:Vcyclicsym} of the interaction vertex $V(k_1,k_2,k_3,k_4)$ to get the basic operation
\begin{align*}
\begin{split}
\{\CS _{\rm int} , \sH(\delta_{x_1}\odot_\star\cdots\odot_\star\delta_{x_n})\}_\star &= -\frac4n \, \sum_{a=1}^n \ \int_{k_1,\dots,k_4} \, V(k_1,\dots,k_4) \, \e^{\,\ii\,k_1\cdot x_a} \,\tilde \sgreen(k_1) \ \delta_{x_1}\odot_\star\cdots\odot_\star\delta_{x_{a-1}} \\
& \hspace{3.5cm} \odot_\star \tte^{k_2}\odot_\star \tte^{k_3}\odot_\star \tte^{k_4}\odot_\star\delta_{x_{a+1}} \odot_\star\cdots\odot_\star\delta_{x_n} \ .
\end{split}
\end{align*}

\begin{figure}[h]%
\centering
\includegraphics[width=0.4\textwidth]{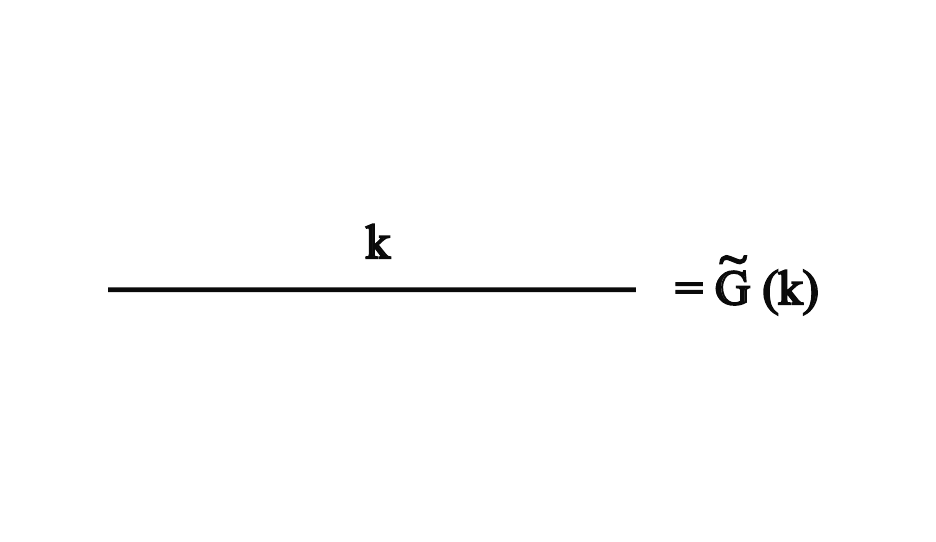} \qquad
\includegraphics[width=0.45\textwidth]{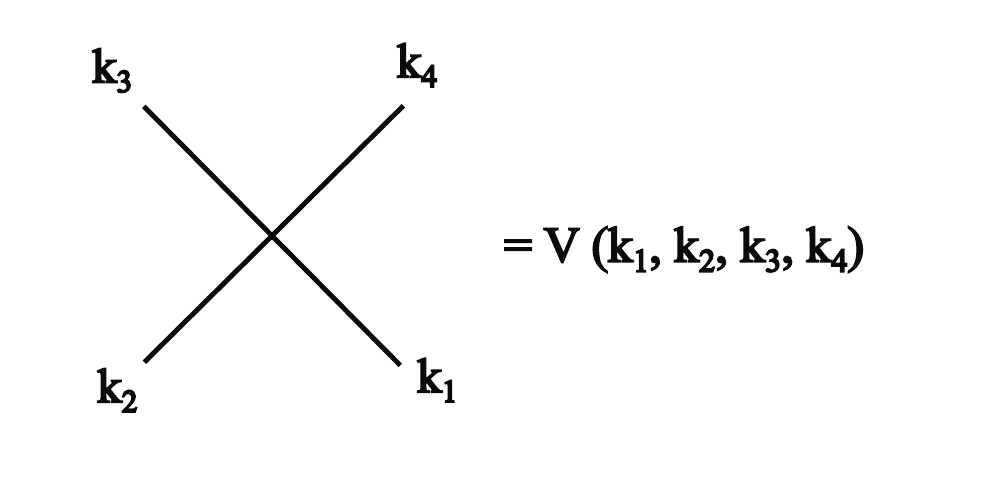}
\vspace{-5mm}
\caption{\small Diagrammatic representation of the propagator (left) and interaction vertex (right) for braided $\lambda\,\phi_4^4$-theory.
}\label{fig:srules}
\end{figure}

The operator $\{\CS _{\rm int} ,-\}_\star$ inserts four legs $\tte^{k_1},\dots,\tte^{k_4}$ and contracts $\tte^{k_1}$ with an external leg $\delta_{x_a}$, at each of the $n$ insertion points $x_1,\dots,x_n$. Thus it changes the symmetric algebra degree from $n$ to $n+2$, and so it follows again that only $n$-point correlation functions with $n=2k$ even are non-vanishing, in which case the sum in \eqref{eq:intcorrelationfn} starts from $p=k$ because only $\sP(1)=1$ is non-zero. The correlation function \eqref{eq:intcorrelationfn} is a formal power series in the parameters $\hbar$ and $\lambda$. The order $\lambda^0$ contribution is just the free $2k$-point function $(\ii\,\hbar\,\BVL\,\sH)^k(\delta_{x_1}\odot_\star\cdots\odot_\star\delta_{x_{2k}})$ discussed earlier. A general order $\lambda^l$ contribution will involve a mixture of loop corrections and both connected as well as disconnected parts; these require at least $k+l$ braided Wick contractions $\ii\,\hbar\,\BVL\,\sH$ in order to produce a non-vanishing result. It follows that the loop expansion parameter is $\kappa:=\hbar\,\lambda$: an $l$-loop contribution to the $2k$-point function is weighted by the factor $\hbar^k\,\kappa^l$. We represent terms in the perturbative expansion \eqref{eq:intcorrelationfn} using standard Feynman diagrammatic techniques; the Feynman rules are depicted in Figure~\ref{fig:srules}.\footnote{See~\cite{SzaboAlex} for a detailed description of how to arrange the computation of \eqref{eq:intcorrelationfn} in terms of a diagrammatic calculus.}

\paragraph{Two-point function at one-loop.}
As an explicit example, let us compute the first non-trivial correction $G_2^\star(x_1,x_2)^{(1)}$ (at order $\lambda$) to the free two-point function, which from \eqref{eq:intcorrelationfn} is given by
\begin{align}\label{eq:2pt1loop}
\begin{split}
& G_2^\star(x_1,x_2)^{(1)} = (\ii\,\hbar\,\BVL\,\sH)^2\,\{\CS _{\rm int},\sH(\delta_{x_1}\odot_\star \delta_{x_2})\}_\star \\[4pt]
& \quad = -2\, \int_{k_1,\dots,k_4} \, V(k_1,\dots,k_4) \, \tilde \sgreen(k_1) \,\Big(\e^{\,\ii\,k_1\cdot x_1} \,  (\ii\,\hbar\,\BVL\,\sH)^2 \big(\tte^{k_2}\odot_\star \tte^{k_3}\odot_\star \tte^{k_4}\odot_\star\delta_{x_2}\big) \\
& \hspace{6.5cm} + \e^{\,\ii\,k_1\cdot x_2} \, \big(\ii\,\hbar\,\BVL\,\sH)^2 (\delta_{x_1}\odot_\star \tte^{k_2}\odot_\star \tte^{k_3}\odot_\star \tte^{k_4}\big)\Big) \ .
\end{split}
\end{align}

The free four-point functions in \eqref{eq:2pt1loop} are evaluated by using the braided Wick expansion \eqref{4PointFull} and the pairing \eqref{eq:eGdeltapair} together with
\begin{align*}
\langle \tte^{k_a},\sgreen(\tte^{k_b})\rangle_\star = \tilde \sgreen(k_a) \, (2\pi)^4 \, \delta(k_a+k_b) \ .
\end{align*}
This gives
\begin{align}\label{eq:G24pt1}
\begin{split}
& (\ii\,\hbar\,\BVL\,\sH)^2 \big(\tte^{k_2}\odot_\star \tte^{k_3}\odot_\star \tte^{k_4}\odot_\star\delta_{x_2}\big) \\[4pt]
& \hspace{1cm} = -\hbar^2 \, \big(\langle \tte^{k_2},\sgreen(\tte^{k_3})\rangle_\star \, \langle \tte^{k_4},\sgreen(\delta_{x_2})\rangle_\star + \langle \tte^{k_2},\sfR_\alpha(\sgreen(\tte^{k_4}))\rangle_\star \, \langle \sfR^\alpha(\tte^{k_3}),\sgreen(\delta_{x_2})\rangle_\star \\ & \hspace{3cm} + \langle \tte^{k_2},\sgreen(\delta_{x_2})\rangle_\star \, \langle \tte^{k_3},\sgreen(\tte^{k_4})\rangle_\star\big) \\[4pt]
& \hspace{1cm} = -\hbar^2 \, (2\pi)^4 \, \tilde \sgreen(k_2) \, \big(\delta(k_2+k_3) \, \tilde \sgreen(k_4) \, \e^{\,\ii\,k_4\cdot x_2}  + \delta(k_2+k_4) \, \tilde \sgreen(k_3) \, \e^{\,\ii\,k_3\cdot x_2} \, \e^{-\ii\, k_3\cdot \theta\,k_4} \\
& \hspace{5cm} + \delta(k_3+k_4) \, \tilde \sgreen(k_3)\, \e^{\,\ii\,k_2\cdot x_2}\big) \ ,
\end{split}
\end{align}
and similarly
\begin{align}\label{eq:G24pt2}
\begin{split}
& \big(\ii\,\hbar\,\BVL\,\sH)^2 (\delta_{x_1}\odot_\star \tte^{k_2}\odot_\star \tte^{k_3}\odot_\star \tte^{k_4}\big) \\[4pt]
& \hspace{1cm} = -\hbar^2 \, (2\pi)^4 \, \tilde \sgreen(k_4) \, \big(\delta(k_3+k_4) \, \tilde \sgreen(k_2) \, \e^{\,\ii\,k_2\cdot x_1}  + \delta(k_2+k_4) \, \tilde \sgreen(k_3) \, \e^{\,\ii\,k_3\cdot x_1} \, \e^{-\ii\, k_2\cdot \theta\,k_3} \\
& \hspace{5cm} + \delta(k_2+k_3) \, \tilde \sgreen(k_2)\, \e^{\,\ii\,k_4\cdot x_1}\big)\ .
\end{split}
\end{align}

We now substitute \eqref{eq:G24pt1} and \eqref{eq:G24pt2} into \eqref{eq:2pt1loop} using the braided symmetry \eqref{eq:Vbraidedsym} of the interaction vertex, and resolve the delta-functions. After relabelling momenta, the noncommutative phase factors in \eqref{eq:Vint} are all unity for the relevant momentum combinations $(k_1,k_2,k_3,k_4)$ given by $(k_1,k_2,-k_2,-k_1)$ and $(k_1,-k_1,k_2,-k_2)$, and one finds that all six contributions are the same. Altogether the one-loop contribution to the two-point function is given by
\begin{align}\label{eq:2pt1loopfinal}
G_2^\star(x_1,x_2)^{(1)} = \frac{\hbar^2\,\lambda}2 \, \int_{k_1,k_2} \, \frac{\e^{-\ii\,k_1\cdot(x_1-x_2)}}{\big(k_1^2-m^2\big)^2\,\big(k_2^2-m^2\big)} \ .
\end{align}
This result is independent of the deformation parameter and coincides with the classical two-point function (at $\nu=0$), including the correct sign and overall combinatorial factor. 

\begin{figure}[h]%
\centering
\includegraphics[width=0.25\textwidth]{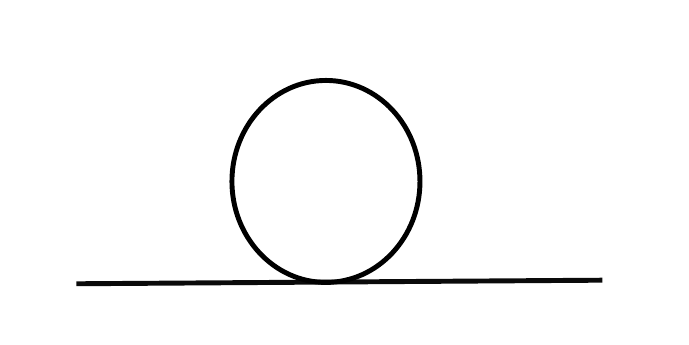} 
\vspace{-5mm}
\caption{\small In braided $\lambda\,\phi_4^4$-theory, the one-loop self-energy receives a contribution only from a planar tadpole diagram.}
\label{fig:stadpole}
\end{figure}

We can recognise the more traditional form by relating the exact two-point function, including all loop corrections, to the dressed propagator in momentum space using the usual Fourier transformation
\begin{align*}
G_2^\star(x_1,x_2)^{\rm int} = -\ii\,\hbar \, \int_{(\FR^{1,3})^*} \, \frac{\dd^4 p}{(2\pi)^4} \ \frac{\e^{-\ii\,p\cdot(x_1-x_2)}}{p^2-m^2-\Pi_\star(p)} \ ,
\end{align*}
where $\Pi_\star(p)$ is the self-energy which is given by a sum over all one-particle irreducible (1PI) diagrams. At order $\lambda$, the result \eqref{eq:2pt1loopfinal} gives
\begin{align}\label{eq:1loopselfenergy}
\frac\ii\hbar \, \Pi_{\star 1} = -\frac\lambda2 \, \int_{(\FR^{1,3})^*} \, \frac{\dd^4k}{(2\pi)^4} \  \frac1{k^2-m^2} \ ,
\end{align}
which leads to the standard one-loop mass renormalization in $\lambda\,\phi_4^4$-theory, represented by the usual tadpole diagram in Figure~\ref{fig:stadpole}.

This result is analogous to the one obtained by~\cite{SzaboAlex} for braided scalar field theory on the fuzzy torus. In particular, it shows that there is no UV/IR mixing in the two-point function at one-loop order, in contrast to the standard noncommutative quantum field theory~\cite{U1Reviewa}, and it seemingly implies the absence of non-planar Feynman diagrams in perturbation theory. This appears to be a consequence of the braided symmetries of the interaction vertex due to the braided $L_\infty$-algebra structure, through its interplay with the braided Wick theorem. It would be interesting to understand to what extent this surprising feature of braided quantum field theory persists at higher loop orders  and in higher point correlation functions. However, having now introduced the main tools from homological algebra, we instead move on to the main theory of interest in the present paper.

\section{Braided electrodynamics}
\label{sec:braidedED}

\subsection{Classical electrodynamics}
\label{sec:classED}

The field content of electrodynamics consists of a $\sU(1)$ gauge field $A=A_\mu (x)\, \d x^\mu\in\Omega^1(\FR^{1,3})$ coupled to a massless\footnote{The zero mass restriction is done for simplicity only, in order to illustrate the general construction. There is no problem with including a mass term for the Dirac spinor as well, and indeed we shall do so later on in Section~\ref{sec:braidedQED}.} Dirac spinor $\psi$ on four-dimensional Minkowski spacetime $\FR^{1,3}$. We use the shorthand notation for partial derivatives $\partial_\mu=\frac\partial{\partial x^\mu}$, where $x^\mu$ are coordinates on $\FR^{1,3}$ with $t=x^0$ the time direction. The infinitesimal $\sU(1)$ gauge variations are given by
\begin{align}
\delta_ c A_\mu &= \tfrac{1}{e}\,\partial_\mu c \ , \quad \delta_ c \psi = \ii\, c\,\psi \qquad \mbox{and} \qquad \delta_ c \bar{\psi} = -\ii\,\bar{\psi}\, c \ ,\nn
\end{align}
where $ c\in\Omega^0(\FR^{1,3})$ is the infinitesimal gauge parameter, $e$ is the electric charge of the fermion, and $\bar{\psi} = \psi^\dagger\, \gamma^0$ is the conjugate Dirac spinor with $\gamma^\mu$ the Dirac matrices in four dimensions (see Appendix~\ref{app:Dirac} for conventions and identities). 

The action functional invariant under these gauge transformations is
\begin{align} \label{eq:EDaction}
S\big(A,\psi,\bar\psi\,\big) = \int_{\FR^{1,3}}\, \d^4 x\ \Big( -\frac{1}{4}\,F^{\mu\nu}\,F_{\mu\nu}+\bar{\psi}\, \ii\,\gamma^\mu\,\big( \partial_\mu\psi -\ii\, e\,A_\mu\,\psi\big) \Big) \ ,
\end{align}
where we introduced the field strength tensor $F=\d A\in\Omega^2(\FR^{1,3})$ whose components are given by $F=  \frac{1}{2}\,F_{\mu\nu}\,\d x^\mu\wedge \d x^\nu = \frac{1}{2}\, (\partial_\mu A_\nu - \partial_\nu A_\mu)\, \d x^\mu\wedge \d x^\nu$. 
The corresponding equations of motion are
\begin{align}
\partial_\nu F^{\mu\nu} = -e\,\bar{\psi}\,\gamma^\mu\,\psi \ , \quad
 \ii\,\gamma^\mu\,\big(\partial_\mu\psi -\ii\,e\,A_\mu\,\psi\big) =0 \eand \big(\partial_\mu\bar{\psi} + \ii\,e\,\bar{\psi}\,A_\mu\big)\,\ii\, \gamma^\mu=0 \ . \nn
\end{align}

Corresponding to the $\sU(1)$ gauge symmetry is the electric matter current $J=J_\mu\,\d x^\mu\in\Omega^1(\FR^{1,3})$ associated with the Dirac spinor $\psi$ given by
\begin{equation}
J^\mu = e\,\bar{\psi}\,\gamma^\mu\,\psi \ . \label{eq:electriccurrent}
\end{equation}
The gauge field $A$ minimally couples to this current and the continuity equation\footnote{We use the notation \smash{$\stackrel{\Phi}{\approx}$} to indicate an equality which holds on-shell when the equation of motion for a field $\Phi$ is imposed. When currents are conserved in this sense, we refer to them as `weakly conserved'.} \smash{$\partial_\mu J^\mu\stackrel{A}{\approx} 0$} follows from the equation of motion for $A$.
It gives the conserved electric charge
\begin{equation}
Q_B = \int_B\, \d B \ J^0 = e \, \int_B\, \d B \  \psi^\dagger \, \psi \label{eq:QB}
\end{equation}
enclosed by any spatial volume $B\subseteq\FR^3\subset\FR^{1,3}$ at fixed time with volume form $\d B$. The conservation law together with Gauss' theorem imply that the time variation of $Q_B$ is cancelled by the net current through the spatial surface $\partial B$ bounding the volume $B$.\footnote{A more general Lorentz invariant definition uses a codimension one hypersurface $B\subset\FR^{1,3}$. The corresponding charge $Q_B$ is then conserved in the normal direction to $B$.}

From the modern perspective of generalized global symmetries~\cite{Gaiotto:2014kfa}, the conserved electric charge is a `zero-form symmetry', which is implemented by codimension one defects $B$ in the field theory, whose natural charged objects are the quanta of the local fermion field $\psi$. Electrodynamics also possesses a one-form magnetic symmetry, implemented by codimension two defects $\Sigma$, with two-form current $\widetilde{F} = \frac12\, g\,\epsilon_{\mu\nu\lambda\sigma}\,F^{\lambda\sigma}\,\d x^\mu\wedge\d x^\nu \in \Omega^2(\FR^{1,3})$. Its conservation law \smash{$\partial_\mu\widetilde{F}^{\mu\nu}=0$} is equivalent to the Bianchi identity for the field strength $F$, or alternatively to the absence of magnetic monopoles. The natural objects that are charged under this symmetry are the non-local 't~Hooft line defects, of magnetic charge $g$ obeying the Dirac--Zwanziger quantization condition $e\,g\in2\pi\,\RZ$ which ensures mutual locality of the electric and magnetic charges. The corresponding conserved quantity is the magnetic flux
\begin{align} \label{eq:magflux}
\varPhi_\Sigma = g \, \int_\Sigma \, F
\end{align}
through a spatial surface $\Sigma\subset\FR^3$. For an open surface, by Stokes' theorem this is equivalently computed by the usual holonomy integral $\varPhi_\Sigma = g\,\int_{\partial\Sigma} \, A$ over the loop $\partial\Sigma$ bounding $\Sigma$. For a closed surface, the magnetic flux vanishes.

The global one-form symmetry can be gauged by minimally coupling the current $\widetilde{F}$ to a background two-form field $b\in\Omega^2(\FR^{1,3})$ with the $\sU(1)$ gauge transformation $\delta_\lambda b = \d \lambda$ for $\lambda\in\Omega^1(\FR^{1,3})$. This shifts \eqref{eq:EDaction} by the action functional of a $BF$-type topological field theory:
\begin{align*}
S_b\big(A,\psi,\bar\psi\,\big) = S\big(A,\psi,\bar\psi\,\big) -g\,\int_{\FR^{1,3}} \, b\wedge F \ .
\end{align*}
Integrating the source term by parts, the shifted action becomes
\begin{align*}
S_b\big(A,\psi,\bar\psi\,\big) = S\big(A,\psi,\bar\psi\,\big) -g\,\int_{\FR^{1,3}} \, A\wedge\d b \ .
\end{align*}
This shows that the field strength $H=\d b\in\Omega^3(\FR^{1,3})$ of the fixed external field $b$ is electrically charged under the gauged zero-form symmetry, thereby introducing a background which modifies the matter current \eqref{eq:electriccurrent} to
\begin{align*}
J_b^\mu = q\,\bar{\psi}\,\gamma^\mu\,\psi + g\, \epsilon^{\mu\nu\lambda\sigma}\,\partial_\nu b_{\lambda\sigma} \ .
\end{align*}
By Gauss' theorem, the  electric charge \eqref{eq:QB} is then shifted to $Q_{B,b} = Q_B+ g\,\int_{\partial B}\,b$ by the flux of the two-form $b$ through the boundary of the spatial volume $B$.

It will prove useful later on, in our general twist deformation formalism, to rewrite this well-known field theory in a basis independent form. Let $\d$ be the exterior differential, $\ast_\hodge$ the Hodge star operator induced by the Minkowski metric, and $\delta = \ast_\hodge\,\d\,\ast_\hodge$ the corresponding codifferential.
Let us introduce a background field $ V $ defined by
\begin{equation}
 V  = \gamma_\mu\,\gamma_\nu\,\gamma_\lambda\, \gamma_5 \,\d x^\mu\wedge\d x^\nu \wedge\d x^\lambda = \ii\,\epsilon_{\mu\nu\lambda\sigma}\,\gamma^\sigma\,\d x^\mu\wedge\d x^\nu\wedge\d x^\lambda \ , \label{V}
\end{equation}
where in the second equality we used the Dirac matrix identity (\ref{Id_1}).
This is a gauge invariant closed three-form valued in the endomorphism algebra ${\sEnd}(\CCS)$ of the complex spinor representation $\CCS$ of $\sSpin(1,3)$ (see Appendix~\ref{app:Dirac}). It can be used to rewrite the spinor action as
\begin{equation}
S_{\psi} = -\frac16\,\int_{\FR^{1,3}}\, \bar{\psi}\, V \wedge (\d \psi - \ii\,e\,A\,\psi)  \ .
\label{SpActionIndexFree}  
\end{equation}
Expanding all the forms,
one arrives at the usual covariant Dirac action\footnote{The covariant  action for a Dirac spinor in four dimensions in an arbitrary basis is usually written as
\begin{equation}
S_{\psi} = -\frac16 \, \int_{\FR^{1,3}}\, \bar{\psi}\, \gamma_5\, {\underline{e}}\wedge{\underline{e}}\wedge{\underline{e}}\wedge (\d \psi - \ii\,e\,A\,\psi) \ ,\nn
\end{equation}
where \smash{${\underline{e}}={\underline{e}}^a_{\,\mu}\,\gamma_a\,\d x^\mu$} is a vierbein one-form valued in $\sEnd(\CCS)$~\cite{PL1}. This action can also describe a Dirac spinor in  curved spacetime, where the gauge field $A$ is then the spin connection $\omega$.
Since we work in a fixed background spacetime, we simply group the three flat vierbeins ${\underline{e}}\wedge{\underline{e}}\wedge{\underline{e}}$ and the chirality matrix $\gamma_5$ into a single background field~$ V $.} 
\begin{equation}
S_\psi = \int_{\FR^{1,3}}\, \d^4 x \ \bar{\psi}\,\ii\,\gamma^\mu\,( \partial_\mu\psi -\ii\,q\,A_\mu\,\psi) \ .\nn
\end{equation}

The full set of equations of motion can now be written in the form
\begin{align}\label{ClsEoM}
\begin{split}
F_A &= \delta\,\d A - \tfrac\ii6\,e\, \bar{\psi}\ast_\hodge V\, \psi =0 \ , \\[4pt]
F_{\bar{\psi}} &= -\tfrac16\,\ast_\hodge\big( V \wedge (\d\psi -\ii\,e\,A\,\psi)\big) =0 \ , \\[4pt]
F_\psi &= -\tfrac16\,\ast_\hodge\big((\d\bar{\psi} + \ii\,e\,\bar{\psi}\,A) \wedge  V\big) =0  \ .
\end{split}
\end{align}

\subsection{$L_\infty$-algebra of electrodynamics}

The graded vector space $L = L^0\oplus L^1\oplus L^2\oplus L^3$ of the $L_\infty$-algebra underlying electrodynamics is given by
\begin{align}\label{eq:electrovector}
\begin{split}
L^0 =L^3&= \Omega^0(\FR^{1,3}) \ , \\[4pt]
L^1 =L^2= \Omega^1(\FR^{1,3}) \, & \oplus \, \Omega^0(\FR^{1,3},\CCS) \, \oplus \, 
\Omega^0(\FR^{1,3},\CCS)  \ .
\end{split}
\end{align}
We arrange the physical fields ${\cal A}\in L^1$ and their duals $\CA^+\in L^2$ as
\begin{equation}
{\cal A} = \left( \begin{array}{c} A\\
{\psi} \\
\bar\psi
\end{array}\right) \eand \CA^+ = \left( \begin{array}{c}
A^+ \\ \bar{{\psi}}^+ \\
{\psi}^+
\end{array}\right) \ . \label{CompositeA_F}
\end{equation}
The elements $\CA^+$ correspond to antifields in the BV formalism containing the equations of motion $F_{\cal A}=0$, where
\begin{align*}
F_{\cal A} = \left( \begin{array}{c}
F_A \\ F_{{\psi}} \\
F_{\bar\psi}
\end{array}\right) \ . 
\end{align*}
Elements $c\in L^0$ correspond to ghosts in the BV--BRST formalism related to the gauge parameters, while elements $c^+\in L^3$ correspond to their antifields containing the Noether identities $\dsf_\CA F_\CA=0$. 

The differential $\ell_1$ is given by
\begin{equation}
\ell_1 (c ) = \left( \begin{array}{c}
\tfrac{1}{e}\, \d c \\[1ex]
0 \\
0 
\end{array}\right) \ ,\quad \ell_1 ({\cal A}) = \left( \begin{array}{c}
\delta\,\d A \\
\tfrac16\,\ast_\hodge (\d \bar{\psi}\wedge V)  \\[1ex]
-\tfrac16\,\ast_\hodge (V \wedge \d \psi)
\end{array}\right) \eand \ell_1 ({\cal A}^+ ) = \tfrac1e\,\delta A^+ \ .\label{l1}
\end{equation}
The non-vanishing $2$-brackets are
\begin{align}\label{l2} 
\begin{split}
 \ell_2 (c, & \, {\cal A} ) = \left( \begin{array}{c}
0 \\
\ii\,c\,\psi \\
-\ii\,\bar{\psi}\,c 
\end{array}\right) \quad  , \quad \ell_2 (c, {\cal A}^+ ) = \left( \begin{array}{c}
0 \\
-\ii\,\bar\psi^+\, c \\
\ii\,c\, {{\psi}}^+
\end{array}\right)  \ , \\[4pt]
\ell_2({\cal A}_1, {\cal A}_2) = \tfrac16\,e \, \ast_\hodge & \left( 
\begin{array}{c}
\ii\,\bar{\psi}_1\, V \,\psi_2 + \ii\,\bar{\psi}_2\, V \,\psi_1 \\ 
-\,\ii\, \bar{\psi}_1\, A_2\wedge V - \ii\, \bar{\psi}_2\, A_1\wedge V   \\ 
-\,\ii\, V \wedge A_1\, \psi_2 -\ii\, V \wedge A_2\, \psi_1  
\end{array}
\right)  \quad  , \quad \ell_2({\cal A}, {\cal A}^+) = \ii\, \bar\psi^+\, \psi + \ii\,\bar{\psi}\, {{\psi}}^+ \ .
\end{split}
\end{align}
All higher brackets $\ell_n$ with $n\geq3$ vanish. 
We checked explicitly that these brackets satisfy the homotopy relations, and thus define an $L_\infty$-algebra $\CCL$ (which is a differential graded Lie algebra in this case).

The underlying cochain complex of the $L_\infty$-algebra $\CCL$ is
\begin{equation}\label{eq:electrocomplex}
\begin{tikzcd}[row sep=0ex,ampersand replacement=\&]
\Omega^0(\mink) \arrow[r,"\frac1e\,\d"] \& \Omega^1(\mink)[-1] \arrow[rr,"\delta\,\d"] \& \& \Omega^1(\mink)[-2] \arrow[r,"\frac1e\,\delta"] \& \Omega^0(\mink)[-3] \\
\& \oplus \& \&  \oplus \& \\
  \& \begin{matrix} \Omega^0(\mink,\CCS)[-1] \\[1ex] \oplus \\[1ex] \Omega^0(\mink,\CCS)[-1] \end{matrix} \ar[rr," \ { \ \Big(\begin{matrix} 0 \!\! & \!\! \ii\overset{\leftarrow}{\slashed\partial} \\[-0.3ex] \ii\,\slashed\partial \!\! & \!\! 0  \end{matrix}\Big) \ } \ "] \&  \& \begin{matrix} \Omega^0(\mink,\CCS)[-2] \\[1ex] \oplus \\[1ex] \Omega^0(\mink,\CCS)[-2] \end{matrix}  \& 
\end{tikzcd}
\end{equation}
where $\ii\,\overset{\leftarrow}{\slashed\partial}\,\bar\psi:=\frac16\,\ast_\hodge(\d\bar\psi\wedge V)$ and $\ii\,\slashed\partial\,\psi:=-\frac16\,\ast_\hodge(V\wedge\d\psi)$.
The cohomology $H^\bullet(L)$ of this complex is given by the free fields in the kernel of $\ell_1$. This defines the minimal model for the $L_\infty$-algebra $\CCL$~\cite{BVChristian,LInfMatter}, which is an $L_\infty$-algebra $H^\bullet(\CCL)$ quasi-isomorphic to $\CCL$ with underlying cochain complex $\big(H^\bullet(L),0\big)$. In particular, fields $(A^{\textrm{\tiny(0)}},\psi^{\textrm{\tiny(0)}},\bar\psi{}^{\textrm{\tiny(0)}})$ in the degree~$1$ cohomology
\begin{align*}
H^1(L) = \frac{\ker(\delta\,\d)}{{\rm im}(\d)} \ \oplus \ \ker\bigg(\begin{matrix} 0 & \ii\,\overset{\leftarrow}{\slashed\partial}\, \\ \ii\,\slashed\partial  & 0  \end{matrix}\bigg)
\end{align*}
correspond to the usual photon states $A^{\textrm{\tiny(0)}}$ of free Maxwell theory on $\mink$ and the conventional description of free spinors as states $(\psi^{\textrm{\tiny(0)}},\bar\psi{}^{\textrm{\tiny(0)}})$ satisfying the Dirac equation. 

One easily checks that the higher brackets of the $L_\infty$-algebra $\CCL$ reproduce the gauge transformations, equations of motion and Noether identity for classical electrodynamics according to the prescription of Section~\ref{sec:Linftyclassical}:
\begin{align} \label{Classical_L_infinity_U(1)}
\begin{split}
\delta_ c {\cal A} &= \ell_1({ c}) + \ell_2( c, {\cal A}) =  \left( \begin{array}{c}
\frac1e\,\d c \\[1ex]
\ii\, c\,\psi \\
-\ii\,\bar{\psi}\, c 
\end{array}\right)  \ , \\[4pt]
F_{\cal A} &= \ell_1({\cal A}) -\frac{1}{2}\, \ell_2({\cal A}, {\cal A}) = \left( \begin{array}{c}
\delta\,  \d A - \tfrac\ii6\,e\,\bar{\psi} \ast_\hodge V\, \psi  \\[1ex]
\tfrac16\,\ast_\hodge \big( (\d\bar\psi +  \ii\,e\, \bar{\psi}\, A )\wedge V\big)  \\[1ex]
-\tfrac16\, \ast_\hodge\big(V \wedge (\d\psi -\ii\,e\,A\, \psi)\big) 
\end{array}\right) \ , \\[4pt]
\dsf_\CA F_{\cal A} &= \ell_1(F_{\cal A}) - \ell_2({\cal A}, F_{\cal A}) = \tfrac1e \, \delta F_A + \ii\,\bar{\psi}\,F_{\bar{\psi}} - \ii\,F_\psi\, \psi \ .
\end{split}
\end{align}
The Noether identity $\dsf_\CA F_{\cal A}=0$ can be understood in different ways. Following~\cite{NoetherII} we interpret it as reproducing the weakly conserved matter current in basis independent form. Indeed, when the equation of motion $F_A=0$ for the gauge field  is inserted, it gives
\begin{equation}
-\tfrac\ii6\,e\, \bar{\psi}\ast_\hodge\big(  V \wedge (\d\psi -\ii\,e\,A\, \psi)\big) - \tfrac\ii6\,e\,\ast_\hodge \big((\d\bar{\psi} + \ii\,e\,\bar{\psi}\,A)\wedge  V\big) \psi  = -\tfrac\ii6\,e\,\delta\ast_\hodge \big( \bar{\psi}\,  V\,  \psi \big) \ \stackrel{A}{\approx} \ 0 \ .\nn
\end{equation}

The $L_\infty$-algebra $\CCL$ is naturally endowed with a cyclic structure defined by the non-vanishing inner products
\begin{align}\label{CyclPairing} 
\begin{split}
\langle c,c^+\rangle = \int_{\FR^{3,1}} \, c \ast_\hodge c^+ \qquad \mbox{and} \qquad
\langle {\cal A}, {\cal A}^+ \rangle = \int_{\FR^{1,3}} \, A\wedge\ast_\hodge\, A^+ -  \bar\psi^+\ast_\hodge \psi + \bar{\psi}\ast_\hodge {{\psi}}^+ \ . 
\end{split}
\end{align}
It is then straightforward to verify that the corresponding homotopy Maurer--Cartan action functional for $\CCL$ coincides with the action functional of electrodynamics:
\begin{align} \label{Classical_L_infinity_U(1)Action}
\begin{split}
S(\CA) &= \tfrac{1}{2}\,\langle {\cal A} ,\ell_1({\cal A})\rangle 
-\tfrac{1}{6}\,\langle  {\cal A},\ell_2({\cal A}, {\cal A})\rangle \\[4pt]
&= \int_{\FR^{1,3}} \,  \frac{1}{2}\,F\wedge *_\hodge\, F - \frac{1}{12}\, \bar{\psi} \, V \wedge \big(\d\psi -\ii\,e\,A\, \psi\big) - \frac{1}{12}\, \big(\d\bar{\psi} + \ii\,e\,\bar{\psi}\,A\big)\wedge  V\, \psi  \\[4pt]
&= \int_{\FR^{1,3}} \,\frac{1}{2}\,F\wedge *_\hodge\, F - \frac16 \, \bar{\psi}\,  V \wedge \big(\d\psi -\ii\,e\,A\, \psi\big) \ ,
\end{split}
\end{align}
where we integrated by parts.

\subsection{Braided $L_\infty$-algebra of electrodynamics}
\label{sec:braidedLinftyspinor}

Following the steps described in Section~\ref{sec:braidedLinfty}, we now deform the classical $L_\infty$-algebra $\CCL$ of electrodynamics to a braided $L_\infty$-algebra $\CCL^\star$.  Like the scalar field theory of Section~\ref{sec:braidedWick}, electrodynamics is only relativistically invariant, so we will work with an abelian Killing twist $\CF$; that is, the vector fields entering the definition of $\CF$ are commuting Killing vectors for Minkowski spacetime $\FR^{1,3}$. The vector spaces \eqref{eq:electrovector} are then modules for the universal enveloping algebra of $\mathfrak{iso}(1,3)=\FR^{1,3}\rtimes\mathfrak{so}(1,3)$.
In our present basis independent formulation, however, we need to further restrict to the isometries of $\FR^{1,3}$ that preserve the closed three-form $V\in\Omega^3\big(\FR^{1,3},\sEnd(\CCS)\big)$, in order to ensure that the brackets \eqref{l1} and \eqref{l2} are equivariant maps. 

Using the Dirac matrix identities from Appendix~\ref{app:Dirac}, it is a straightforward calculation to show that the only Poincar\'e transformations which leave $V$ invariant are translations $\FR^{1,3}\subset\mathfrak{iso}(1,3)$. This forces us to work in a suitably chosen local frame where the abelian twist reduces to the Moyal--Weyl twist \eqref{eq:MWtwist}~\cite{Aschieri:2009qh}; then the actions of the twist and of the $\RR$-matrix on the background field $ V $ are trivial. The typical example, which we will eventually restrict to later on, is the Moyal--Weyl twist itself, which acts trivially on basis vector fields of  the holonomic coordinate frame $\partial_\mu$ and on basis one-forms of the dual coframe $\d x^\mu$; in particular, the star-products among basis one-forms, as well as between functions and basis one-forms, are trivial in this case.

The underlying graded vector space of the braided $L_\infty$-algebra $\CCL^\star$ is again given by \eqref{eq:electrovector},  with the same cochain complex \eqref{eq:electrocomplex}. The non-trivial $2$-brackets are modified to
\begin{align} \label{l2Star} 
\begin{split}
\ell^\star_2 (c, {\cal A} ) = & \, \left( \begin{array}{c}
 0 \\
\ii\,c \star \psi \\
-\ii\,\sfR_\alpha(\bar{\psi})\star \sfR^\alpha(c) 
\end{array}\right) \qquad  , \qquad \ell^\star_2 (c, {\cal A}^+ ) = \left( \begin{array}{c}
0 \\
-\ii\, \sfR_\alpha(\bar\psi^+) \star \sfR^\alpha(c) \\
\ii\,c \star {{\psi}}^+ 
\end{array}\right)  \ , \\[4pt]
\ell^\star_2({\cal A}_1, {\cal A}_2) = & \, \tfrac16\,e\ast_\hodge \left( 
\begin{array}{c}
\ii\,\bar{\psi}_1 \star V \star\psi_2 
+\, \ii\,\sfR_\alpha(\bar{\psi}_2) \star V \star \sfR^\alpha (\psi_1)\\
-\,\ii\, \bar{\psi}_1 \star A_2\wedge_\star  V -\, \ii\, \sfR_\alpha(\bar{\psi}_2) \star \sfR^\alpha(A_1) \wedge_\star  V
\\
-\,\ii\, V\wedge_\star A_1 \star\psi_2 -\ii\, V\wedge_\star \sfR_\alpha(A_2)\star \sfR^\alpha(\psi_1)
\end{array} \right) \ , \\[4pt]
 \ell^\star_2({\cal A}, {\cal A}^+) =& \, \ii\, \sfR_\alpha (\bar\psi^+)\star \sfR^\alpha (\psi) + \ii\,\bar{\psi} \star {{\psi}}^+ \ .
\end{split}
\end{align}
By construction, the braided homotopy relations (in this case the braided graded Jacobi identities) are satisfied.

Finally, the cyclic inner product on $\CCL^\star$ is defined by the non-trivial pairings
\begin{align}\label{CyclPairingStar} 
\begin{split}
\langle c,c^+\rangle_\star = \int_{\FR^{1,3}} \, c\wedge_\star \ast_\hodge\, c^+ \quad , \quad
\langle {\cal A}, {\cal A}^+ \rangle_\star = \int_{\FR^{1,3}} \, A\wedge_\star \ast_\hodge\, A^+   - \bar\psi^+ \wedge_\star \ast_\hodge\, \psi + \bar{\psi} \wedge_\star \ast_\hodge\,{{\psi}}^+ \ .
\end{split}
\end{align}

\subsection{Braided electrodynamics}

The braided $L_\infty$-algebra $\CCL^\star$ defined by \eqref{eq:electrocomplex} and (\ref{l2Star}) defines a new noncommutative deformation of classical electrodynamics that we call `braided electrodynamics'. The braided $\sU(1)$ gauge transformations follow from
\begin{equation}
\delta^\star_ c {\cal A} = \ell_1({ c}) + \ell^\star_2( c, {\cal A}) \nn
\end{equation}
and are given by
\begin{align}
\delta^\star_ c A  &= \tfrac{1}{e}\,\d c \ , \label{BrGTrA} \\[4pt]
\delta^\star_ c \psi = \ii\, c\star \psi \quad , \quad & \delta^\star_ c \bar{\psi} = -\ii\,\sfR_\alpha(\bar{\psi})\star \sfR^\alpha( c)  \ . \label{BrGTrPsi} 
\end{align}
Note that in the case of braided $\sU(1)$ gauge symmetry, the braided Lie algebra brackets also vanish. Therefore the braided $\sU(1)$ gauge theory retains its abelian nature (\ref{BrGTrA}). This is very different from the usual noncommutative deformation of $\sU(1)$ gauge theory with star-gauge symmetry (see e.g.~\cite{U1Reviewa,U1Reviewb}), which becomes a non-abelian gauge theory after deformation.

The equations of motion $F_\CA^\star=0$ of braided electrodynamics follow from braided homotopy Maurer--Cartan equations
\begin{equation}
F^\star_{\cal A}  = \left( \begin{array}{c}
F^\star_A\\
F^\star_{{\psi}} \\
F^\star_{\bar\psi} 
\end{array}\right) = \ell_1({\cal A}) -\tfrac{1}{2}\, \ell^\star_2({\cal A}, {\cal A}) \ , \nn 
\end{equation}
since all higher brackets vanish. Inserting the corresponding brackets given in (\ref{l1}) and (\ref{l2Star}), these equations can be written as
\begin{align*}
F^\star_{A} &= \delta\,\d A - \tfrac{\ii}{12}\,e\ast_\hodge\big(  \bar{\psi}\star V \star\psi  + \sfR_\alpha(\bar{\psi})\star V \star \sfR^\alpha(\psi) \big) \ , \\[4pt]
F^\star_{\psi} &= \tfrac16\,\ast_\hodge\big[  \big( \d\bar\psi + \tfrac{\ii}{2}\, e\,\bar\psi\star A + \tfrac{\ii}{2}\, e\,\sfR_\alpha(\bar\psi)\star \sfR^\alpha(A) \big)\wedge_\star V\big] = \tfrac{1}{12}\ast_\hodge\big[\big( D^{\mbox{\tiny L}}\bar{\psi} + D^{\mbox{\tiny R}}\bar{\psi} \big)\wedge_\star  V\big] \ , \\[4pt]
F^\star_{\bar{\psi}} &= -\tfrac{1}{12}\ast_\hodge\big[ V \wedge_\star \big( D^{\mbox{\tiny L}}\psi + D^{\mbox{\tiny R}}\psi \big)\big] \ , 
\end{align*}
where we introduced left and right braided covariant derivatives
\begin{equation}
D^{\mbox{\tiny L}}\psi = \d\psi - \ii\,e\,A\star \psi \eand D^{\mbox{\tiny R}}\psi = \d\psi - \ii\,e\,\sfR_\alpha(A)\star \sfR^\alpha(\psi) = \d\psi - \ii\,e\,\psi\star A \ .\nn 
\end{equation}

Using the braided Leibniz rule
\begin{equation}
\delta^\star_ c (\bar{\psi}\star V\star\psi ) =  \delta^\star_ c \bar{\psi}\star V\star\psi + \sfR_\alpha(\bar{\psi})\star V\star \delta^\star_{\sfR^\alpha( c)}\psi\nn
\end{equation}
one can easily check that the noncommutative fermion bilinear observable $\bar{\psi}\star V\star\psi$ is invariant under braided gauge transformations and that 
\begin{eqnarray} \nn
\delta^\star_ c D^{\mbox{\tiny L,R}}\psi =  \ii\, c\star D^{\mbox{\tiny L,R}}\psi \ .
\end{eqnarray}
Furthermore, one can explicitly show that all equations of motion transform covariantly, as expected on general grounds from the braided $L_\infty$-algebra formulation, namely
\begin{eqnarray}
\delta^\star_ c F_{A}^\star = 0 \ , \quad \delta^\star_ c F^\star_\psi = -\ii\, \sfR_\alpha(F^\star_\psi) \star \sfR^\alpha( c) \qquad \mbox{and} \qquad
\delta^\star_ c F_{\bar{\psi}}^\star = \ii\, c \star F_{\bar{\psi}}^\star \ . \nn
\end{eqnarray}

The action functional for braided electrodynamics follows from the braided homotopy Maurer--Cartan action functional for $\CCL^\star$ which is given by
\begin{equation}\label{eq:MCspinor}
S_\star(\CA) = \tfrac{1}{2}\,\langle {\cal A}, \ell_1({\cal A}) \rangle_\star - \tfrac{1}{6}\,\langle {\cal A}, \ell^\star_2({\cal A}, {\cal A}) \rangle_\star =: S_0(\CA) + S_{\rm int}(\CA)\ .
\end{equation}
Inserting the corresponding brackets with the cyclic inner product \eqref{CyclPairingStar} gives 
\begin{equation}
S_\star(\CA) = \int_{\FR^{1,3}}\,  \frac{1}{2}\, F\wedge_\star *_\hodge\, F - \frac{1}{12}\,\bar{\psi}\star  V \wedge_\star \big(D^{\mbox{\tiny L}}\psi + D^{\mbox{\tiny R}}\psi\big) \ . \label{MWBrAction}
\end{equation}
Since $F=\d A$ and one star-product can be removed under the integral, the action functional for the gauge field $A$ is just the usual classical Maxwell action functional. 

It is easy to check that $S_\star(\CA)$ is gauge invariant, and also real for a Hermitian twist. Explicitly, using the braided Leibniz rule
one can easily check that  
\begin{align*}
\delta^\star_ c (\bar{\psi}\star  V \wedge_\star D^{\mbox{\tiny L,R}}\psi) & = \delta^\star_ c \bar{\psi}\star  V \wedge_\star D^{\mbox{\tiny L,R}}\psi +  \sfR_\alpha(\bar{\psi})\star V\wedge_\star \delta^\star_{ \sfR^\alpha( c)}D^{\mbox{\tiny L,R}}\psi\nn\\[4pt]
&= -\ii\,\sfR_\alpha(\bar{\psi})\star \sfR^\alpha( c)\star V \wedge_\star D^{\mbox{\tiny L,R}}\psi \\
& \quad \, + \ii\,\sfR_\alpha(\bar{\psi})\star \big(\sfR^\alpha( c)\star  V  - V\star \sfR^\alpha( c) \big)\wedge_\star D^{\mbox{\tiny L,R}}\psi\nn\\
& \quad \,  +\ii\, \sfR_\alpha(\bar{\psi})\star V\wedge_\star \sfR^\alpha( c)\star D^{\mbox{\tiny L,R}}\psi \ = \ 0 \ . \nn
\end{align*}
Compared to standard noncommutative electrodynamics with star-gauge symmetry, one direct consequence of our model is the absence of photon self-interaction vertices. Another is the curtailing of the problem of charge quantization \cite{U1ChargeQuantization}.

Looking more closely at the equations of motion $F_A^\star=0$, we immediately observe the conservation law 
\begin{eqnarray*}
\tfrac12\,e\,\delta\ast_\hodge \big( \bar{\psi}\star  V \star \psi + \sfR_\alpha(\bar{\psi})\star  V  \star \sfR^\alpha(\psi) \big) = \partial_\mu J_\star^\mu \  \stackrel{A}{\approx} \ 0
\end{eqnarray*}
for the braided matter current
\begin{eqnarray}
\quad J_\star^\mu = \tfrac12\,e\,\big( \bar{\psi}\star \gamma^\mu\, \psi + \sfR_\alpha (\bar{\psi})\star \gamma^\mu\, \sfR^\alpha(\psi)\big) \ . \label{JStar}
\end{eqnarray}
The corresponding conserved charge is
\begin{equation}
Q_B^\star =  \int_B\,\d B \ J_\star^0 = \frac e2\,\int_B\, \d B \  \big( \psi^\dagger\star \psi + \sfR_\alpha(\psi^\dagger)\star \sfR^\alpha(\psi)\big)\label{QStar}
\end{equation}
for a spatial volume $B\subseteq\FR^3$ at fixed time. From the current conservation \smash{$\partial_\mu J_\star^\mu \stackrel{A}{\approx} 0$} and Gauss' theorem it follows that
\begin{equation}
\frac{\d Q_B^\star}{\d t} - \int_\Sigma\,\d\vec\Sigma\cdot\vec J_\star \ \stackrel{A}{\approx} \ 0 \ , \nn
\end{equation}
where $\Sigma=\partial B$ with oriented area form $\d\vec\Sigma$ and $\vec J_\star = J_\star^i\,\partial_i$ is the spatial current; in particular, if $B$ is large enough to have no current flux through $\Sigma$, then the charge $Q_B^\star$ is a constant of the motion. 

Although our choice of twist is compatible with the cyclic inner product \eqref{CyclPairingStar}, the second term in \eqref{QStar} has a non-trivial contribution to the conserved charge, because the integration in (\ref{QStar}) is only taken over the three-dimensional spatial volume $B$. Hence $Q_B^\star$ generally differs from the electric charge not only in the classical theory but also in the standard noncommutative theory. Only when $B$ is closed or $B=\FR^3$, and the time component of the twist vanishes (e.g.  $\theta^{0i}=0$ as in Section~\ref{sec:braidedWick}), do we recover the usual electric charge. 

The conserved matter current (\ref{JStar}) also follows from the braided Noether identity in the braided $L_\infty$-algebra description. Let us show this explicitly. The braided Noether identity $\dsf_\CA^\star F_\CA^\star=0$ in $\Omega^0(\FR^{1,3})[[\nu]]$ is given by
\begin{eqnarray}
\ell_1(F^\star_{\cal A}) + \tfrac{1}{2}\,\big( \ell_2^\star (F^\star_{\cal A}, {\cal A}) - \ell_2^\star ({\cal A}, F^\star_{\cal A}) \big)
 + \tfrac{1}{4}\,\big( \ell_2^\star (\ell^\star_2({\cal A}, {\cal A}), {\cal A}) - \ell_2^\star ({\cal A}, \ell^\star_2({\cal A}, {\cal A})) \big) = 0 \ . \nn
\end{eqnarray}
Inserting the corresponding brackets, we find that most of the terms cancel and we are left with
\begin{equation}
\tfrac1e\,\delta F^\star_A - \tfrac{\ii}{12}\,e\,\delta\ast_\hodge \big( \bar{\psi}\star  V \star \psi + \sfR_\alpha(\bar{\psi})\star  V  \star \sfR^\alpha(\psi)\big) =0 \ .\nn
\end{equation}
As previously, we understand this as expressing  the weak conservation of the electric matter current, leading to the conserved charge (\ref{QStar}).

The description of the one-form magnetic symmetry in braided electrodynamics evidently follows the classical discussion from Section~\ref{sec:classED} with no changes.

\section{Braided quantum electrodynamics}
\label{sec:braidedQED}

\subsection{Braided Batalin--Vilkovisky  functional}

Throughout this section we will work with the Moyal--Weyl twist \eqref{eq:MWtwist} for definiteness. Expanding the action functional (\ref{MWBrAction}) in the coordinate basis we find
\begin{align}
\begin{split}
S_\star(\CA) &= \int_{\FR^{1,3}} \,  \d^4x \  \Big( -\frac{1}{4}\,F^{\mu\nu}\star F_{\mu\nu} + \bar{\psi} \star\big( \ii\,\slashed\partial - m\big)\,\psi \\
& \hspace{3cm}+ \frac{e}{2}\,\big(\bar{\psi}\star A_\mu\, \gamma^\mu\star\psi + \bar{\psi}\star \sfR_\alpha(A_\mu)\, \gamma^\mu\star\sfR^\alpha(\psi)
\big) \Big) \ ,
\end{split}
\label{MWLagranzijani}
\end{align}
where we gave the fermion field $\psi$ a mass $m$ in order to avoid spurious  infrared divergences in the following. This action functional follows from the action functional (\ref{MWBrAction}) by assigning a non-zero mass term to the fermion field $\psi$.

In contrast to the scalar field theory of Section~\ref{sec:braidedBV}, the BV action functional here differs from the classical action functional \eqref{MWLagranzijani} by terms involving ghosts and antifields, due to the local $\sU(1)$ gauge symmetry of the theory. It can be computed by promoting $c\in L^0$ to a ghost field and inserting the `superfield' $c+\CA+\CA^++c^+$, regarded as an element of the shifted vector space $L[1]$, into the braided Maurer--Cartan action functional \eqref{eq:MCspinor}. Using the non-zero brackets from \eqref{l1} and \eqref{l2Star}, together with the cyclic inner products from \eqref{CyclPairingStar},  we get
\begin{align*}
\begin{split}
S_\BV(c,\CA,\CA^+) &= S_\star(\CA) + \langle\CA^+,\ell_1(c)\rangle_\star - \tfrac12\,\langle c,\ell_2^\star(\CA,\CA^+) + \ell_2^\star(\CA^+,\CA)\rangle_\star \\[4pt]
& = S_\star(\CA) + \int_{\FR^{1,3}} \, \d^4x \ \Big( \frac1e \, \partial_\mu c\star A^{+\mu} + \frac\ii2\, c\star\big(\bar\psi^+\star\psi + \sfR_\alpha(\bar\psi^+)\star\sfR^\alpha(\psi) \\
& \hspace{8cm} - \bar\psi\star\psi^+ - \sfR_\alpha(\bar\psi)\star\sfR^\alpha(\psi^+)\big)\Big) \ .
\end{split}
\end{align*} 

Since the action functional (\ref{MWLagranzijani}) is invariant under the local (braided) $\sU(1)$ gauge symmetry, we have to perform gauge fixing before quantization. In the present case this alters only the free part of the gauge sector, so we may follow the standard treatment. In the BV formalism it is implemented by extending the field content of the braided $L_\infty$-algebra $\CCL^\star$ with the antighost $\bar c\in\Omega^0(\mink)[-2]$ of the ghost field $c\in\Omega^0(\mink)$, together with its antifield $\bar c^+\in\Omega^0(\mink)[-1]$, and a Nakanishi--Lautrup multiplier field $b\in\Omega^0(\mink)[-1]$ along with its antifield $b^+\in\Omega^0(\mink)[-2]$. 

This extends the cochain complex \eqref{eq:electrocomplex} to
\begin{equation*}
\begin{tikzcd}[row sep=0ex,ampersand replacement=\&]
\Omega^0(\mink) \arrow[r,"\frac1e\,\d"] \& \Omega^1(\mink)[-1] \arrow[rrr,"\delta\,\d"] \& \& \& \Omega^1(\mink)[-2] \arrow[r,"\frac1e\,\delta"] \& \Omega^0(\mink)[-3] \\
\& \oplus \& \& \&  \oplus \& \\
  \& \begin{matrix} \Omega^0(\mink,\CCS)[-1] \\[1ex] \oplus \\[1ex] \Omega^0(\mink,\CCS)[-1] \end{matrix} \ar[rrr," \ { \ \Big(\begin{matrix} 0 \!\! & \!\! \ii\,\overleftarrow{\slashed\partial+m} \\[-0.3ex] \ii\,\slashed\partial-m \!\! & \!\! 0  \end{matrix}\Big) \ } \ "] \&  \& \& \begin{matrix} \Omega^0(\mink,\CCS)[-2] \\[1ex] \oplus \\[1ex] \Omega^0(\mink,\CCS)[-2] \end{matrix}  \& \\
\& \oplus \& \& \& \oplus \& \\
\& \begin{matrix} \Omega^0(\mink)[-1] \\[1ex] \oplus \\[1ex] \Omega^0(\mink)[-1] \end{matrix} \ar[rrr," \ { \ \frac12\,\big(\begin{matrix} 0 \!\! & \!\! -1 \\[-1ex] 1 \!\! & \!\! 0  \end{matrix}\big) \ } \ "] \&  \& \& \begin{matrix} \Omega^0(\mink)[-2] \\[1ex] \oplus \\[1ex] \Omega^0(\mink)[-2] \end{matrix}  \&
\end{tikzcd}
\end{equation*}
where the extended pairs are $(b,\bar c^+)$ and $(b^+,\bar c)$ in degrees~$1$ and~$2$, respectively. This complex has the same cohomology $H^\bullet(L)$ as \eqref{eq:electrocomplex}. There are no further non-zero higher brackets, while the additional non-trivial inner products are $\langle (b,\bar c^+)\,,\,(b^+,\bar c)\rangle_\star = \int_{\mink}\,\dd^4x \ b\star b^+ - \bar c^+\star\bar c$. The corresponding superfield Maurer--Cartan action functional shifts the BV functional to $S_\BV -\int_{\FR^{1,3}}\,\d^4x \ b\star\bar c^+$. 

A general $R_\xi$ gauge, for some gauge parameter $\xi\in\FR$~\cite{Cheng:1984vwu}, is then implemented by applying the isometry of $\CCL^\star$ defined as $(A^+,b^+,\bar c^+)\mapsto \big(A^++\dd\bar c,b^+-\frac\xi2\,\bar c, \bar c^+-\delta A-\frac\xi2\,b\big)$ in the shifted BV functional. This yields the one-parameter family of gauge fixed BV action functionals
\begin{align}\label{eq:BVspinorQED}
\begin{split}
S^{{\rm gf},\xi}_\BV(b,c,\bar c,\bar c^+,\CA,\CA^+) &= S_\BV(c,\CA,\CA^+) - \int_\mink \, \d^4x \ \frac1e \, \bar c \star\square\,c -\frac\xi2\,b\star b - b\star\delta A+b\star\bar c^+ \ .
\end{split}
\end{align}
Note that the ghost field $c$ completely decouples from the gauge field $A$ due to the abelian nature of the gauge symmetry.

\subsection{Braided homological perturbation theory}

As in Section~\ref{sec:braidedBV}, we quantize this theory by applying braided homological perturbation theory to compute correlation functions, following~\cite{Macrelli:2019afx,LInfMatter,SzaboAlex}.

\paragraph{Propagators.}
In order to transfer the homotopy algebra structure to the cohomology $H^\bullet(\CCL^\star)$ of the braided $L_\infty$-algebra $\CCL^\star$ of electrodynamics from Section~\ref{sec:braidedLinftyspinor}, we need to specify a $U\FR^3$-equivariant projection $\sfp:L\to H^\bullet(L)$ of degree~$0$ and a $U\FR^3$-invariant contracting homotopy $\sfh:L\to L$ of degree~$-1$. From \eqref{eq:electrocomplex} it follows that the cochain complex underlying $H^\bullet(\CCL^\star)$ is
\begin{equation*}
\begin{tikzcd}[row sep=0ex,ampersand replacement=\&]
\ker(\d) \arrow[r,"0"] \& \displaystyle\frac{\ker(\delta\,\d)}{{\rm im}(\d)}[-1] \arrow[r,"0"]  \& \displaystyle\frac{\ker(\delta\,\d)}{{\rm im}(\d)}[-2] \arrow[r,"0"] \& \ker(\d)[-3] \\
\& \oplus \&   \oplus \& \\
  \&  \ker\bigg(\begin{matrix} 0 & \ii\,\overleftarrow{\slashed\partial+m} \\ \ii\,\slashed\partial-m & 0 \end{matrix}\bigg)[-1] \arrow[r,"0"] \&   {\rm coker}\bigg(\begin{matrix} 0 & \ii\,\overleftarrow{\slashed\partial+m} \\ \ii\,\slashed\partial-m & 0 \end{matrix}\bigg)[-2]  \& 
\end{tikzcd}
\end{equation*}
where we used the fact that the codifferential $\delta$ is the Hodge adjoint of the differential $\d$.

The same arguments that we gave in Section~\ref{sec:braidedBV} carry through to the present case to show that, for the purposes of computing correlation functions of the physical fields, one should take the trivial projections 
\begin{align*}
\sfp^\swone = 0 = \sfp^\swtwo
\end{align*}
to the vacua $\CA=0=\CA^+$; that is, the kernel and cokernel of the Maxwell operator $\delta\,\d$ and of the massive Dirac operators $\ii\,\slashed\partial\pm m$ become trivial on Wick rotation to Euclidean signature. Of course, this is not true for the exterior differential $\d$, and we take the remaining components to be the natural projections \smash{$\sfp^\swzero,\sfp^\swthree:\Omega^0(\mink)\to\ker(\d)$} to the space of constant functions.

For the components of the contracting homotopy $\sfh^\swk:L^k\to L^{k-1}$, which satisfy $\ell_1\circ\sfh^\swk=\id_{L^k}$, we consider the massless Feynman propagator $\mgreen:\Omega^0(\mink)\to\Omega^0(\mink)$ given by
\begin{align*}
\mgreen = \frac1\square \qquad \mbox{with} \quad \tilde\mgreen(k)=-\frac1{k^2} \ ,
\end{align*}
where we indicated the action of the Green operator on plane waves of the form $\e^{\,\pm\,\ii\,k\cdot x}$. We extend the massless propagator to a map $\mgreen:\Omega^1(\mink)\to\Omega^1(\mink)$ by using the $\Omega^0(\mink)$-module structure of the space of one-forms $\Omega^1(\mink)$. 

The massive spinor Feynman propagator $\fgreen:\Omega^0(\mink,\CCS\oplus \CCS)\to\Omega^0(\mink,\CCS\oplus \CCS)$ acts on fermion antifields as
\begin{align*}
\fgreen\bigg(\begin{matrix} \bar\psi^+ \\ \psi^+ \end{matrix} \bigg)(x)
=-\frac1{\square+m^2} \, \bigg(\begin{matrix} 0 & \ii\,\slashed\partial+m \\ \ii\,\overleftarrow{\slashed\partial-m} & 0 \end{matrix}\bigg) \bigg(\begin{matrix} \bar\psi^+ \\ \psi^+ \end{matrix} \bigg)(x) = \int_\mink\,\d^4y \ \int_k\, \e^{-\ii\,k\cdot(x-y)} \ \tilde\fgreen(k)\,\bigg(\begin{matrix} \bar\psi^+(y) \\ \psi^+(y) \end{matrix} \bigg)
\end{align*}
where
\begin{align*}
\tilde\fgreen(k) = \frac1{k^2-m^2} \, \bigg(\begin{matrix} 0 & \slashed k+m \\ \overleftarrow{\slashed k-m} & 0 \end{matrix}\bigg) =: \bigg(\begin{matrix}
0 & \tilde \fgreen^+(k) \\ \tilde \fgreen^-(k) & 0
\end{matrix}\bigg) \ ,
\end{align*}
with $\tilde\fgreen^-(-k)= -\tilde\fgreen^+(k)$, and we defined $\overleftarrow{(\slashed k-m)}\,\bar\psi^+ :=\bar\psi^+\,(\slashed k-m)$.

Then the non-zero components of the contracting homotopy are
\begin{align}\label{eq:homotopyspQED}
\begin{split}
\sfh^\swone = \big(\begin{matrix}
e\,\mgreen\circ\delta & 0
\end{matrix}\big) &: L^1 \longrightarrow L^0 \ , \\[4pt]
\sfh^\swtwo = \bigg(\begin{matrix}
\mgreen\circ{\sf\Pi} & 0 \\ 0 & \fgreen
\end{matrix}\bigg) &: L^2 \longrightarrow L^1 \ , \\[4pt]
\sfh^\swthree = \bigg(\begin{matrix}
e\,\d\circ\mgreen \\ 0 
\end{matrix}\bigg) &: L^3\longrightarrow L^2 \ ,
\end{split}
\end{align}
where ${\sf\Pi}:\Omega^1(\mink)\to\Omega^1(\FR^{1,3})$ is the projector onto the image of the Maxwell operator $\delta\,\d$ with eigenvalues
\begin{align*}
\tilde{\sf\Pi}_{\mu\nu}(k) = \eta_{\mu\nu} - \frac{k_\mu\,k_\nu}{k^2}
\end{align*}
when acting on plane waves of the form $\e^{\,\ii\,k\cdot x}\,\d x^\mu$. Explicitly
\begin{align*}
\begin{split}
\sfh^\swone(\CA) = e\,\frac{\partial_\mu}\square \, A^\mu \quad , \quad
\sfh^\swtwo(\CA^+) = \begin{pmatrix}
\displaystyle \frac1\square\,\Big(\eta_{\mu\nu} - \frac{\partial_\mu\,\partial_\nu}{\square}\Big)\,A^{+\nu} \, \d x^\mu \\[2ex]
\displaystyle -\frac{\ii\,\slashed\partial+m}{\square+m^2}\,\psi^+ \\[2ex]
\displaystyle -\bar\psi^+\,\frac{\ii\,\overset{\leftarrow}{\slashed\partial}-m}{\square+m^2}
\end{pmatrix} \quad , \quad
\sfh^\swthree(c^+) = \begin{pmatrix} \displaystyle e\,\frac{\partial_\mu}{\square} \, c^+ \, \d x^\mu \\[2ex] 0 \\ 0 \end{pmatrix}
\end{split}
\end{align*}
for $\CA\in L^1$, $\CA^+\in L^2$ and $c^+\in L^3$. 

The propagator $\mgreen$ is the right inverse of the differential that appears in the ghost field kinetic term of \eqref{eq:BVspinorQED}. Gauge fixing is implemented by setting $\sfh^\swone(\CA)=0$. By our choice of representatives of the cohomology in \eqref{eq:homotopyspQED}, this corresponds to imposing the Lorenz gauge condition $\delta A=\partial_\mu A^\mu=0$ on the gauge fields $A\in\Omega^1(\mink)$. Different gauge choices correspond to different choices of contracting homotopy, or equivalently different choices of propagators in the field theory. In particular, from \eqref{eq:BVspinorQED} it follows that, after eliminating the auxiliary field $b$ through its equation of motion, a general $R_\xi$ gauge corresponds to the choice of photon propagator with
\begin{align*}
\tilde{\sf\Pi}_{\mu\nu}(k) = \eta_{\mu\nu} - (1-\xi) \, \frac{k_\mu\,k_\nu}{k^2} \ .
\end{align*}
Only two choices are associated with projectors: the Landau gauge $\xi=0$, which projects onto the image of $\delta\,\d$ as above, and the Feynman gauge $\xi=1$, which projects onto the kernel of $\delta\,\d$.

\begin{figure}[h]%
\centering
\includegraphics[width=0.45\textwidth]{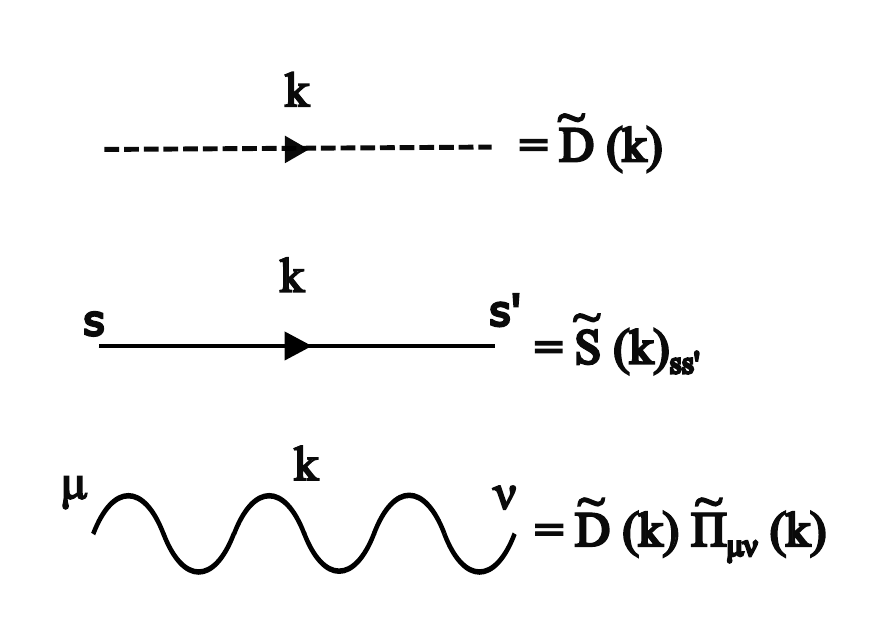} 
\vspace{-5mm}
\caption{\small From top to bottom: The ghost, spinor and photon propagators.}
\label{fig:frules}
\end{figure}

As in Section~\ref{sec:braidedBV}, we extend the projection $\sfp$ and the contracting homotopy $\sfh$ to maps $\sP$ and $\sH$ on the braided symmetric algebra $\Sym_\RR L[2]$ of polynomial observables of electrodynamics. Again on observables involving solely physical fields $\CA\in L^1$, only $\sP(1)=1$ is non-trivial. The role of $\sH$ is to insert propagators in correlation functions of the braided quantum field theory, which we depict diagrammatically as shown in Figure~\ref{fig:frules}. On the other hand, the braided BV Laplacian $\BVL$ implements the braided Wick theorem in braided homological perturbation theory. 

\paragraph{Vertices.} 
We now proceed to construct the interaction vertices that need to be inserted. We start by introducing the basis of plane waves $\tte_k$ for $L^0$ and its dual basis $\tte^k$ for $L^3$ as in Section~\ref{sec:braidedBV}. To construct corresponding bases for $L^1$ and $L^2$, we first identify the space of one-forms $\Omega^1(\mink)\subset L^1$ with $(\mink)^*\otimes\Omega^0(\FR^{1,3})$. A gauge field $A\in\Omega^1(\mink)$ can then be expanded as\footnote{Throughout we omit tensor products among basis elements from different vector spaces for simplicity, e.g. we abbreviate $\ttv^\mu\otimes\tte_k$ by $\ttv^\mu\,\tte_k$, and so on.} $A=\int_k\, a_\mu(k)\,\ttv^\mu\, \tte_k$ where $\ttv^\mu$ are covectors of the standard basis for $(\mink)^*$, and the Lorenz gauge fixing condition imposes the constraint $k^\mu\,a_\mu(k)=0$ on the photon polarization vectors. Similarly, we identify $\Omega^1(\mink)\subset L^2$ with $\mink\otimes\Omega^0(\mink)$ and expand the antifields $A^+\in\Omega^1(\mink)$ as $A^+=\int_k\,a^{+\mu}(k)\,\ttv_\mu\, \tte^k$, where $\ttv_\mu$ are vectors of a dual basis for $\mink$ in the sense that
\begin{align*}
\langle\ttv^\mu\, \tte_k,\ttv_\nu\, \tte^p\rangle_\star = \delta_\nu^\mu \, \langle \tte_k,\tte^p\rangle_\star \ .
\end{align*}

For the fermion fields $(\psi,\bar\psi)$, we identify the vector space $\Omega^0(\mink,\CCS \oplus \CCS)\subset L^1$ with the space $\big(\CCS\,\otimes\,\Omega^0(\mink)\big)\oplus\big(\CCS\,\otimes\,\Omega^0(\mink)\big)$ and basis $(\ttu^s,\bar \ttu^s)$, $s=1,2,3,4$ for $\CCS\oplus \CCS$. A Dirac spinor can then be expanded as $\psi = \int_k\, u_s(k) \, \ttu^s \, \tte_k $, while a Dirac adjoint spinor is expanded as $\bar\psi = \int_k\, \bar v_s(k)\,\bar\ttu^s\,\tte_k $. For the antifields $(\bar\psi^+,\psi^+)$, we identify $\Omega^0(\mink,\CCS \oplus \CCS)\subset L^2$ with the space $\big(\CCS\,\otimes\,\Omega^0(\mink)\big)\oplus\big(\CCS\,\otimes\,\Omega^0(\mink)\big)$ and dual basis $(\bar \ttu_s,\ttu_s)$ for $\CCS\oplus \CCS$. The non-zero cyclic pairings among the spinor bases are given by
\begin{align*}
\begin{split}
\langle\bar \ttu^s\, \tte_k ,\ttu_{s'}\, \tte^p\rangle_\star = \delta_{s'}^s \, \langle \tte_k,\tte^p\rangle_\star = \langle\ttu^s\, \tte_k , \bar \ttu_{s'}\, \tte^p\rangle_\star \ .
\end{split}
\end{align*}

With these conventions we introduce the contracted coordinate functions $\mbf\xi\in(\Sym_\RR L[2])\otimes L[[\nu]]$ as the degree~$1$ elements
\begin{align}\label{eq:mbfxispinorQED}
\begin{split}
\mbf\xi &= \int_k \, \big(\tte_k\otimes \tte^k + \tte^k\otimes \tte_k + \ttv^\mu\,\tte_k\otimes\ttv_\mu\,\tte^k + \ttv_\mu\,\tte^k\otimes \ttv^\mu\,\tte_k \\
& \hspace{3cm} +\ttu^s\,\tte_k \otimes\bar \ttu_s\,\tte^k + \bar \ttu^s\,\tte_k \otimes \ttu_s\,\tte^k \\
& \hspace{6cm} + \bar \ttu_s\,\tte^k\otimes \ttu^s\,\tte_k  + \ttu_s\,\tte^k\otimes\bar \ttu^s\,\tte_k \big) \ .
\end{split}
\end{align}
Using the braided Maurer--Cartan action functional \eqref{eq:MCspinor} and the extended cyclic braided $L_\infty$-algebra structure on $(\Sym_\RR L[2])\otimes L[[\nu]]$, we introduce the interacting part of the BV action functional $\CS _{\rm int}\in\Sym_\RR L[2]$ as the degree~$0$ element
\begin{align}\label{eq:CSintspinorQED}
\CS_{\rm int} := - \tfrac{1}{6}\,\langle\!\!\langle {\mbf\xi}, \mbf\ell^\star_2({\mbf\xi}, {\mbf\xi}) \rangle\!\!\rangle_\star \ . 
\end{align}

In this paper we are solely interested in correlation functions of \emph{physical} fields $\CA\in L^1$, so we can simplify the computation by keeping only those terms of \eqref{eq:mbfxispinorQED} involving antifields which will generate non-zero braided Wick contractions and antibrackets with elements of $L^1$. Likewise, from \eqref{eq:BVspinorQED} it follows that we may drop all ghost terms from \eqref{eq:mbfxispinorQED}. In other words, for our purposes it suffices to compute \eqref{eq:CSintspinorQED} when evaluated on the element
\begin{align*}
\mbf\xi^{\rm phys} = \int_k \, \big( \ttv_\mu\,\tte^k\otimes \ttv^\mu\,\tte_k + \bar \ttu_s\,\tte^k\otimes \ttu^s\,\tte_k  + \ttu_s\,\tte^k\otimes\bar \ttu^s\,\tte_k \big) \ .
\end{align*}

With this simplification in mind, we use the non-zero $2$-brackets from \eqref{l2Star} and cyclic inner products from \eqref{CyclPairingStar}, together with braided cyclicity of the extended inner product, and perform analogous calculations to those of Section~\ref{sec:braidedBV}. Additionally, we keep in mind that the contracted coordinate functions ${\mbf\xi}$ are $U\frv$-invariant elements of $(\Sym_\RR L[2])\otimes L[[\nu]]$ and hence all appearances of $\RR$-matrices in the properties
of the extended brackets $\mbf\ell_n$ and of the extended pairing $\langle\!\!\langle\>,\> \rangle\!\!\rangle_\star$ disappear when evaluated on tensor powers of the $U\frv$-invariant element ${\mbf\xi}$. This yields
\begin{align*}
\begin{split}
\CS_{\rm int} &= -\frac1{6} \, \int_{k_1,k_2,k_3} \, 3\, \langle\!\!\langle \ttv_\mu\,\tte^{k_1}\otimes \ttv^\mu\,\tte_{k_1}\,,\, \mbf\ell_2^\star(\bar\ttu_s\,\tte^{k_2}\otimes\ttu^s\,\tte_{k_2} , \ttu_{s'}\,\tte^{k_3}\otimes\bar\ttu^{s'}\,\tte_{k_3}) \\
& \hspace{7cm} + \mbf\ell_2^\star(\ttu_{s'}\,\tte^{k_3}\otimes\bar\ttu^{s'}\,\tte_{k_3},\bar\ttu_s\,\tte^{k_2}\otimes\ttu^s\,\tte_{k_2})  \rangle\!\!\rangle_\star \\[4pt]
&= \frac1{2} \, \int_{k_1,k_2,k_3} \, 2\, \langle\!\!\langle \ttv_\mu\,\tte^{k_1}\otimes \ttv^\mu\,\tte_{k_1} , \big(\bar\ttu_s\,\tte^{k_2}\odot_\star\ttu_{s'}\,\sfR_\alpha(\tte^{k_3})\big) \otimes \ell_2^\star\big(\ttu^s\,\sfR^\alpha(\tte_{k_2}),\bar\ttu^{s'}\,\tte_{k_3}\big)\rangle\!\!\rangle_\star \\[4pt]
&= \int_{k_1,k_2,k_3} \, \e^{\,\ii\,k_2\cdot\theta\,k_3} \, \langle\!\!\langle \ttv_\mu\,\tte^{k_1}\otimes \ttv^\mu\,\tte_{k_1} , (\bar\ttu_s\,\tte^{k_2}\odot_\star\ttu_{s'}\,\tte^{k_3}) \otimes \ell_2^\star(\ttu^s\,\tte_{k_2},\bar\ttu^{s'}\,\tte_{k_3})\rangle\!\!\rangle_\star \\[4pt]
&= \int_{k_1,k_2,k_3} \, \e^{\,\ii\,k_2\cdot\theta\,k_3} \, \ttv_\mu\,\tte^{k_1} \odot_\star \sfR_\alpha(\bar\ttu_s\,\tte^{k_2}\odot_\star\ttu_{s'}\,\tte^{k_3}) \, \langle  \ttv^\mu\,\sfR^\alpha(\tte_{k_1}) , \ell_2^\star(\ttu^s\,\tte_{k_2},\bar\ttu^{s'}\,\tte_{k_3})\rangle_\star \\[4pt]
&= -e \, \int_{k_1,k_2,k_3} \, \e^{\,\ii\,\sum\limits_{a<b}\,k_a\cdot\theta\,k_b} \,\ttv_\mu\,\tte^{k_1} \odot_\star \bar\ttu_s\,  \tte^{k_2}\odot_\star\ttu_{s'}\,\tte^{k_3} \, \langle  \ttv^\mu\,\tte_{k_1} , \ttv_\nu\,(\bar\ttu^{s'}\,\gamma^\nu \, \ttu^s) \, \sfR_\alpha(\tte_{k_3}) \star\sfR^\alpha(\tte_{k_2})\rangle_\star \\[4pt]
&= -e \, \int_{k_1,k_2,k_3} \, \e^{\,\ii\,\sum\limits_{a<b}\,k_a\cdot\theta\,k_b} \, (\bar\ttu^{s'}\,\gamma^\mu \, \ttu^s) \ \ttv_\mu\,\tte^{k_1} \odot_\star \bar\ttu_s\,\tte^{k_2}\odot_\star\ttu_{s'}\,\tte^{k_3} \ \langle \tte_{k_1}, \tte_{k_2}\star \tte_{k_3}\rangle_\star \\[4pt]
&=: \int_{k_1,k_2,k_3} \, V^{\mu;s,s'}(k_1,k_2,k_3) \ \ttv_\mu\,\tte^{k_1} \odot_\star\ttu_{s}\,\tte^{k_2}\odot_\star \bar\ttu_{s'}\,\tte^{k_3} \ .
\end{split}
\end{align*}

\begin{figure}[h]%
\centering
\includegraphics[width=0.45\textwidth]{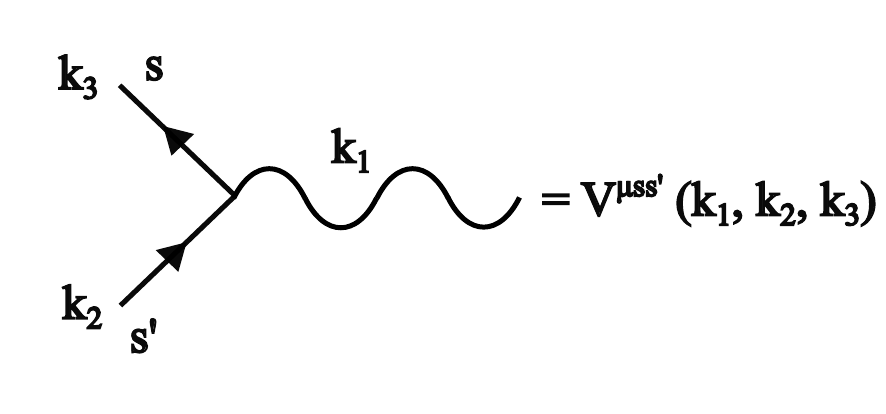} 
\vspace{-5mm}
\caption{\small Diagrammatic representation of the photon-fermion vertex.}
\label{fig:fvertex}
\end{figure}

The photon-fermion vertex is given by
\begin{align}\label{eq:fermionphotonvertex}
V^{\mu;s,s'}(k_1,k_2,k_3) = -e \, (\gamma^\mu)^{ss'} \, \e^{\,\frac\ii2\,\sum\limits_{a<b} \, k_a\cdot\theta\, k_b} \ (2\pi)^4 \, \delta(k_1+k_2+k_3) \ ,
\end{align}
where $(\gamma^\mu)^{ss'} := \bar\ttu^{s}\,\gamma^\mu\,\ttu^{s'}$ are the Dirac matrix elements in our chosen basis for the complex spinor representation $\CCS$. Its diagrammatic representation is shown in Figure~\ref{fig:fvertex}. Surprisingly, the result (\ref{eq:fermionphotonvertex}) is different from the vertex that follows from a naive application of standard Feynman rules to the action (\ref{MWLagranzijani}), \cite{CiricDimitrijevic:2022eei}. The naive application of standard Feynman rules results in a vertex with a noncommutative contribution 
$\cos\big(\tfrac12\,k_2\cdot\theta\,k_3\big)$. This vertex is completely (strictly) symmetric under interchange of any pair of momenta, as well as under reflection $k_a\to -k_a$ of all three momenta, while the vertex (\ref{eq:fermionphotonvertex}) has braided symmetry similarly to (\ref{eq:Vbraidedsym}) and (\ref{eq:Vcyclicsym}).

Note that (\ref{eq:fermionphotonvertex}) is the same vertex as in the standard noncommutative QED with star-gauge symmetry, where the noncommutative correction to the vertex also consists of solely a phase factor in the momenta. Unlike the standard noncommutative QED, here there are no three-photon or four-photon vertices and no photon-ghost vertex, as anticipated from \eqref{eq:BVspinorQED}.

\paragraph{Correlation functions.} The correlation functions of braided QED are now defined and computed analogously to Section~\ref{sec:braidedBV} by the formula
\begin{align}\label{eq:Gspinorgen}
\begin{split}
G_{\CA_1,\dots,\CA_n}^\star(x_1,\dots,x_n)&=\langle 0|{\rm T}[\CA_1(x_1)\star\cdots\star\CA_n(x_n)]|0\rangle_\star \\[4pt]
:\!&= \sum_{p=1}^\infty \, \sP\big((\ii\,\hbar\,\BVL\,\sH + \{\CS_{\rm int},-\}_\star\,\sH)^p(\delta^{\CA_1}_{x_1}\odot_\star\cdots\odot_\star\delta^{\CA_n}_{x_n})\big) \ ,
\end{split}
\end{align}
where $\delta_{x_a}^{\CA_a}\in L^2[2]$ are again Dirac distributions supported at the insertion points $x_a\in\FR^{1,3}$ of the chosen physical fields $\CA_a\in L^1$. With our conventions these distributions are given by
\begin{align*}
\delta_{x_a}^{A_{\mu_a}}(x) := \ttv_{\mu_a} \, \delta(x-x_a) \ , \quad \delta_{x_a}^{\psi_{s_a}}(x) := \bar\ttu_{s_a} \, \delta(x-x_a) \qquad \mbox{and} \qquad \delta_{x_a}^{\bar\psi_{s_a}}(x) := \ttu_{s_a} \, \delta(x-x_a) \ ,
\end{align*}
for the various species of photon field $A\in\Omega^1(\mink)$ and fermions $(\psi,\bar\psi\,)\in\Omega^0(\mink,\CCS\oplus \CCS)$. 

Arguing similarly to Section~\ref{sec:braidedBV}, the operator $\{\CS_{\rm int},-\}_\star$ increases the symmetric algebra degree $n$ by $1$. Thus if $n=2k$ is even, the series \eqref{eq:Gspinorgen} starts at $p=k$ and a non-vanishing contribution requires an even number $2l$ of photon-fermion vertex insertions; in this case $k+l$ braided Wick contractions $\ii\,\hbar\,\BVL\,\sH$ are needed. On the other hand, if $n=2k+1$ is odd then an odd number $2l+1$ of vertex insertions is required together with $k+l+1$ applications of $\ii\,\hbar\,\BVL\,\sH$. In both instances the expansion parameter is $\kappa:=\hbar\,e^2$: the contribution to a $2k$-point function is weighted by $\hbar^k\,\kappa^l$ and to a $2k{+}1$-point function by $\hbar^{k+1}\, e\, \kappa^l$. Note that only correlation functions involving equal numbers of Dirac fields $\psi$ and $\bar\psi$ are non-vanishing, as these are the only ones that will have non-trivial braided Wick expansions under iterated applications of $\ii\,\hbar\,\BVL\,\sH$. This is the statement of charge conjugation symmetry of braided quantum electrodynamics.

Using this formalism one can calculate various correlation functions and also check the ultraviolet behaviour of the quantum field theory. We will analyse the photon and fermion self-energies at one-loop order in Sections~\ref{sec:photonselfenergy} and~\ref{sec:fermionselfenergy}, respectively.

\subsection{Braided Wick's theorem for Dirac spinors}

We first consider the theory of a free Dirac field. In the classical case, Wick's theorem expresses the non-zero correlation functions as the Pfaffian of the two-point correlation matrix:
\begin{align}\label{eq:Pfaff}
\langle0|{\rm T}[\psi(1)\cdots\psi(2k)]|0\rangle^{(0)} = \frac1{k!\,2^k} \, \sum_{\sigma\in S_{2k}} \, {\rm sgn}(\sigma) \ \prod_{a=1}^k\,\langle0|{\rm T}\big[\psi\big(\sigma(2a-1)\big) \, \psi\big(\sigma(2a)\big)\big]|0\rangle^{(0)} \ ,
\end{align}
where for compactness we have written in the argument of $\psi(a)$ the spacetime coordinate $x_a$, the spinor index $s_a$, and a $\RZ_2$-valued index which distinguishes $\psi$ from $\bar\psi$. The non-vanishing two-point functions are determined by the free fermion propagator
\begin{align*}
\frac\ii\hbar \, \langle0|{\rm T}[\psi_s(x)\,\bar\psi_r(y)]|0\rangle^{(0)} = \int_k \, \e^{-\ii\,k\cdot(x-y)} \ \tilde\fgreen_{sr}^+(k) \ .
\end{align*}
In the following we abbreviate the non-zero Wick contractions as
\begin{align}\label{eq:Wickanticomm}
\bcontraction{}{\psi_a}{}{\bar\psi_b} \psi_a \, \bar\psi_b
:=\langle0| {\rm T}[\psi_{s_a}(x_a) \, \bar\psi_{s_b}(x_b) ]|0\rangle^{(0)} = -\bcontraction{}{\bar\psi_b}{}{\psi_a}{} \bar\psi_b \, \psi_a \ .
\end{align}

Arguing similarly to Section~\ref{sec:braidedWick}, in the free braided quantum field theory we expect the correlation functions to be determined by a formula analogous to \eqref{eq:G2kstar}, with the Hafnian replaced by the Pfaffian formula \eqref{eq:Pfaff}. For our later calculations we will need to carefully understand how this braided Wick expansion is implemented in homological perturbation theory, where the $2k$-point functions are computed by applying the operator $(\ii\,\hbar\,\BVL\,\sH)^k$ to braided symmetric products of $k$ distributions of the form \smash{$\delta^{\psi_{s_a}}_{x_a}$} with $k$ distributions of the form \smash{$\delta^{\bar\psi_{s_{a}}}_{x_{a}}$}. To properly implement Fermi statistics, we need to carefully define both the action of the extended contracting homotopy $\sH$ and the braided BV Laplacian $\BVL$ when acting on spinor fields.\footnote{The inclusion of fermions into the $L_\infty$-algebra framework is discussed in e.g.~\cite{Saberi,Grigoriev:2023lcc}. However, we could not find a detailed treatment of fermions in homological perturbation theory in the literature.} There are two important extra signs which arise in this case: an extra weight $(-1)^{a-1}$ in each term of the sum \eqref{eq:sfH} defining the extended contracting homotopy $\sH$, and an extra weight $(-1)^{(a-1)+(b-1-a)}=(-1)^b$ in each term of the sum in \eqref{eq:BVL} defining the braided BV Laplacian $\BVL$. The only non-trivial pairings (\ref{CyclPairingStar}) are between $\psi, \bar\psi$ and their corresponding antifields $\bar\psi^+, \psi^+$. This definition also fixes the antibrackets of spinor fields through the relation \eqref{eq:BVLantibracket}.

In the following we consider some illustrative examples of how these rules work.

\paragraph{Two-point functions.}
Consider the free braided two-point Green's function defined by
\begin{align*}
G_{\psi_{s_1},\bar\psi_{s_2}}^\star(x_1,x_2)^{(0)} = \langle0|{\rm T}[\psi_{s_1}(x_1)\star\bar\psi_{s_2}(x_2)]|0\rangle^{(0)}_\star := \ii\,\hbar \, \BVL\,\sH\,\big(\delta_{x_1}^{\psi_{s_1}}\odot_\star \delta_{x_2}^{\bar\psi_{s_2}}\big) \ .
\end{align*}
Applying \eqref{eq:sfH} and \eqref{eq:BVL} we obtain
\begin{align}\label{eq:free2ptferm}
G_{\psi_{s_1},\bar\psi_{s_2}}^\star(x_1,x_2)^{(0)} = \frac{\ii\,\hbar}2 \, \Big(\big\langle\fgreen\big(\delta_{x_1}^{\psi_{s_1}}\big),\delta_{x_2}^{\bar\psi_{s_2}}\big\rangle_\star - \big\langle\delta_{x_1}^{\psi_{s_1}},\fgreen\big(\delta_{x_2}^{\bar\psi_{s_2}}\big)\big\rangle_\star\Big) \ ,
\end{align}
where the minus sign comes from the fermionic Koszul sign rules discussed above. The first inner product gives
\begin{align*}
\big\langle\fgreen\big(\delta_{x_1}^{\psi_{s_1}}\big),\delta_{x_2}^{\bar\psi_{s_2}}\big\rangle_\star &= \int_\mink\,\d^4x \ \delta(x-x_2) \ \int_\mink\,\d^4y \ \int_k\,\e^{-\ii\,k\cdot(x-y)} \ \big(\bar\ttu_{s_1}\,\tilde\fgreen^-(k)\,\ttu_{s_2}\big) \ \delta(y-x_1) \\[4pt]
&= \int_k\,\e^{-\ii\,k\cdot(x_2-x_1)} \ \tilde\fgreen^-_{s_1s_2}(k) = -\int_k\,\e^{-\ii\,k\cdot(x_1-x_2)} \ \tilde\fgreen^+_{s_1s_2}(k) \ ,
\end{align*}
where in the last step we changed integration variable $k\to-k$ and used $\tilde\fgreen^\pm(-k) = -\tilde\fgreen^\mp(k)$. Similarly, the second inner product gives
\begin{align*}
\big\langle\delta_{x_1}^{\psi_{s_1}},\fgreen\big(\delta_{x_2}^{\bar\psi_{s_2}}\big)\big\rangle_\star = \int_k \, \e^{-\ii\,k\cdot(x_1-x_2)} \ \tilde\fgreen^+_{s_1s_2}(k) \ .
\end{align*}
Finally, substituting into \eqref{eq:free2ptferm} yields the free braided two-point function
\begin{align*}
G_{\psi_{s_1},\bar\psi_{s_2}}^\star(x_1,x_2)^{(0)} = -\ii\,\hbar \, \int_k \, \e^{-\ii\,k\cdot(x_1-x_2)} \ \tilde\fgreen^+_{s_1s_2}(k) = \bcontraction{}{\psi_1}{}{\bar\psi_2} \psi_1 \, \bar\psi_2 \ ,
\end{align*}
which is exactly the Feynman propagator for the Dirac spinor field, as anticipated.

In our BV formalism, the anticommutation relation \eqref{eq:Wickanticomm} follows from the calculation
\begin{align*}
\bcontraction{}{\bar\psi_2}{}{\psi_1}{} \bar\psi_2 \, \psi_1 &= G_{\bar\psi_{s_2},\psi_{s_1}}^\star(x_2,x_1)^{(0)} = \ii\,\hbar \, \BVL\,\sH\,\big(\delta_{x_2}^{\bar\psi_{s_2}}\odot_\star \delta_{x_1}^{\psi_{s_1}}\big) \\[4pt]
& = \frac{\ii\,\hbar}2 \, \Big(\big\langle\fgreen\big(\delta_{x_2}^{\bar\psi_{s_2}}\big),\delta_{x_1}^{\psi_{s_1}}\big\rangle_\star - \big\langle\delta_{x_2}^{\bar\psi_{s_2}},\fgreen\big(\delta_{x_1}^{\psi_{s_1}}\big)\big\rangle_\star\Big) \\[4pt]
& = \frac{\ii\,\hbar}2 \, \int_\mink\,\d^4x \ \int_\mink\,\d^4y \ \int_k\,\e^{-\ii\,k\cdot(x-y)} \, \Big( \big(\bar\ttu_{s_1}\,\tilde\fgreen^+(k)\,\ttu_{s_2}\big) \, \delta(y-x_2) \, \delta(x-x_1) \\
& \hspace{8cm} - \big(\bar\ttu_{s_1}\,\tilde\fgreen^-(k)\,\ttu_{s_2}\big) \, \delta(y-x_1) \, \delta(x-x_2)\Big) \\[4pt]
& = \ii\,\hbar \, \int_k \, \e^{-\ii\,k\cdot(x_1-x_2)} \ \tilde\fgreen^+_{s_1s_2}(k) = - \bcontraction{}{\psi_1}{}{\bar\psi_2} \psi_1 \, \bar\psi_2 \ .
\end{align*}

\paragraph{Four-point functions.}
Consider the free four-point Green's function defined by
\begin{align*}
G^\star_{\psi_{s_1},\bar\psi_{s_2},\psi_{s_3},\bar\psi_{s_4}}(x_1,x_2,x_3,x_4)^{(0)} =& \ \langle0|{\rm T}[\psi_{s_1}(x_1)\star\bar\psi_{s_2}(x_2)\star\psi_{s_3}(x_3)\star\bar\psi_{s_4}(x_4)]|0\rangle^{(0)}_\star \\[4pt]
:=& \ (\ii\,\hbar\,\BVL\,\sH)^2\big(\delta_{x_1}^{\psi_{s_1}}\odot_\star \delta_{x_2}^{\bar\psi_{s_2}}\odot_\star \delta_{x_3}^{\psi_{s_3}}\odot_\star \delta_{x_4}^{\bar\psi_{s_4}}\big) \ .
\end{align*}
This is evaluated using 
\begin{align*}
\ii\,\hbar\,\BVL\,\sH\big(\delta_{x_a}^{\psi_{s_a}}\odot_\star \delta_{x_b}^{\psi_{s_b}}\big) = 0 = \ii\,\hbar\,\BVL\,\sH\big(\delta_{x_a}^{\bar\psi_{s_a}}\odot_\star \delta_{x_b}^{\bar\psi_{s_b}}\big)
\end{align*}
and
\begin{align*}
\ii\,\hbar\,\BVL\,\sH\big(\delta_{x_a}^{\psi_{s_a}}\odot_\star \delta_{x_b}^{\bar\psi_{s_b}}\big) = \bcontraction{}{\psi_a}{}{\bar\psi_b} \psi_a \, \bar\psi_b = -\bcontraction{}{\bar\psi_b}{}{\psi_a}{} \bar\psi_b \, \psi_a =- \ii\,\hbar\,\BVL\,\sH\big(\delta_{x_b}^{\bar\psi_{s_b}}\odot_\star \delta_{x_a}^{\psi_{s_a}}\big) \ ,
\end{align*}
together with the fermionic Koszul sign rules discussed above.

We start from
\begin{align*}
& \ii\,\hbar\,\BVL\,\sH\big(\delta_{x_1}^{\psi_{s_1}}\odot_\star \delta_{x_2}^{\bar\psi_{s_2}}\odot_\star \delta_{x_3}^{\psi_{s_3}}\odot_\star \delta_{x_4}^{\bar\psi_{s_4}}\big) \\[4pt]
& \hspace{4cm} = \tfrac12\,\big(\bcontraction{}{\psi_1}{}{\bar\psi_2} \psi_1  \, \bar\psi_2 \ \delta_{x_3}^{\psi_{s_3}}\odot_\star\delta_{x_4}^{\bar\psi_{s_4}} +  \bcontraction{}{\psi_1}{}{\sfR_\alpha\,\sfR_\beta(\bar\psi_4)} \psi_1 \, \sfR_\alpha\,\sfR_\beta(\bar\psi_4) \ \sfR^\alpha(\delta_{x_2}^{\bar\psi_{s_2}})\odot_\star\sfR^\beta(\delta_{x_3}^{\psi_{s_3}}) \\
& \hspace{8cm} + \bcontraction{}{\bar\psi_2}{}{\psi_3}{} \bar\psi_2 \, \psi_3 \ \delta_{x_1}^{\psi_{s_1}}\odot_\star\delta_{x_4}^{\bar\psi_{s_4}} + \bcontraction{}{\psi_3}{}{\bar\psi_4} \psi_3 \, \bar\psi_4 \ \delta_{x_1}^{\psi_{s_1}}\odot_\star\delta_{x_2}^{\bar\psi_{s_2}} \big) \ .
\end{align*}
Applying $\ii\,\hbar\,\BVL\,\sH$ to this expression gives the four-point function
\begin{align*}
G^\star_{\psi_{s_1},\bar\psi_{s_2},\psi_{s_3},\bar\psi_{s_4}}(x_1,x_2,x_3,x_4)^{(0)} = \tfrac12\, \big( 2\, \bcontraction{}{\psi_1}{}{\bar\psi_2} \psi_1  \, \bar\psi_2 \, \bcontraction{}{\psi_3}{}{\bar\psi_4} \psi_3  \, \bar\psi_4 + \bcontraction{}{\psi_1}{}{\sfR_\alpha\,\sfR_\beta(\bar\psi_4)} \psi_1 \, \sfR_\alpha\,\sfR_\beta(\bar\psi_4) \, \bcontraction{}{\sfR^\alpha(\bar\psi_2)}{}{\sfR^\beta(\psi_3)} \sfR^\alpha(\bar\psi_2) \, \sfR^\beta(\psi_3) + \bcontraction{}{\psi_1}{}{\bar\psi_4}{} \psi_1 \, \bar\psi_4 \, \bcontraction{}{\bar\psi_2}{}{\psi_3}{} \bar\psi_2 \, \psi_3\big) \ .
\end{align*}
Employing the same $\RR$-matrix manipulations as in Section~\ref{sec:braidedWick} shows that the last two terms are equal, and we arrive finally at
\begin{align}\label{eq:braided4pt}
G^\star_{\psi_{s_1},\bar\psi_{s_2},\psi_{s_3},\bar\psi_{s_4}}(x_1,x_2,x_3,x_4)^{(0)} = \bcontraction{}{\psi_1}{}{\bar\psi_2} \psi_1  \, \bar\psi_2 \, \bcontraction{}{\psi_3}{}{\bar\psi_4} \psi_3  \, \bar\psi_4 + \bcontraction{}{\psi_1}{}{\bar\psi_4}{} \psi_1 \, \bar\psi_4 \, \bcontraction{}{\bar\psi_2}{}{\psi_3}{} \bar\psi_2 \, \psi_3 \ .
\end{align}
This also agrees with the classical free fermion four-point function.

The other non-zero four-point functions can be similarly obtained. For example
\begin{align*}
G^\star_{\psi_{s_1},\psi_{s_2},\bar\psi_{s_3},\bar\psi_{s_4}}(x_1,x_2,x_3,x_4)^{(0)} = - \bcontraction{}{\psi_1}{}{\sfR_\alpha(\bar\psi_3)} \psi_1 \, \sfR_\alpha(\bar\psi_3) \, \bcontraction{}{\sfR^\alpha(\psi_2)}{}{\bar\psi_4} \sfR^\alpha(\psi_2) \, \bar\psi_4 + \bcontraction{}{\psi_1}{}{\bar\psi_4}{} \psi_1 \, \bar\psi_4 \, \bcontraction{}{\psi_2}{}{\bar\psi_3}{} \psi_2 \, \bar\psi_3 \ ,
\end{align*}
where we simplified the last term by using the fact that the $\RR$-matrix acts trivially on the two-point functions, analogously to \eqref{2Pointcalc}.

\subsection{Photon self-energy at one-loop}
\label{sec:photonselfenergy}

The two-point photon correlation function is obtained from \eqref{eq:Gspinorgen} by setting $n=2$, $\CA_1=A_\mu$ and $\CA_2=A_\nu$. It is easy to see that the only non-vanishing contributions at one-loop order are given by
\begin{align}\label{eq:photon2ptdef}
\begin{split}
G^\star_{A_\mu,A_\nu}(x_1,x_2)^{(1)} &= \langle0|{\rm T}[A_\mu(x_1)\star A_\nu(x_2)]|0\rangle_\star^{(1)} \\[4pt]
:\!&= (\ii\,\hbar\,\BVL\,\sH)^2\,\big\{\CS_{\rm int},\sH\,\big\{\CS_{\rm int},\sH\big(\delta_{x_1}^{A_\mu}\odot_\star\delta_{x_2}^{A_\nu}\big)\big\}_\star\big\}_\star \\
& \quad \, + \ii\,\hbar\,\BVL\,\sH \,\big\{\CS_{\rm int}, \sH\,(\ii\,\hbar\,\BVL\,\sH)\big\{\CS_{\rm int},\sH\big(\delta_{x_1}^{A_\mu}\odot_\star\delta_{x_2}^{A_\nu}\big) \big\}_\star\big\}_\star \\[4pt]
&=: \CG^1_{\mu\nu}(x_1,x_2) + \CG^2_{\mu\nu}(x_1,x_2) \ .
\end{split}
\end{align}

We start from the non-zero inner products
\begin{align}\label{eq:non0DPidelta}
\big\langle\ttv_\mu\,\tte^{k_a},\mgreen\,\sPi\big(\delta_{x_b}^{A_\nu}\big) \big\rangle_\star = \e^{\,\ii\,k_a\cdot x_b} \, \tilde\mgreen(k_a)\,\tilde\sPi_{\mu\nu}(k_a)
\end{align}
to compute the basic operation
\begin{align}\label{eq:basicop}
\begin{split}
& \big\{\CS_{\rm int} , \sH\big(\delta_{x_1}^{A_\mu}\odot_\star\delta_{x_2}^{A_\nu} \big) \big\}_\star \\[4pt]
& \hspace{1cm} = \frac12\,\int_{k_1,k_2,k_3} \, V^{\lambda;s,s'}(k_1,k_2,k_3) \, \big(\e^{\,\ii\,k_1\cdot x_1} \, \tilde\mgreen(k_1)\,\tilde\sPi_{\mu\lambda}(k_1) \, \ttu_s\,\tte^{k_2}\odot_\star \bar\ttu_{s'} \, \tte^{k_3}\odot_\star\delta_{x_2}^{A_\nu} \\
& \hspace{6.5cm} + \e^{\,\ii\,k_1\cdot x_2} \, \tilde\mgreen(k_1)\,\tilde\sPi_{\lambda\nu}(k_1) \, \delta_{x_1}^{A_\mu} \odot_\star \ttu_s\,\tte^{k_2}\odot_\star \bar\ttu_{s'} \, \tte^{k_3} \big) \ ,
\end{split}
\end{align}
where we used momentum conservation to trivialize the phase factors arising from $\RR$-matrix insertions in the braided derivation property of the antibracket. 
We will now evaluate each contribution $\CG^1_{\mu\nu}(x_1,x_2)$ and $\CG^2_{\mu\nu}(x_1,x_2)$ to \eqref{eq:photon2ptdef} separately in turn. 

\paragraph{Evaluation of $\boldsymbol{\CG^1_{\mu\nu}(x_1,x_2)}$.}
We start with another iteration of the expression \eqref{eq:basicop} by applying the operator $\{\CS_{\rm int},-\}_\star\circ\sH$ again, which generates pairings involving both photon propagators on the remaining external legs and fermion propagators for the internal loops. In this case, momentum conservation does not always trivialize $\RR$-matrix factors arising from the braided derivation property of the antibracket. In particular, we encounter the basic braided symmetric product relations
\begin{align}\label{eq:deltaeR}
\delta_{x_a}^{A_\mu}\odot_\star\tte^{k_b} = 
\sfR_\alpha(\tte^{k_b})\odot_\star \sfR^\alpha(\delta_{x_a}^{A_\mu}) = \tte^{k_b} \odot_\star \delta^{A_\mu}_{x_a+\theta\,k_b}
\end{align}
in two of the antibrackets with $\CS_{\rm int}$ coming from the second line after applying the extended contracting homotopy $\sH$. 

Since the only non-zero pairings occur between elements in complementary degrees~$1$ and~$2$, the only non-trivial antibrackets that survive are those which come from a dual pairing generated by $\sH$, which lowers the degree from~$2$ to~$1$, and so always involve propagators. For example, a non-zero pairing involving the spinor basis elements is 
\begin{align}\label{eq:spinorpropinner1}
\begin{split}
\langle\bar\ttu_{s}\,\tte^{k_a},\fgreen(\ttu_{s'}\,\tte^{k_b})\rangle_\star &= \int_\mink\,\d^4x \ \e^{\,\ii\,k_a\cdot x} \ \int_\mink\,\d^4y \ \int_p\,\e^{-\ii\,p\cdot(x-y)} \ \big(\bar\ttu_s\,\tilde\fgreen^+(p)\,\ttu_{s'}\big) \ \e^{\,\ii\,k_b\cdot y} \\[4pt] &= (2\pi)^4\,\delta(k_a+k_b) \, \tilde\fgreen^+_{ss'}(-k_b) = -(2\pi)^4 \, \delta(k_a+k_b) \, \tilde\fgreen^-_{ss'}(k_b) \ ,
\end{split}
\end{align}
where we used $\tilde\fgreen^\pm(-k)= - \tilde\fgreen^\mp(k)$. Similarly, the only other non-zero spinor pairing is given by
\begin{align}\label{eq:spinorpropinner2}
\begin{split}
\langle\fgreen(\bar\ttu_{s'}\,\tte^{k_b}),\ttu_s\,\tte^{k_a}\rangle_\star 
= -(2\pi)^4 \, \delta(k_a+k_b) \, \tilde\fgreen^+_{s's}(k_b)  \ .
\end{split}
\end{align}

The first term that arises after applying $\sH$ to \eqref{eq:basicop} involves the non-zero antibracket
\begin{align*}
& \big\{\ttv_\sigma\,\tte^{p_1}\odot_\star\ttu_r\,\tte^{p_2}\odot_\star \bar\ttu_{r'}\,\tte^{p_3} \,,\, \fgreen(\ttu_{s}\,\tte^{k_2})\odot_\star\bar\ttu_{s'}\,\tte^{k_3}\odot_\star \delta^{A_\nu}_{x_2}\big\}_\star \\[4pt]
& \hspace{5cm} = -\ttv_\sigma\,\tte^{p_1} \odot_\star \ttu_r\,\tte^{p_2} \odot_\star \big\langle \bar\ttu_{r'}\,\tte^{p_3}\,,\,\fgreen(\ttu_{s}\,\tte^{k_2})\big\rangle_\star\,\bar\ttu_{s'}\,\tte^{k_3} \odot_\star \delta^{A_\nu}_{x_2} \\[4pt]
& \hspace{5cm} = (2\pi)^4\,\delta(p_3+k_2) \, \tilde\fgreen_{r's}^-(k_2)  \, \ttv_\sigma\,\tte^{p_1} \odot_\star \ttu_{r}\,\tte^{p_2} \odot_\star \bar\ttu_{s'}\,\tte^{k_3} \odot_\star \delta_{x_2}^{A_\nu} \ ,
\end{align*}
and similarly for the other five terms generated.

After some calculation using momentum conservation, as well as relabellings of spinor momenta, one arrives at
\begin{align}\label{eq:Sintcompspinor}
\begin{split}
& \big\{\CS_{\rm int},\sH\,\big\{\CS_{\rm int} , \sH\big(\delta_{x_1}^{A_\mu}\odot_\star\delta_{x_2}^{A_\nu} \big) \big\}_\star\big\}_\star \\[4pt]
& \quad = \frac16\,\int_{p_1,p_2,p_3} \, V^{\sigma;r,r'}(p_1,p_2,p_3) \ \int_{k_1,k_2,k_3} \, V^{\lambda;s,s'}(k_1,k_2,k_3) \\
&\hspace{2cm} \times \Big[
(2\pi)^4\, \e^{\,\ii\,k_1\cdot x_1}\,\tilde\mgreen(k_1) \, \tilde\sPi_{\lambda\mu}(k_1)\Big(
 \delta(p_3+k_2) \, \tilde\fgreen_{r's}^-(k_2) \, \ttv_\sigma\,\tte^{p_1} \odot_\star \ttu_{r}\,\tte^{p_2} \odot_\star \bar\ttu_{s'}\,\tte^{k_3} 
 \\
& \hspace{3.5cm} + \e^{\ii\,(p_3+k_2)\cdot\theta\,p_2} \, 
\delta(p_2+k_3)\,
\tilde\fgreen_{s'r}^-(p_2)  \, \ttv_\sigma\,\tte^{p_1} \odot_\star \bar\ttu_{r'}\,\tte^{p_3} \odot_\star \ttu_s\,\tte^{k_2} \Big)\, \odot_\star
\delta^{A_\nu}_{x_2} \\
& \hspace{2.5cm} + (2\pi)^4 \e^{\,\ii\,k_1\cdot x_2}\tilde\mgreen(k_1) \, \tilde\sPi_{\lambda\nu}(k_1)\, \delta^{A_\mu}_{x_1}\\
& \hspace{4.5cm}\odot_\star\Big(
 \delta(p_3+k_2) \, \tilde\fgreen_{r's}^-(k_2)  \ttv_\sigma\,\tte^{p_1} \odot_\star \ttu_{r}\,\tte^{p_2} \odot_\star \bar\ttu_{s'}\,\tte^{k_3}
 \\
& \hspace{5.5cm} + \e^{\ii\,p_3\cdot\theta\,p_2} \, 
\delta(p_2+k_3)\,
\tilde\fgreen_{s'r}^-(p_2)  \,  \ttu_s\,\tte^{k_2}\odot_\star\, \ttv_\sigma\,\tte^{p_1} \odot_\star \bar\ttu_{r'}\,\tte^{p_3} \Big)\\
& \hspace{4cm} + 2 \, \e^{\,\ii\,p_1\cdot x_1 + \ii\,k_1\cdot x_2} \, \tilde\mgreen(p_1) \, \tilde\sPi_{\mu\sigma}(p_1) \, \tilde\mgreen(k_1) \, \tilde\sPi_{\lambda\nu}(k_1) \\
& \hspace{8cm} \times \ttu_r\,\tte^{p_2} \odot_\star \bar\ttu_{r'}\,\tte^{p_3} \odot_\star \ttu_s\,\tte^{k_2} \odot_\star \bar\ttu_{s'}\,\tte^{k_3}\Big]
 \ .
\end{split}
\end{align}
Finally, applying the operator $(\ii\,\hbar\,\BVL\,\sH)^2$ to \eqref{eq:Sintcompspinor} generates free four-point functions, which we evaluate using the braided Wick expansions \eqref{4PointFull} and \eqref{eq:braided4pt}. Many pairings vanish in this case.

Let us start with the last line of \eqref{eq:Sintcompspinor}, whose Wick expansion gives
\begin{align}\label{eq:lastlineSintcomp}
\begin{split}
& (\ii\,\hbar\,\BVL\,\sH)^2 \big(\ttu_r\,\tte^{p_2}\odot_\star\bar\ttu_{r'}\,\tte^{p_3}\odot_\star\ttu_s\,\tte^{k_2}\odot_\star\bar\ttu_{s'}\,\tte^{k_3}\big) \\[4pt]
& \hspace{2cm} = -\hbar^2\,\big(\langle\fgreen(\bar\ttu_{r'}\,\tte^{p_3}),\ttu_r\,\tte^{p_2}\rangle_\star \,\langle\fgreen(\bar\ttu_{s'}\,\tte^{k_3}),\ttu_s\,\tte^{k_2}\rangle_\star \\
& \hspace{7cm} + \langle\fgreen(\bar\ttu_{s'}\,\tte^{k_3}),\ttu_r\,\tte^{p_2}\rangle_\star \, \langle\bar\ttu_{r'}\,\tte^{p_3},\fgreen(\ttu_s\,\tte^{k_2})\rangle_\star \big) \\[4pt]
& \hspace{2cm} = -\hbar^2\,(2\pi)^8 \,\big(\tilde\fgreen_{r'r}^+(p_3)\,\tilde\fgreen_{s's}^+(k_3) \, \delta(p_2+p_3) \, \delta(k_2+k_3) \\
& \hspace{7cm} + \tilde\fgreen_{r's} ^-(k_2)\,\tilde\fgreen^+_{s'r}(k_3)\, \delta(p_2+k_3) \, \delta(k_2+p_3)\big) \ .
\end{split}
\end{align}

\begin{figure}[h]%
\centering
\includegraphics[width=0.33\textwidth]{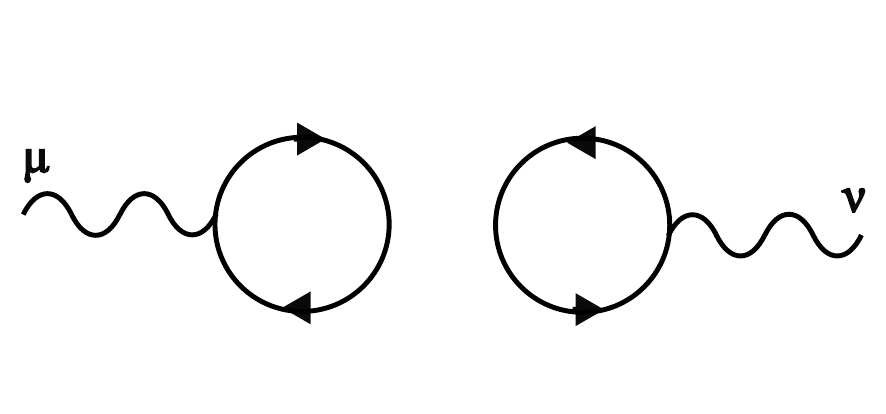} \qquad
\includegraphics[width=0.35\textwidth]{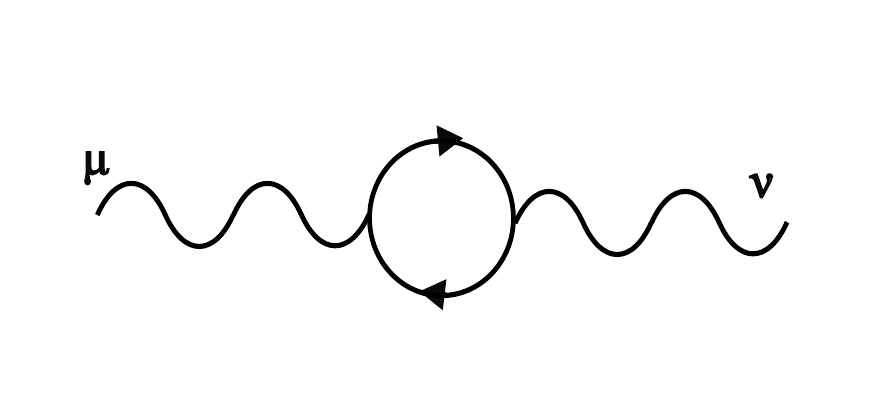}
\vspace{-5mm}
\caption{\small The disconnected (left) and connected (right) contributions to the photon two-point function at one-loop order. The photon one-point functions each vanish.}
\label{fig:fdisconn}
\end{figure}

Substituting into \eqref{eq:Sintcompspinor} and relabelling momenta, it is straightforward to see that the contribution of the first term in the last line of \eqref{eq:lastlineSintcomp} can be written as
\begin{align}\label{eq:disconn}
\begin{split}
& \frac13\,\Big(\ii\,\hbar\,e\,\int_{k_1,p_1} \, (2\pi)^4 \, \delta(p_1) \, \e^{\,\ii\,p_1\cdot x_1}\,\tilde\mgreen(p_1)\,\tilde\sPi_{\mu\sigma}(p_1) \, \Tr\big(\gamma^\sigma\,\tilde\fgreen^+(k_1)\big)\Big) \\
& \hspace{4cm} \times \Big(\ii\,\hbar\,e\,\int_{k_2,p_2} \, \Tr\big(\gamma^\lambda\,\tilde\fgreen^+(k_2)\big) \, (2\pi)^4 \, \delta(p_2) \, \e^{\,\ii\,p_2\cdot x_2} \, \tilde\mgreen(p_2) \, \tilde\sPi_{\lambda\nu}(p_2)\Big) \\[4pt]
& \hspace{2cm} = \frac13\,\Big(\ii\,\hbar\,\BVL\,\sH\,\big\{\CS_{\rm int},\sH\big(\delta_{x_1}^{A_\mu}\big)\big\}_\star\Big) \, \Big(\ii\,\hbar\,\BVL\,\sH\,\big\{\CS_{\rm int},\sH\big(\delta_{x_2}^{A_\nu}\big)\big\}_\star\Big) \ ,
\end{split}
\end{align}
where $\Tr$ is the trace over spinor indices (see Appendix~\ref{app:Dirac}), and the additional delta-functions have set all $\theta$-dependent quantities to unity, including the contributions from the vertex \eqref{eq:fermionphotonvertex}. This is recognised as a disconnected contribution to the photon two-point function from the fermion tadpoles at this order, as depicted diagrammatically in Figure~\ref{fig:fdisconn}. As in the standard noncommutative QED, the tadpole contributions coincide with their commutative counterparts. Each one-point function is superficially divergent, but vanishes after cutoff regularization using $\Tr(\gamma^\mu)=0$ and $\int_{|k|<\Lambda} \, k^\mu/(k^2-m^2)=0$; this is a special case of Furry's theorem in usual~QED.

Similarly, the contribution to \eqref{eq:Sintcompspinor} from the second term in the last line of \eqref{eq:lastlineSintcomp} can be easily manipulated to the form
\begin{align}\label{eq:secondlineSintcomp}
-\frac{\hbar^2\,e^2}3 \, \int_{k,p} \, \e^{-\ii\, p\cdot(x_1-x_2)} \,\tilde\mgreen(p)^2 \, \tilde\sPi_{\mu\sigma}(p) \, \tilde\sPi_{\lambda\nu}(p) \, \Tr\big(\tilde\fgreen^-(p-k) \, \gamma^\sigma\,\tilde\fgreen^+(k)\,\gamma^\lambda\big) \ .
\end{align}
This is recognised as a contribution to the connected two-point function from the fermion bubble at this order, as depicted diagrammatically in Figure~\ref{fig:fdisconn}. We immediately realize that there are no noncommutative contributions in this term.

For the first four terms in \eqref{eq:Sintcompspinor}, we compute the braided Wick expansion
\begin{align*}
(\ii\,\hbar\,\BVL\,\sH)^2\big(\ttv_\sigma\,\tte^{p_1}\odot_\star \ttu_r\,\tte^{p_2}\odot_\star \bar\ttu_{s'}\, \tte^{k_2}\odot_\star \delta_{x_2}^{A_\nu}\big) &= -\hbar^2\,\langle\ttv_\sigma\,\tte^{p_1},\mgreen\,\sPi(\delta_{x_2}^{A_\nu})\rangle_\star\,\langle\fgreen(\ttu_r\,\tte^{p_2}),\bar\ttu_{s'}\,\tte^{k_2}\rangle_\star \\[4pt]
&  = -\hbar^2\,(2\pi)^4\,\delta(p_2+k_2)\,\e^{\,\ii\,p_1\cdot x_2} \, \tilde\mgreen(p_1)\,\tilde\sPi_{\sigma\nu}(p_1) \, \tilde\fgreen_{s'r}^+(k_2)
\end{align*}
and similarly
\begin{align*}
& (\ii\,\hbar\,\BVL\,\sH)^2\big(\ttv_\sigma\,\tte^{p_1}\odot_\star \bar\ttu_{r'}\,\tte^{p_2}\odot_\star \ttu_{s}\, \tte^{k_2}\odot_\star \delta_{x_2}^{A_\nu}\big) \\[4pt] & \hspace{7cm} = -\hbar^2\,(2\pi)^4\,\delta(p_2+k_2)\,\e^{\,\ii\,p_1\cdot x_2} \, \tilde\mgreen(p_1)\,\tilde\sPi_{\sigma\nu}(p_1) \, \tilde\fgreen_{r's}^-(k_2) \ ,
\end{align*}
along with the corresponding expressions for the remaining terms with $\delta_{x_1}^{A_\mu}$. After substitution into \eqref{eq:Sintcompspinor} and resolving the delta-functions, we find that all four terms contribute equally and give
\begin{align}\label{eq:first4linesSintcomp}
-\frac{2\,\hbar^2\,e^2}3 \, \int_{k,p} \, \e^{-\ii\, p\cdot(x_1-x_2)} \,\tilde\mgreen(p)^2 \, \tilde\sPi_{\mu\sigma}(p) \, \tilde\sPi_{\lambda\nu}(p) \, \Tr\big(\tilde\fgreen^-(p-k) \, \gamma^\sigma\,\tilde\fgreen^+(k)\,\gamma^\lambda\big) \ .
\end{align}

Altogether, summing \eqref{eq:secondlineSintcomp} and \eqref{eq:first4linesSintcomp} we arrive at
\begin{align}\label{eq:CG1}
\CG^1_{\mu\nu}(x_1,x_2) = -\hbar^2\,e^2 \, \int_{k,p} \, \e^{-\ii\, p\cdot(x_1-x_2)} \ & \tilde\mgreen(p)^2 \, \tilde\sPi_{\mu\sigma}(p) \, \tilde\sPi_{\lambda\nu}(p)\, \Tr\big(\tilde\fgreen^-(p-k) \, \gamma^\sigma\,\tilde\fgreen^+(k)\,\gamma^\lambda\big) \ .
\end{align}
As in (\ref{eq:secondlineSintcomp}), all noncommutative contributions vanish.

\paragraph{Evaluation of $\boldsymbol{\CG^2_{\mu\nu}(x_1,x_2)}$.}
Now we start by applying the operator $\ii\,\hbar\,\BVL\,\sH$ to \eqref{eq:basicop} using
\begin{align*}
\ii\,\hbar\,\BVL\,\sH\big(\ttu_s\,\tte^{k_2}\odot_\star \bar\ttu_{s'} \, \tte^{k_3}\odot_\star\delta_{x_2}^{A_\nu}\big) = \tfrac23\,\ii\,\hbar\,(2\pi)^4\,\delta(k_2+k_3)\,\tilde\fgreen_{s's}^+(k_3) \ \delta_{x_2}^{A_\nu}
\end{align*}
and the corresponding expression for the terms with $\delta_{x_1}^{A_\mu}$. Upon further application of the operator $\ii\,\hbar\,\BVL\,\sH\circ\{\CS_{\rm int},-\}_\star\circ \sH$, it is easy to see that these expressions contribute to \eqref{eq:photon2ptdef} with the same momentum and spinor index structure as that which led to the disconnected contribution \eqref{eq:disconn}. Hence
\begin{align}\label{eq:CG2}
\CG^2_{\mu\nu}(x_1,x_2) = 0 \ .
\end{align}

\paragraph{Vacuum polarization.}
From \eqref{eq:CG1} and \eqref{eq:CG2} it follows that the one-loop photon two-point function \eqref{eq:photon2ptdef} is given by
\begin{align}\label{eq:photon2ptfinal}
\begin{split}
G^\star_{A_\mu,A_\nu}(x_1,x_2)^{(1)} = -\hbar^2\,e^2 \, \int_{k,p} \, \e^{-\ii\, p\cdot(x_1-x_2)} \ & \tilde\mgreen(p)^2 \, \tilde\sPi_{\mu\sigma}(p) \, \tilde\sPi_{\lambda\nu}(p) \\
& \times \, \Tr\big(\tilde\fgreen^-(p-k) \, \gamma^\sigma\,\tilde\fgreen^+(k)\,\gamma^\lambda\big) \ .
\end{split}
\end{align}
As before, we relate the all orders photon two-point function to the dressed photon propagator through the self-energy $\Pi^{\mu\nu}_{\star}(p)$ by
\begin{align*}
G^\star_{A_\mu,A_\nu}(x_1,x_2) = -\ii\,\hbar\,\int_{(\FR^{1,3})^*} \, \frac{\dd^4p}{(2\pi)^4} \ \e^{-\ii\,p\cdot(x_1-x_2)} \ \bigg(\frac{1}{p^2\,\big(\eta-(1-\xi)\,\frac{p\,\otimes\, p}{p^2}\big)^{-1}-\Pi_\star(p)}\bigg)_{\mu\nu} \ .
\end{align*}
At order $e^2$, the result \eqref{eq:photon2ptfinal} leads to
\begin{align}\label{eq:sameascorfu}
\hspace{-0.1cm} \frac\ii\hbar \, \Pi^{\mu\nu}_{\star 2}(p) = e^2 \, \int_{(\FR^{1,3})^*} \, \frac{\dd^4k}{(2\pi)^4} \ \frac{\Tr\big((\slashed p - \slashed k - m) \, \gamma^\mu \, (\slashed k + m) \, \gamma^\nu\big)}{\big((p-k)^2-m^2\big)\,\big(k^2-m^2\big)}  \ .
\end{align}

The vacuum polarization tensor \eqref{eq:sameascorfu} is {identical} to the result of the photon self-energy calculation in the standard commutative electrodynamics, with precisely the same sign and overall combinatorial factor. As in the scalar field theory of Section~\ref{sec:braidedBV}, the braided Wick theorem contributes phase factors which cancel the noncommutative contributions of the interaction vertex~\eqref{eq:fermionphotonvertex}. 

This result is different from the result of~\cite{CiricDimitrijevic:2022eei}, which was based on a naive application of the conventional Feynman rules following from the action functional \eqref{MWLagranzijani}. There the noncommutative contribution of the form $\cos^2\big(\frac12\,p\cdot \theta\,k\big)$ appears and in this way introduces the UV/IR mixing. The result (\ref{eq:sameascorfu}) suggests that the algebraic quantization using the braided BV formalism is the correct way to quantize braided field theories.  

The result \eqref{eq:sameascorfu} further agrees from the corresponding contribution to the one-loop vacuum polarization in standard noncommutative QED, which coincides with that found in ordinary QED~\cite{U1ChargeQuantization}: in that case each photon-fermion vertex contributes a phase factor which cancel each other in the diagram. Non-trivial corrections in the standard approach come from three more contributing diagrams due to the non-abelian nature of the star-gauge symmetry: the photon bubble, the photon tadpole and the ghost bubble. These involve non-planar diagrams which result in UV/IR mixing~\cite{U1ChargeQuantization,U1UsefulFormula}.

\subsection{Fermion self-energy at one-loop}
\label{sec:fermionselfenergy}

The two-point fermion correlation function is obtained from \eqref{eq:Gspinorgen} by setting $n=2$, $\CA_1=\psi_{s_1}$ and $\CA_2=\bar\psi_{s_2}$. Again the only non-vanishing contributions at one-loop order are given by
\begin{align}\label{eq:fermion2ptdef}
\begin{split}
G^\star_{\psi_{s_1},\bar\psi_{s_2}}(x_1,x_2)^{(1)} &= \langle0|{\rm T}[\psi_{s_1}(x_1)\star \bar\psi_{s_2}(x_2)]|0\rangle_\star^{(1)} \\[4pt]
:\!&= (\ii\,\hbar\,\BVL\,\sH)^2\,\big\{\CS_{\rm int},\sH\,\big\{\CS_{\rm int},\sH\big(\delta_{x_1}^{\psi_{s_1}}\odot_\star\delta_{x_2}^{\bar\psi_{s_2}}\big)\big\}_\star\big\}_\star \\
& \quad \, + \ii\,\hbar\,\BVL\,\sH \,\big\{\CS_{\rm int}, \sH\,(\ii\,\hbar\,\BVL\,\sH)\big\{\CS_{\rm int},\sH\big(\delta_{x_1}^{\psi_{s_1}}\odot_\star\delta_{x_2}^{\bar\psi_{s_2}}\big) \big\}_\star\big\}_\star \\[4pt]
&=:\CG_{s_1s_2}^1(x_1,x_2) + \CG_{s_1s_2}^2(x_1,x_2) \ .
\end{split}
\end{align}

The computation proceeds in a completely analogous way to that of Section~\ref{sec:photonselfenergy}. We start from the non-zero inner products
\begin{align*}
\big\langle\fgreen\big(\delta_{x_1}^{\psi_{s_1}}\big),\ttu_r\,\tte^{k_a}\big\rangle_\star = \e^{\,\ii\, k_a\cdot x_1} \, \tilde\fgreen_{s_1r}^-(k_a) \qquad \mbox{and} \qquad \big\langle\bar\ttu_r\,\tte^{k_a},\fgreen\big(\delta_{x_2}^{\bar\psi_{s_2}}\big)\big\rangle_\star = \e^{\,\ii\, k_a\cdot x_2} \, \tilde\fgreen_{rs_2}^+(k_a) \ ,
\end{align*}
to get the basic operation
\begin{align}\label{eq:basicopspin}
\begin{split}
& \big\{\CS_{\rm int} ,\sH\big(\delta_{x_1}^{\psi_{s_1}}\odot_\star\delta_{x_2}^{\bar\psi_{s_2}} \big) \big\}_\star  \\[4pt]
& \hspace{0.5cm}= \frac12\,\int_{k_1,k_2,k_3} \, V^{\nu;r,r'}(k_1,k_2,k_3) \, \big(-\e^{\,\ii\,k_3\cdot x_2} \, \tilde\fgreen_{r's_2}^+(k_3) \, \delta_{x_1}^{\psi_{s_1}}\odot_\star\ttv_\nu\,\tte^{k_1} \odot_\star \ttu_r\,\tte^{k_2} \\
& \hspace{6cm} + \e^{\,\ii\,k_2\cdot (x_1-\theta\,k_3)} \, \tilde\fgreen_{s_1r}^-(k_2) \, \ttv_\nu\,\tte^{k_1} \odot_\star \bar\ttu_{r'} \, \tte^{k_3} \odot_\star \delta_{x_2}^{\bar\psi_{s_2}} \big) \ .
\end{split}
\end{align}

\paragraph{Evaluation of $\boldsymbol{\CG^1_{s_1s_2}(x_1,x_2)}$.}
After some algebraic manipulations, the operator $\{\CS_{\rm int},-\}_\star\circ\sH$ applied to \eqref{eq:basicopspin} yields
\begin{align}\label{eq:Sintspin2pt}
\begin{split}
& \hspace{-0.3cm} \big\{\CS_{\rm int},\sH\big\{\CS_{\rm int},\sH\big(\delta_{x_1}^{\psi_{s_1}} \odot_\star\delta_{x_2}^{\bar\psi_{s_2}}\big)\big\}_\star\big\}_\star \\[4pt]
& \hspace{-0.3cm} \quad = \frac16\,\int_{p_1,p_2,p_3} \, V^{\mu;t,t'}(p_1,p_2,p_3) \ \int_{k_1,k_2,k_3} \, V^{\nu;r,r'}(k_1,k_2,k_3) \ \Big[(2\pi)^4\,\delta(p_1+k_1) \\
& \hspace{-0.3cm} \hspace{1cm} \times \Big(\tilde\mgreen(p_1)\,\tilde\sPi_{\mu\nu}(p_1) \, \ttu_t\,\tte^{p_2}\odot_\star\bar\ttu_{t'}\,\tte^{p_3} \odot_\star\big( -\e^{\,\ii\,k_3\cdot x_2} \, \tilde\fgreen_{r's_2}^+(k_3) \, \ttu_r\,\tte^{k_2} \odot_\star \delta_{x_1-\theta\,k_3}^{\psi_{s_1}} \\
& \hspace{-0.3cm} \hspace{8cm} + \e^{\,\ii\,k_2\cdot x_1} \, \e^{-\,\ii\,k_1\cdot\theta\, k_2} \, \tilde\fgreen_{s_1r}^-(k_2) \, \bar\ttu_{r'}\,\tte^{k_3}\odot_\star \delta_{x_2}^{\bar\psi_{s_2}}\big) \\
& \hspace{-0.3cm} \hspace{2cm} + \ttv_\mu\,\tte^{p_2}\odot_\star\ttv_\nu\,\tte^{k_3} \odot_\star \big(  \e^{\,\ii\,k_2\cdot x_1} \, \e^{\,-\ii\,k_3\cdot\theta\,p_2 } \, \tilde\fgreen_{s_1r}^-(k_2)\,\tilde\fgreen_{r't}^-(p_1) \, \bar\ttu_{t'}\,\tte^{p_3}\odot_\star\delta_{x_2}^{\bar\psi_{s_2}} \\
& \hspace{-0.3cm} \hspace{5cm} - \e^{\,\ii\,k_2\cdot x_2} \, \e^{\,-\ii\,k_3\cdot\theta\,p_2} \, \tilde\fgreen_{r's_2}^+(k_2) \, \tilde\fgreen_{t'r}^+(p_1)\, \ttu_t\,\tte^{p_3}\odot_\star \delta_{x_1-\theta\,k_2}^{\psi_{s_1}} \big)\Big) \\
& \hspace{-0.3cm} \hspace{2cm} - 2\,\e^{\,\ii\,k_3\cdot x_2+\ii\,p_2\cdot x_1} \, \e^{\ii\,k_2\cdot\theta\,(k_1-p_3)}\,  \tilde\fgreen^+_{r's_2}(k_3)\,\tilde\fgreen^-_{s_1t}(p_2) \\
& \hspace{-0.3cm} \hspace{8cm} \times \ttu_r\,\tte^{k_2} \odot_\star \bar\ttu_{t'}\,\tte^{p_3} \odot_\star  \ttv_\mu\,\tte^{p_1}\odot_\star\ttv_\nu\,\tte^{k_1}\Big] \ .
\end{split}
\end{align}
Next we apply the operator $(\ii\,\hbar\,\BVL\,\sH)^2$ to each term, and use the braided Wick expansions \eqref{4PointFull} and \eqref{eq:braided4pt}.

\begin{figure}[h]%
\centering
\includegraphics[width=0.3\textwidth]{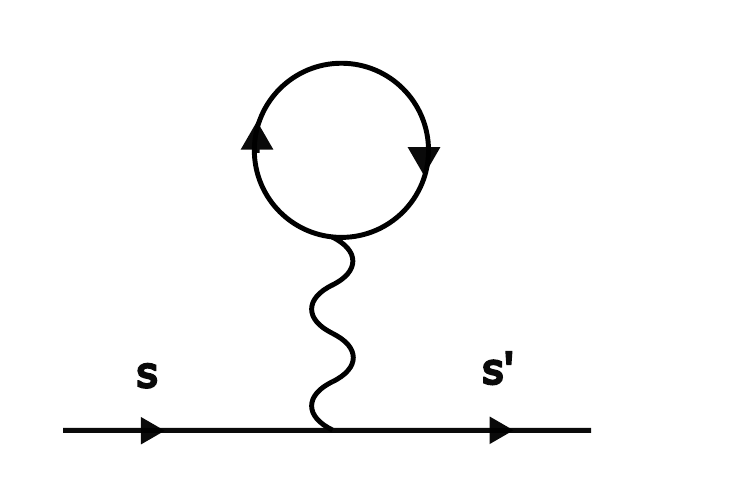} \qquad
\includegraphics[width=0.4\textwidth]{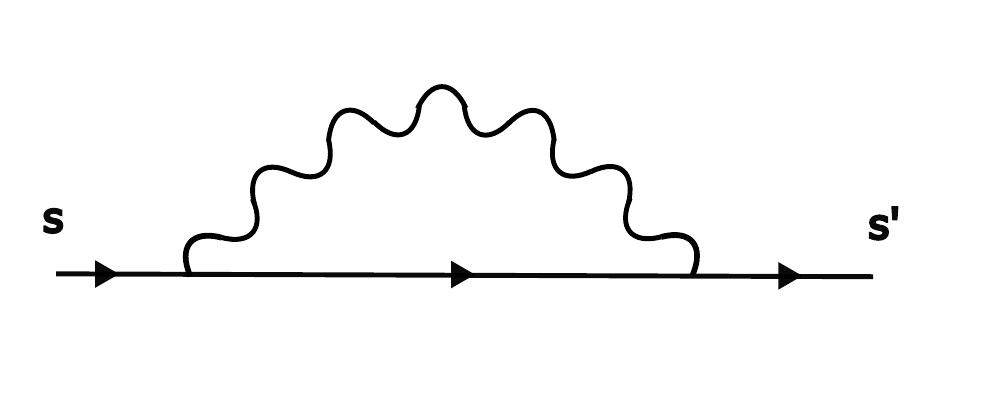}
\vspace{-5mm}
\caption{\small The fermion tadpole (left) and one-particle irreducible (right) contributions to the fermion two-point function at one-loop order. The tadpole contribution vanishes.}
\label{fig:ferm2pt}
\end{figure}

Looking at the first term of \eqref{eq:Sintspin2pt}, we find two contributions
\begin{align}\label{eq:spin2pt1stterm}
\begin{split}
& (\ii\,\hbar\,\BVL\,\sH)^2\big( \ttu_t\,\tte^{p_2}\odot_\star\bar\ttu_{t'}\,\tte^{p_3} \odot_\star\ttu_r\,\tte^{k_2} \odot_\star \delta_{x_1-\theta\,k_2}^{\psi_{s_1}}\big) \\[4pt]
& \ \ = \hbar^2\,(2\pi)^4\,\big(\e^{\,\ii\,k_2\cdot x_1} \, \delta(p_2+p_3) \, \tilde\fgreen_{t't}^+(p_3)\,\tilde\fgreen_{s_1r}^-(k_2) + \e^{\,\ii\,p_2\cdot (x_1-\theta\,k_2)} \, \delta(p_3+k_2) \, \tilde\fgreen^+_{t'r}(k_2)\, \tilde\fgreen^-_{s_1t}(p_2) \big) \ .
\end{split}
\end{align}
Upon substitution into \eqref{eq:Sintspin2pt}, the spinor index structure of the first term in \eqref{eq:spin2pt1stterm} is easily seen to give a factor
\begin{align}\label{eq:tadpoleferm}
\Tr\big(\gamma^\mu\,\tilde\fgreen^+(p_3)\big) \ \tilde\mgreen(p_1) \, \tilde\sPi_{\mu\nu}(p_1) \ \big(\tilde\fgreen^-(k_2)\,\gamma^\nu\,\tilde\fgreen^+(k_3)\big)_{s_1s_2} \ .
\end{align}
This is recognised as a contribution to the spinor two-point function from the fermion tadpole, which vanishes by an argument similar to that which led to the vanishing of the photon one-point function in Section~\ref{sec:photonselfenergy}. The second term in \eqref{eq:spin2pt1stterm} instead contributes to the 1PI two-point function. These are depicted diagrammatically in Figure~\ref{fig:ferm2pt}. 

Similary, the second term of \eqref{eq:Sintspin2pt} splits into a tadpole contribution (which vanishes) and a 1PI contribution. The remaining terms in \eqref{eq:Sintspin2pt} all yield spinor index structures that contribute only to the 1PI correlator. For example, the braided Wick expansion of the third term is given by
\begin{align*}
& (\ii\,\hbar\,\BVL\,\sH)^2\big(\ttv_\mu\,\tte^{p_2} \odot_\star \ttv_\nu\,\tte^{k_3} \odot_\star \bar\ttu_{t'}\,\tte^{p_3} \odot_\star \delta_{x_2}^{\bar\psi_{s_2}}\big) \\[4pt]
& \hspace{7cm} = \hbar^2 \, (2\pi)^4 \, \delta(p_2+k_3)  \, \e^{\,\ii\,p_3\cdot x_2} \, \tilde\mgreen(k_3) \, \tilde\sPi_{\mu\nu}(k_3) \, \tilde\fgreen_{t's_2}^+(p_3) \ ,
\end{align*}
and similarly for the remaining terms.

After some tedious algebraic manipulations and resolution of all delta-functions, one finds that all non-vanishing contributions are equal and result in
\begin{align}\label{eq:fermCG1}
\begin{split}
\CG^1_{s_1s_2}(x_1,x_2) &= -\hbar^2\,e^2 \, \int_{k,p}\, \e^{-\ii\,p\cdot(x_1-x_2)}  \, \tilde\mgreen(k)\,\tilde\sPi_{\mu\nu}(k) \, \big(\tilde \fgreen^+(p) \, \gamma^\mu \, \tilde \fgreen^-(k-p) \, \gamma^\nu \, \tilde \fgreen^+(p)\big)_{s_1s_2} \ .
\end{split}
\end{align}

\paragraph{Evaluation of $\boldsymbol{\CG^2_{s_1s_2}(x_1,x_2)}$.}
Now we start by applying $\ii\,\hbar\,\BVL\,\sH$ to \eqref{eq:basicopspin} using
\begin{align*}
\ii\,\hbar\,\BVL\,\sH\big(\delta_{x_1}^{\psi_{s_1}}\odot_\star \ttv_\nu\,\tte^{k_1} \odot_\star \ttu_r\,\tte^{k_2} \big) = \tfrac23\,\ii\,\hbar\,\e^{\,\ii\,k_2\cdot x_1} \, \e^{-\ii\,k_1\cdot\theta\,k_2} \, \tilde\fgreen^-_{s_1r}(k_2) \, \ttv_\nu\,\tte^{k_1} \ ,
\end{align*}
and similarly for the second term in \eqref{eq:basicopspin}. Applying the operator $\ii\,\hbar\,\BVL\,\sH\circ\{\CS_{\rm int},-\}_\star\circ \sH$ to these expressions, it is easy to see that the momentum and spinor index structures of their contributions to \eqref{eq:basicopspin} are of the same form as that in \eqref{eq:tadpoleferm}, and so lead to vanishing fermion tadpole contributions. It therefore follows as in Section~\ref{sec:photonselfenergy} that
\begin{align}\label{eq:fermCG2}
\CG^2_{s_1s_2}(x_1,x_2) = 0 \ .
\end{align}

\paragraph{Self-energy.}
From \eqref{eq:fermCG1} and \eqref{eq:fermCG2} it follows that the fermion two-point function \eqref{eq:fermion2ptdef} is given by
\begin{align}\label{eq:ferm2ptfinal}
\begin{split}
G^\star_{\psi_{s_1},\bar\psi_{s_2}}(x_1,x_2)^{(1)} &= -\hbar^2\,e^2 \, \int_{k,p} \, \e^{-\ii\,p\cdot(x_1-x_2)} \,\tilde\mgreen(k)\,\tilde\sPi_{\mu\nu}(k) \\
& \hspace{5cm} \times \big(\tilde \fgreen^+(p) \, \gamma^\mu \, \tilde \fgreen^-(k-p) \, \gamma^\nu \, \tilde \fgreen^+(p)\big)_{s_1s_2} \ .
\end{split}
\end{align}
The exact two-point function defines the dressed fermion propagator through the Fourier transformation
\begin{align*}
G^\star_{\psi_{s_1},\bar\psi_{s_2}}(x_1,x_2)  = -\ii\,\hbar \, \int_{(\mink)^*} \, \frac{\dd^4p}{(2\pi)^4} \ \e^{-\ii\,p\cdot(x_1-x_2)} \ \bigg(\frac1{\slashed p-m-\Sigma_\star(p)}\bigg)_{s_1s_2} \ ,
\end{align*}
where $\Sigma_\star(p)$ is the fermion self-energy. At order $e^2$, the result \eqref{eq:ferm2ptfinal} in Feynman gauge gives
\begin{align}\label{eq:fermselfenergy1loop}
\frac\ii\hbar \, \Sigma_{\star 2}(p) = e^2 \, \int_{(\FR^{1,3})^*} \, \frac{\dd^4k}{(2\pi)^4} \, \frac{\gamma^\mu\, (\slashed p-\slashed k - m) \, \gamma_\mu}{k^2\,\big((p-k)^2 - m^2\big)}  \ .
\end{align}

As with the braided correction to the photon self-energy, the fermion self-energy \eqref{eq:fermselfenergy1loop} agrees with the standard result in the commutative electrodynamics: it has no noncommutative contributions and no UV/IR mixing. It also agrees with the calculation in standard noncommutative QED, where the phases cancel between the two vertices and there is no difference from the fermion self-energy in ordinary QED~\cite{U1ChargeQuantization}. In fact, also in the standard approach the fermion self-energy does not receive any noncommutative corrections and so does not display UV/IR mixing.

\section{Conclusions and outlook}
\label{sec:outlook}

Braided noncommutative field theories are defined by deforming the $L_\infty$-structure of a field theory to a braided $L_\infty$-algebra~\cite{BraidedLinf}. Given that in the standard noncommutative quantum field theories the problem of UV/IR mixing arises, and that as a result these field theories cannot be renormalized, in this paper we started looking at the quantum properties of braided noncommutative field theories. Lacking any other currently available quantization scheme for braided fields, we exploited a braided deformation of the BV formalism \cite{SzaboAlex} and homological perturbation theory to calculate quantum correlation functions. This framework is what we call braided quantum field theory.

As a first example, in Section~\ref{sec:braidedBV} we discussed braided $\phi^4$ scalar field theory. In the free field theory, correlation functions follow from the braided Wick theorem. In the commutative limit they reduce to the free correlation functions for the undeformed theory. The interaction vertex (\ref{eq:Vint}) is the same as in the standard noncommutative $\phi^4$-theory. However, the combination of the noncommutative interaction vertex with the braided Wick theorem conspire to render the interacting two-point function at one-loop order (\ref{eq:1loopselfenergy}) free of noncommutative corrections, and in that way also free of UV/IR mixing. We also notice that, contrary to the standard noncommutative field theories, no non-planar diagrams appear. In order to decide whether or not braided scalar quantum field theories are renormalizable, one needs to extend our analysis beyond one-loop order and also to include higher-point correlation functions. In future work we plan to investigate these problems in more detail.

As a first step towards understanding  gauge theories, in Section \ref{sec:braidedQED} we studied an example of a braided $\sU(1)$ gauge theory coupled to a Dirac fermion. We call this model braided QED. Unlike the standard noncommutative $\sU(1)$ gauge theory \cite{U1ChargeQuantization, U1UsefulFormula} where the photon is self-interacting and non-planar diagrams appear, braided QED is an abelian gauge theory. The photon does not interact with itself and the only interaction vertex is the photon-fermion vertex. As a consequence, no non-planar diagrams appear and the noncommutative contributions come from the interaction vertex (\ref{eq:fermionphotonvertex}) and the braided Wick theorem. Similarly to braided scalar field theory, the photon two-point function at one-loop (\ref{eq:sameascorfu}) has no noncommutative contribution and it is free of the UV/IR mixing. The fermion two-point function at one-loop (\ref{eq:fermselfenergy1loop}) also has no noncommutative contribution and shows no UV/IR mixing. We discussed these two-point functions in detail in Sections~\ref{sec:photonselfenergy} and~\ref{sec:fermionselfenergy}, respectively. To gain some further insight into the ultraviolet behaviour of braided QED, the 3-point function at one-loop (vertex correction) and the corresponding beta-function at one-loop should be calculated. This is also slated into our future work plans.

The present paper clearly shows that braided noncommutative quantum field theories have better ultraviolet behaviour than the corresponding standard noncommutative field theories. We plan to extend our analysis to non-abelian gauge theories. Besides correlation functions, we also plan to discuss scattering amplitudes, following the approach of~\cite{Arvanitakis:2019ald, Macrelli:2019afx} but using the braided $L_\infty$-algebra structure of our theories and braided homological perturbation theory.

\appendix

\renewcommand{\theequation}{\Alph{section}.\arabic{equation}}
\setcounter{equation}{0}

\section{Drinfel'd twist deformation theory}
\label{app:Drinfeld}

In this appendix we briefly summarise the basic theory of Drinfel'd
twists as needed in this paper, more details can be found in \cite{MajidBook, SLN}.

We start from the Lie
algebra of vector fields $\frv:=\Gamma(TM)$ on a manifold $M$, which
generate infinitesimal diffeomorphisms of $M$. The enveloping algebra $U\frv$ is naturally a cocommutative Hopf
algebra with coproduct $\Delta:U\frv\to U\frv\otimes U\frv$, counit
$\varepsilon:U\frv\to\FC $ and antipode $S:U\frv\to U\frv$ defined on
generators by
\begin{align*}
\Delta(\xi)&= \xi\otimes 1 + 1\otimes \xi\qquad \mbox{and} \qquad
             \Delta(1)=1\otimes 1 \ , \\[4pt]
  \varepsilon(\xi)&=0 \qquad \mbox{and} \qquad \varepsilon(1)=1 \ ,
  \\[4pt]
S(\xi)&=-\xi \qquad \mbox{and} \qquad S(1)=1 \ ,
\end{align*}
for all $\xi\in \frv$. The maps $\Delta$ and $\varepsilon$ are
extended as algebra homomorphisms, and $S$ as an algebra
antihomomorphism to all of $U\frv$. We adopt the standard
Sweedler notation $\Delta(X)=:X_{\textrm{\tiny(1)}}\otimes
X_{\textrm{\tiny(2)}}$ (with summations understood) to abbreviate the
coproduct of $X\in U\frv$.

A Drinfel'd twist is a normalized two-cocycle of the Hopf
algebra $U\frv[[\nu]]$, the formal power series in a deformation parameter $\nu$ with coefficients valued in $U\frv$. By this we mean an invertible element
$\CF\in U\frv[[\nu]]\otimes U\frv[[\nu]]$ satisfying the cocycle
condition
\begin{align*}
\CF_{12}\,(\Delta\otimes \id)\CF=\CF_{23}\,(\id\otimes \Delta)\CF \ ,
\end{align*}
where $\CF_{12}=\CF\otimes 1$ and $\CF_{23}=1\otimes \CF$, together
with the normalization condition
\begin{align*}
  (\varepsilon\otimes \id)\CF=1=(\id\otimes \varepsilon)\CF \ .
\end{align*}
We write the power series expansion of the twist as $\CF=:\sff^\alpha\otimes
\sff_\alpha\in U\frv[[\nu]]\otimes U\frv[[\nu]]$, with the sum over
$\alpha$ understood. Then the cocycle
condition may be written in Sweedler notation as
\begin{align}\label{eq:cocyclesw}
\sff^\alpha\,\sff^\beta_{\textrm{\tiny(1)}}\otimes\sff_\alpha\,\sff^\beta_{\textrm{\tiny(2)}}\otimes\sff_\beta
  =
  \sff^\beta\otimes\sff^\alpha\,{\sff_\beta}_{\textrm{\tiny(1)}}\otimes\sff_\alpha\,{\sff_\beta}_{\textrm{\tiny(2)}}
  \ ,
\end{align}
and the normalization condition as
\begin{align}\label{eq:twistnorm}
\varepsilon(\sff^\alpha)\,\sff_\alpha = 1 = \sff^\alpha\,\varepsilon(\sff_\alpha) \ .
\end{align}
As a consequence, the inverse twist $\CF^{-1}=: \bar{\sff}^\alpha\otimes
\bar{\sff}_\alpha\in U\frv[[\nu]]\otimes U\frv[[\nu]]$ satisfies
similar conditions.

A Drinfel'd twist $\CF$ defines a new Hopf algebra structure on the
universal enveloping algebra $U\frv[[\nu]]$, which we denote by
$U_\CF\frv$. As algebras, $U_\CF\frv=U\frv[[\nu]]$ and also the
counit of $U_\CF\frv$ is the same as the counit $\varepsilon$ of
$U\frv[[\nu]]$. The new coproduct $\Delta_\CF$ and antipode $S_\CF$
of $U_\CF\frv$ are given by
\begin{align*}
\Delta_{\CF}(X):= \CF \, \Delta(X) \, \CF^{-1} \qquad \mbox{and}
  \qquad S_{\CF}(X):=\sff^\alpha\,S(\sff_\alpha)\, S(X)\,
  S(\bar{\sff}^\beta)\,\bar{\sff}_\beta \ ,
\end{align*}
for all $X\in U\frv[[\nu]]$. For $X\in U_\CF\frv$, we adopt the Sweedler notation $\Delta_\CF(X) =: X_{\bar\swone}\otimes X_{\bar\swtwo}$ to distinguish the twisted and untwisted coproducts. The invertible $\RR$-matrix $\RR\in
U\frv[[\nu]]\otimes U\frv[[\nu]]$ is induced by the twist as
\begin{align}
\RR=\CF_{21}\, \CF^{-1}=:\sfR^\alpha\otimes\sfR_\alpha \ ,
\end{align}
where $\CF_{21}=\tau(\CF)=\sff_\alpha\otimes\sff^\alpha$ is the twist with its legs swapped. It is easy to see that the $\RR$-matrix is triangular, that is
\begin{align*}
  \RR_{21} = \RR^{-1} = \sfR_\alpha\otimes\sfR^\alpha \ ,
\end{align*}
  and moreover that 
\begin{align*}
(\Delta_{\CF}\otimes \id) \RR = \RR_{13}\, \RR_{23} \qquad \mbox{and} \qquad
(\id\otimes \Delta_{\CF})\RR= \RR_{13}\, \RR_{12} \ , 
\end{align*}
where $\RR_{13}=\sfR^\alpha\otimes 1 \otimes\sfR_\alpha$, or in Sweedler notation
\begin{align}\label{eq:Rmatrixidsw}
\sfR^\alpha_{\bar{\textrm{\tiny(1)}}}\otimes \sfR^\alpha_{\bar{\textrm{\tiny(2)}}}\otimes \sfR_\alpha= \sfR^\beta\otimes \sfR^\alpha\otimes
  \sfR_\beta\, \sfR_\alpha \qquad \mbox{and} \qquad \sfR^\alpha\otimes
  {\sfR_\alpha}_{\bar{\textrm{\tiny(1)}}}\otimes {\sfR_\alpha}_{\bar{\textrm{\tiny(2)}}} =
  \sfR^\beta\,\sfR^\alpha\otimes \sfR_\alpha \otimes \sfR_\beta \ .
\end{align}
The $\RR$-matrix also satisfies the Yang--Baxter equation 
\begin{align}\label{eq:YangBaxter}
\RR_{12} \, \RR_{13} \, \RR_{23}= \RR_{23} \, \RR_{13} \, \RR_{12} \ .
\end{align}

Drinfel'd twist deformation quantization consists in twisting the enveloping
Hopf algebra $U\frv$ to a non-cocommutative Hopf algebra $U_\CF\frv$,
while simultaneously twisting all of its modules~\cite{SLN}. A $U\frv$-module algebra is an algebra $(\CA,\mu)$ with a
$U\frv$-action $\triangleright: U\frv\otimes \CA
\rightarrow \CA$ which is compatible with the algebra multiplication
via the coproduct $\Delta$, that is
\begin{align*}
X\mu(a\otimes b)
  = \mu\big(\Delta(X)(a\otimes b)\big)
\end{align*}
for all $X\in U\frv$ and $a,b\in\CA$, where $\mu:\CA\otimes\CA\to\CA$
is the product on $\CA$. We will usually drop the symbol
$\triangleright$ to simplify the notation. This condition means in particular that
vector fields $\xi\in\frv$ act on $(\CA,\mu)$ as:
\begin{align*}
  \xi\big(\mu(a\otimes b)\big) = \mu\big(\xi(
  a)\otimes b\big) + \mu\big(a\otimes\xi( b)\big) \ .
\end{align*}

If $(\CA,\mu)$ is a (left) $U\frv$-module algebra, then we
can deform the product $\mu$ on $\CA$ by precomposing it with the
inverse of the twist
$\CF$ to get a new product
\begin{align}\label{eq:mustar}
\mu_\star (a\otimes b) = \mu\circ\CF^{-1}(a\otimes b) =
  \mu\big(\bar\sff^\alpha(a)\otimes\bar\sff_\alpha(b)\big) \ ,
\end{align}
for $a,b\in\CA$, where on the right-hand side we extend $\mu$ to
$\CA[[\nu]]\otimes\CA[[\nu]] \cong(\CA\otimes\CA)[[\nu]]$ by
applying it term by term to the coefficients of a formal power series.
The cocycle condition on $\CF$ guarantees that this
produces an associative star-product $\mu_\star$ on $\CA[[\nu]]$, and it generally defines a
noncommutative $U_\CF\frv$-module algebra $(\CA[[\nu]],\mu_\star)$:
\begin{align*}
X\big(\mu_\star(a\otimes b)\big) =
  \mu_\star\big(\Delta_\CF(X)(a\otimes b)\big) \ ,
\end{align*}
for all $X\in U\frv$ and $a,b\in\CA$.
We denote $(\CA,\mu)$ by $\CA$ and
$(\CA[[\nu]],\mu_\star)$ by $\CA_\star$ for brevity. 

If the algebra $\CA$ is commutative,
then $\CA_\star$ is braided commutative: the noncommutativity of
$\CA_\star$ is controlled by the $\RR$-matrix as
\begin{align*}
\mu_\star(a\otimes b) = \mu_\star\big(\sfR_\alpha(b)\otimes\sfR^\alpha(a)\big) \ ,
\end{align*}
which is easily proven by recalling that $\RR=\CF_{21}\,
\CF^{-1}$. 

\section{Dirac matrices in four dimensions}
\label{app:Dirac}

Let $\CCS$ denote the complex spinor representation of $\sSpin(1,3)$. The $16$-dimensional  $\FC$-algebra $\sEnd(\CCS)$ can be identified with the complex Clifford algebra ${\sf Cl}(\FR^{1,3})\otimes\FC$. For this, we construct complex $4{\times}4$ matrices $\gamma^\mu$ with $\mu=0,1,2,3$ which satisfy the anticommutation relations
\begin{equation*}
\{\gamma^\mu,\gamma^\nu\}=2\,\eta^{\mu\nu} \, \unit_4 \ .\nn
\end{equation*}
Then a basis of the complex vector space $\sEnd(\CCS)$ consists of the matrices $\unit_4$, $\gamma_\mu$, $\gamma_\mu\,\gamma_\nu$ with $\mu<\nu$ and $\gamma_\mu\,\gamma_\nu\,\gamma_\lambda$ with $\mu<\nu<\lambda$, together with the chirality matrix
\begin{equation}
\gamma^5= \ii\,\gamma ^0\,\gamma ^1\,\gamma ^2\,\gamma ^3 = -\tfrac{\ii}{4!}\,\epsilon_{\mu\nu\lambda\rho}\,
    \gamma^\mu\,\gamma^\nu\,\gamma^\lambda\,\gamma^\rho \nn 
\end{equation}
which satisfies $\{\gamma^5,\gamma^\mu\}=0$, where $\epsilon^{\mu\nu\lambda\rho}$ is the Levi--Civita symbol with the convention $\epsilon^{0123} = +1$.

For any covector $k\in(\FR^{1,3})^*$, we use the Feynman slash notation to write the corresponding element of $\sEnd(\CCS)$ as $\slashed k :=  k_\mu\,\gamma^\mu = \eta_{\mu\nu}\,k^\mu\,\gamma^\nu$. The Dirac operator can be expressed as $\ii\,\slashed\partial := \ii\,\gamma^\mu \, \partial_\mu$. 

The Dirac representation of $\sEnd(\CCS)$ is defined by
\begin{equation}
 \gamma _0=\bigg(\begin{matrix}
\unit_2 & 0 \\
0 & -\unit_2
\end{matrix}\bigg) \qquad \mbox{and} \qquad \gamma^i = \bigg(\begin{matrix}
0 & \sigma^i \\
-\sigma^i & 0
\end{matrix}\bigg) \ , \nn
\end{equation}
where $\sigma^i$ for $i=1,2,3$ are the standard $2{\times}2$ Pauli spin matrices 
\begin{align*}
\sigma^1 = \bigg( \begin{matrix}
0 & 1 \\ 1 & 0 
\end{matrix} \bigg) \ , \quad \sigma^2 = \bigg( \begin{matrix}
0 & -\ii \\ \ii & 0 
\end{matrix} \bigg) \qquad \mbox{and} \qquad \sigma^3 = \bigg( \begin{matrix}
1 & 0 \\ 0 & -1 
\end{matrix} \bigg)
\end{align*}
satisfying $\sigma^i\,\sigma^j = \delta^{ij} \, \unit_2 + \ii\,\epsilon^{0ijk}\, \sigma_k$. In this basis the Dirac matrices have the reality properties
\begin{equation}
\gamma_\mu^{\dagger}=\gamma^0\,\gamma_\mu \, \gamma^0  \qquad \mbox{and} \qquad \gamma_5^{\dagger}=\gamma_5=\gamma^5=\gamma_5^{-1} \ .\nn
\end{equation}
One also easily derives the product formula
\begin{equation}
\gamma_\mu\,\gamma_\nu\,\gamma_\lambda
= (\eta_{\mu\nu}\,\eta_{\lambda\rho} -\eta_{\mu\lambda}\,\eta_{\nu\rho}
+\eta_{\mu\rho}\,\eta_{\nu\lambda} )\,\gamma^\rho
+ \ii\,\epsilon_{\mu\nu\lambda\rho}\,\gamma^\rho\,\gamma_5  \ . \label{Id_1} 
\end{equation}

Let $\Tr$ denote the trace in the Dirac representation. Then one easily derives the trace identities
\begin{align}
\begin{split}
\Tr (\gamma^\mu) &=0 = \Tr (\gamma^\mu\,\gamma^\nu\,\gamma^\lambda) \ , \\[4pt]
\Tr (\gamma^\mu\,\gamma^\nu) &= 4\,\eta^{\mu\nu} \ , \\[4pt]
\Tr (\gamma^\mu\,\gamma^\nu\,\gamma^\lambda\,\gamma^\rho) &= 4\,\big( \eta^{\mu\nu}\,\eta^{\lambda\rho} - \eta^{\mu\lambda}\,\eta^{\nu\rho} + \eta^{\mu\rho}\,\eta^{\nu\lambda} \big) \ .\label{B3}
\end{split}
\end{align}

\end{document}